\documentclass[10pt,english,tightenlinesletterpaper,groupaddress,nofootinbib]{revtex4-2}
\usepackage{courier}

\usepackage[T1]{fontenc}
\usepackage[latin9]{inputenc}
\usepackage[letterpaper]{geometry}
\usepackage{color}
\usepackage{babel}
\usepackage{mathrsfs}
\usepackage{amsmath}
\usepackage{amssymb}
\usepackage{graphicx}
\usepackage{esint}
\usepackage[unicode=true,pdfusetitle,
 bookmarks=true,bookmarksnumbered=false,bookmarksopen=false,
 breaklinks=true,pdfborder={0 0 1},backref=false,colorlinks=true]
 {hyperref}
\hypersetup{
 linkcolor=blue,urlcolor=blue,citecolor=blue,linktoc=all}

\makeatletter

\newcommand{\lyxdot}{.}

\makeatother

\begin{document}
\title{Computation of Gravitational Particle Production Using Adiabatic Invariants}
\author{Edward E. Basso}
\email{ebasso@wisc.edu}

\affiliation{Department of Physics, University of Wisconsin-Madison, Madison, WI
53706, USA}
\author{Daniel J. H. Chung}
\email{danielchung@wisc.edu}

\affiliation{Department of Physics, University of Wisconsin-Madison, Madison, WI
53706, USA}
\begin{abstract}
Analytic and numerical techniques are presented for computing gravitational
production of scalar particles in the limit that the inflaton mass
is much larger than the Hubble expansion rate at the end of inflation.
These techniques rely upon adiabatic invariants and time modeling
of a typical inflaton field which has slow and fast time variation
components. A faster computation time for numerical integration is
achieved via subtraction of slowly varying components that are ultimately
exponentially suppressed. The fast oscillatory remnant results in
production of scalar particles with a mass larger than the inflationary
Hubble expansion rate through a mechanism analogous to perturbative
particle scattering. An improved effective Boltzmann collision equation
description of this particle production mechanism is developed. This
model allows computation of the spectrum using only adiabatic invariants,
avoiding the need to explicitly solve the inflaton equations of motion.
\end{abstract}
\maketitle
{\footnotesize{}\tableofcontents{}}{\footnotesize\par}

\section{Introduction}

Gravitational production of superheavy hidden sector particles during
the transition out of the inflationary quasi-de Sitter (quasi-dS)
era remains a plausible mechanism of dark matter production (e.g.\,\citep{Chung:1998zb,Kuzmin:1998uv,Chung:2001cb,Shiu:2003ta,Aloisio:2006yi,Berezinsky:2008bg,Fedderke:2014ura,Ema:2018ucl,Hashiba:2018tbu,Hashiba:2018iff,Ema:2019yrd,Li:2019ves,Modak:2019jbg,Herring:2020cah,Ling:2021zlj}).\footnote{In addition, see refs.\,\citep{Benakli:1998ut,Han:1998pa,Hamaguchi:1998wm,Leontaris:1999ce,Hamaguchi:1999cv,Uehara:2001wd,Csaki:1999mp,Coriano:2001mg,Shiu:2003ta,Berezinsky:2008bg,Park:2013bza,Kannike:2016jfs,Harigaya:2016vda,Redi:2020ffc,Gross:2020zam}
for examples of superheavy dark matter motivated by UV model constructions.} However, gravitational production of super-Hubble-mass particles
was largely considered phenomenologically uninteresting due to an
exponential suppression during inflation.\footnote{Here super-Hubble-mass refers to a mass that is larger than the Hubble
expansion rate during the quasi-dS era.} This outlook has changed after it was found \citep{Ema:2015dka,Ema:2016hlw,Ema:2018ucl}
that unsuppressed production can take place \emph{after} inflation
via 
\begin{equation}
\mbox{coherent oscillating inflatons }\phi\rightarrow\mbox{graviton}\rightarrow\chi\mbox{ particles }\label{eq:intro eq 1}
\end{equation}
for some inflationary models if $m_{\phi}>m_{\chi}\gg H_{\mathrm{inf}}$,
where $H_{\mathrm{inf}}$ is the inflationary Hubble expansion rate,
and $m_{\phi}$ and $m_{\chi}$ are the masses of the respective particles.
For example, this can occur in hilltop models such as those of refs.\,\citep{Boubekeur:2005zm,Kohri:2007gq,Lin:2008ys,Armillis:2012bs,Ling:2021zlj}.

Intuitively, the fast oscillatory component in an otherwise slow Hubble
expansion rate $H$ after the quasi-dS era leads to the production
of particles. In contrast with preheating related particle production
(e.g.~ \citep{Traschen:1990sw,Shtanov:1994ce,Kofman:1994rk,Kofman:1997yn,Giudice:1999fb,Bai:2020ttp}),
this scenario requires mediation by the massless gravitational sector,
which cannot be integrated out. Also, unlike the scenarios of refs.\,\citep{Redi:2020ffc,Mambrini:2021zpp,Tang:2017hvq,Haro:2019umj},
the phase space structure of the initial inflaton degrees of freedom
is fixed by the Bunch-Davies/adiabatic vacuum prescription during
inflation. It is in this sense that the particle production considered
in ref.\,\citep{Ema:2018ucl} might naively be thought of as part
of the gravitational particle production resulting from the transition
out of the quasi-dS era.

In ref.\,\citep{Chung:2018ayg}, an analytic formalism for computing
the particle production in this scenario was constructed based on
a double expansion in $H/m_{\phi}\ll1$ and $H/m_{\chi}\ll1$. In
using the formalism, a Gaussian spectral model was introduced to obtain
a compact formula. It was based on the assumptions that the Fourier
spectrum of the fast time variation of the inflaton field is approximately
Gaussian, and that most of the particle production takes place during
the first few oscillations. A production rate was then estimated using
a naive matching to the Fermi's golden rule, and the Boltzmann equation
was integrated to obtain the final number density.

However, there were at least two problems with this. Firstly, the
matching to Fermi's golden rule was done in a non-systematic manner,
and neglected the long time behavior of the dynamics, although it
certainly is correct to $O(1)$ accuracy. Secondly, the high $k$
part of the particle production spectrum requires the fast oscillation
component of $H$ to be tracked for longer than a few oscillation
periods. This is because the time $t_{*}$ at which the time-integral
in the formalism picks up the dominant contribution to the particle
production amplitude is when 
\begin{equation}
\frac{k^{2}}{a^{2}(t_{*})}+m_{\chi}^{2}\approx m_{\phi}^{2},\label{eq:onshellstart}
\end{equation}
corresponding to the on-shell condition of eq.\,(\ref{eq:intro eq 1}).

One of the main results of this paper is to correct these shortcomings.
More specifically, using the technique of adiabatic invariants, we
construct an accurate time model instead of a spectral model. We will
see that these will lead to a 2+$O(1)$ factor correction to the final
spectrum compared to that of ref.\,\citep{Chung:2018ayg}.

It is important to note that the adiabatic invariant formalism of
this paper is distinct from the adiabatic vacuum of refs.\,\citep{Habib:1999cs,Kluger:1998bm,Birrell:1982ix}.
In particular, the latter is essentially the WKB approximation, and
applies only to linear wave equations, while the former requires oscillatory
motion, and can be applied to non-linear Hamiltonian dynamics. Furthermore,
in the context of gravitational particle production, the WKB approach
is used to solve the $\chi_{k}$ mode equation, and is controlled
by an expansion in $H/m_{\chi}$. In contrast, the adiabatic invariant
formalism of this paper is used to analyze the dynamics of the inflaton
$\phi$ and the scale factor $a$, and is controlled by $H/m_{\phi}$.
This then determines the $\chi$ mode dispersion, which controls its
particle production.

A second main result of this paper demonstrates that one can use the
formalism of ref.\,\citep{Chung:2018ayg} to compute the spectrum
numerically more efficiently than the brute force methods used in
the literature. We note that what is expensive to compute numerically
is the slow time-varying component weighted integration over fast
oscillating functions which converge slowly to a negligible contribution.\footnote{The suppression behaves as $e^{-m_{\chi}/H}$.}
Instead, the formalism of ref.\,\citep{Chung:2018ayg} already isolates
the fast time variation such that its numerical integration can be
more than $O(1000)$ faster in convergence. The relative gain in efficiency
depends on the amplitude of the spectrum.

It is important to note that although the original motivation for
this scenario is the production of dark matter, the computational
tools presented in this work are useful for constraining top down
constructions as well. For example, in contrast with moduli problem
scenarios such as those of refs.\,\citep{Felder:1999wt,Giudice:1999yt},
moduli much heavier than the inflationary expansion rate can be gravitationally
produced as well during the first few oscillations out of inflation
\citep{Nakayama:2019yjv} (where the largest number is produced in
the first oscillation). Furthermore, even if the particles decay too
rapidly to be dark matter, they can have observable consequences as
long as they are sufficiently long lived, as discussed for example
in refs.\,\citep{Kuzmin:1998uv,Blasi:2001hr,Aloisio:2006yi,Albuquerque:2010bt,Rott:2014kfa,Marzola:2016hyt,Kachelriess:2018rty,Hooper:2019ytr,Bauer:2020jay,Kalashev:2020hqc,Anchordoqui:2021crl,Ishiwata:2019aet}.\footnote{In the case of the gravity wave probe presented in \citep{Gouttenoire:2019rtn},
the present mechanism for dark matter production apparently does not
lead to an early enough matter domination in the plausible scenarios
they considered.}

The order of presentation will be as follows. In section \ref{sec:Previous-work-on},
we summarize the previous work of separating the $H$ time scale from
the $m_{\phi}$ time scale and explain its shortcomings \citep{Chung:2018ayg}.
Here, we also establish the notation for this paper. In section \ref{sec:Adiabatic-invariant-1},
we explain how the adiabatic invariant can be constructed and how
to use it to compute the long time behavior of slowly varying fields.
This tool will play a crucial role in section \ref{sec:Integration-approximation-of},
where we present a more accurate approximation of the Boltzmann equation,
which is one of the main goals of this paper. We parameterize the
associated errors based on a coarse graining over short time scales.
In section \ref{sec:Time-model}, we introduce an explicit time model
of the inflaton dynamics that separates the short and long time scale
dependences.

In section \ref{sec:Amplitude}, we derive an analytic formula for
the coarse grained particle production amplitude that includes $\{1\rightarrow2,2\rightarrow2,3\rightarrow2\}$
resonance contributions. The formulae of this section are analogous
to the S-matrix amplitude, which will need to be supplemented with
an analog of phase space integrals to express a physically measurable
number density. The coarse graining time scale is computed in section
\ref{sec:Estimation-of-Deltat} by minimizing the competing errors
associated with quantum interference and general sum to integral approximations.
We calculate the spectrum and number density estimates in section
\ref{sec:Final-particle-number} by utilizing the explicit solutions
obtained from the adiabatic invariant tool. In section \ref{sec:Comparison-of-different},
we present our second main result: numerical integration advantage
of the ``fast only'' formula of eq.\,(\ref{eq:beta fast def})
over the brute force Bogoliubov formulation of eq.\,(\ref{eq:beta exact definition}).
In that section, we also generalize the Gaussian spectral model of
ref.\,\citep{Chung:2018ayg} by incorporating the adiabatic invariant
formalism, and compare it with other computational methods.

We summarize our results in section \ref{sec:Summary}. The appendices
include a review of the adiabatic invariant formalism used in this
paper and a derivation of an explicit formula for the time dependence
of the slowly varying Hubble expansion rate after the end of inflation
for a generic inflationary potential.

\section{\label{sec:Previous-work-on}Aspects of previous work on the topic}

Here we review our conventions and previous work on this topic. Consider
the action in the metric signature (1,-1,-1,-1):
\begin{equation}
S=\int d^{4}x\sqrt{-g}\left(-\frac{M_{P}^{2}}{2}R+\frac{1}{2}g^{\mu\nu}\partial_{\mu}\phi\partial_{\nu}\phi-V(\phi)+\frac{1}{2}g^{\mu\nu}\partial_{\mu}\chi\partial_{\nu}\chi-\frac{1}{2}\left(m_{\chi}^{2}-\xi R\right)\chi^{2}\right),
\end{equation}
where $M_{P}=1/\sqrt{8\pi G}$ is the reduced Planck mass, $R$ is
the Ricci scalar, $\phi$ denotes the inflaton field with potential
$V(\phi)$, and $\chi$ denotes a real scalar field. We assume that
$\chi$ interacts only through gravity controlled by a non-minimal
coupling $\xi$. Pure Einstein gravity (i.e.\,minimal coupling) corresponds
to $\xi=0$ while conformal coupling corresponds to $\xi=1/6$. The
standard background cosmological metric is written as
\begin{equation}
ds^{2}=g_{\mu\nu}dx^{\mu}dx^{\nu}=dt^{2}-a^{2}(t)\left|d\vec{x}\right|^{2}=a^{2}(\tau)\left(d\tau^{2}-\left|d\vec{x}\right|^{2}\right),
\end{equation}
where the scalar factor $a$ is a function of either coordinate time
$t$ or conformal time $\tau$. The Hubble parameter $H$ and Ricci
scalar $R$ have the conventions
\begin{equation}
H=\frac{\dot{a}}{a}=\frac{a'}{a^{2}}\qquad\text{and}\qquad R=-6\left(\frac{\ddot{a}}{a}+\frac{\dot{a}^{2}}{a^{2}}\right)=-6\frac{a''}{a^{3}},
\end{equation}
where $\dot{a}=\partial_{t}a$ and $a'=\partial_{\tau}a$. The energy-momentum
tensor is assumed to arise primarily from a homogeneous inflaton field
$\phi$, and therefore Einstein's equation can be written as
\begin{equation}
3M_{P}^{2}H^{2}=\rho\simeq\frac{1}{2}\dot{\phi}^{2}+V(\phi)\qquad\text{and}\qquad-M_{P}^{2}R=\rho-3P\simeq-\dot{\phi}^{2}+4V(\phi).
\end{equation}

It is well known that the  number density $n_{\chi}$ can be written
as
\begin{equation}
a^{3}(t)n_{\chi}(t)=\int\frac{d^{3}k}{(2\pi)^{3}}\,f_{\chi}(k,t),\qquad f_{\chi}(k,t)\equiv\left|\beta_{k}(t)\right|^{2},\label{eq:numdensitydef}
\end{equation}
where the Bogoliubov coefficient expression for gravitational production
of $\chi$ particles with mass $m_{\chi}$ is 
\begin{equation}
\beta_{k}(t)=\int_{-\infty}^{t}dt'\frac{\dot{\omega}_{k}(t')}{2\omega_{k}(t')}e^{-2i\int_{-\infty}^{t'}dt''\frac{\omega_{k}(t'')}{a(t'')}},\label{eq:beta exact definition}
\end{equation}
with $\omega_{k}^{2}\equiv k^{2}+a^{2}m_{\chi}^{2}+\frac{1}{6}(1-6\xi)a^{2}R$
as long as $|\beta_{k}|\ll1$ \citep{Brezin:1970xf,Kofman:1997yn}.\footnote{We will restrict $\xi$ to values such that $\omega_{k}^{2}>0$ for
all times.} After separating out the time scale much faster than $H^{-1}$,
the expression can be written as
\begin{equation}
\beta_{k}(t)\approx\int_{t_{\mathrm{end}}}^{t}dt'\,B_{k}(t')e^{-2i\Omega_{k}(t')},\label{eq:betak massaged}
\end{equation}
where $t_{\mathrm{end}}$ is the time when inflation ends (when the
slow-roll parameter $\epsilon$ is unity), with
\begin{align}
E_{k}^{2}(t) & \equiv\frac{k^{2}}{a_{\mathrm{slow}}^{2}(t)}+m_{\chi}^{2},\label{eq:eksqdef}\\
\Omega_{k}(s) & \equiv\int_{t_{\mathrm{end}}}^{s}E_{k}(s')ds',\label{eq:Omega_k definition}\\
B_{k}(t) & =\left(\frac{H_{\mathrm{UV}}(t)}{2}+\frac{1}{24}\left(1-6\xi\right)\frac{\dot{R}_{\mathrm{UV}}(t)}{m_{\chi}^{2}}\right)\frac{m_{\chi}^{2}}{E_{k}^{2}(t)},\label{eq:B_k(t)}
\end{align}
and the subscript UV denotes the quantities with the low frequency
components filtered out \citep{Chung:2018ayg}.\footnote{More explicitly, low frequency components refer to time dependent
quantities that vary on a time scale of order $H_{\mathrm{end}}^{-1}$
or longer. Also, by filtered out, we mean subtracted out from the
total.} In the work of ref.\,\citep{Chung:2018ayg}, a Gaussian Fourier
spectrum function with a high frequency peak was fit to the properties
of the potential to effectively define these UV components. Although
this did allow one to analytically compute the dominant contribution
to the particle production in a clean formula, there were some shortcomings:
\begin{enumerate}
\item Although the largest amplitude of inflaton oscillations occurs at
$t_{\mathrm{end}}$, particle production remains significant at later
times when the amplitude is appreciably smaller. Indeed, the final
number density was computed in that reference by converting the first
few oscillations computation into a rate using 
\begin{equation}
\frac{d\gamma(t_{\mathrm{end}})}{dk}\equiv\frac{\sigma}{\sqrt{2\pi}}\left.\frac{dn_{\chi}}{dk}\right|_{t_{\mathrm{end}}+\frac{\sqrt{2\pi}}{\sigma}},\label{eq:original}
\end{equation}
(where $\sigma$ is related to the Gaussian spectral model), setting
$t_{\mathrm{end}}\rightarrow t$, and using a Boltzmann equation to
compute the production rate. However, no formalism was provided on
how to solve for the long time scale dependence such that the replacement
$t_{\mathrm{end}}\rightarrow t$ can be made for an inflaton potential
that is not purely quadratic. One expects this long time scale to
be particularly important for the $k$ modes where
\begin{equation}
\frac{k}{a_{\mathrm{end}}}\gg\sqrt{m_{\phi}^{2}-m_{\chi}^{2}}\gg H,\qquad a_{\mathrm{end}}\equiv a(t_{\mathrm{end}}),\label{eq:long-time-scale}
\end{equation}
since for these cases, $\beta_{k}$ settles to its asymptotic final
value only for $t\gg t_{\mathrm{end}}$.
\item The choice of the coarse graining time scale identification $\left(\Delta t\right)^{-1}=\sigma/\sqrt{2\pi}$
in equation 6.11 of ref.\,\citep{Chung:2018ayg} was an estimate
for the coarse graining time scale based on the naive expectation
of the Gaussian frequency model:
\begin{align}
\delta(E_{*}-E_{*}) & =\lim_{\sigma^{2}\rightarrow0}\frac{1}{\sqrt{2\pi\sigma^{2}}}\exp[-(E_{*}-E_{*})^{2}/(2\sigma^{2})],\\
\frac{\Delta t}{2\pi} & =\lim_{\sigma^{2}\rightarrow0}\frac{1}{\sqrt{2\pi\sigma^{2}}}.\label{eq:naive}
\end{align}
However, since the actual Gaussian model comes from the square of
$\dot{\phi}^{2}$ whereas eq.\,(\ref{eq:naive}) came from the $\phi$
spectral representation itself (see equation 5.23 of ref.\,\citep{Chung:2018ayg}),
there is an ambiguity in this identification.
\end{enumerate}
Both of these shortcomings will be addressed in this paper through
a novel time dependent model based on an adiabatic invariant construction
replacing the Gaussian Fourier spectral model of ref.\,\citep{Chung:2018ayg}.
The use of the adiabatic invariant also makes more precise exactly
which $\dot{\phi}^{2}$ component is being kept when evaluating $\beta_{k}$
contributions from ``fast varying'' (UV) modes.

\section{\label{sec:Adiabatic-invariant-1} Constructing the adiabatic invariant}

One of the issues outlined near eq.\,(\ref{eq:long-time-scale})
is the long time scale behavior of the inflaton oscillation amplitude
necessary for the accurate determination of the amplitude for the
high $k$ modes. To compute this, we use the adiabatic invariant technique
reviewed in appendix \ref{sec:Adiabatic-invariant}. This technique
allows one to construct an approximate invariant $Q$ because of the
combination of the properties that Hamiltonian equations of motion
preserving phase space and energy being conserved in the limit that
an external time-translation breaking function becomes a constant.

To use the adiabatic invariant to isolate the slowly varying part
from the fast varying part of $T^{\mu\nu}$ cleanly, we decompose
the FLRW scale factor as $a(t)=a_{\mathrm{slow}}(t)+a_{\mathrm{fast}}(t)$.
Let the slowly varying source function for the adiabatic invariant
construction be $\lambda(t)=a_{\mathrm{slow}}(t)$ and let the coordinate
of the phase space $q$ (in the notation of the appendix \ref{sec:Adiabatic-invariant})
be the inflaton field $\phi$. An interesting part of the procedure
we are going to follow is that $a_{\mathrm{slow}}(t)$ does not need
to be specified to obtain the adiabatic invariant.

We start by carrying out the integral in eq.\,(\ref{eq:charge}).
We choose $C$ to be the $\phi$ path that begins and ends at $\phi(v)$.
We can set the initial condition $\phi(v)$ at which $\pi(v)=0$.
This can typically be done since the motion is oscillatory by assumption
and the turning point typically corresponds to $\pi=0$. We can therefore
write $E(v)$ from eq.\,(\ref{eq:Eidentity}) as
\begin{equation}
E(v)=H\left(\phi',\Pi\left[\phi',E(v),a_{\mathrm{slow}}(v)\right],a_{\mathrm{slow}}(v)\right)=H(\phi',0,a_{\mathrm{slow}}(v)),\label{eq:E(v) in terms of H}
\end{equation}
where $H$ is the Hamiltonian and $\phi'$ is any field value in the
path $C$. This results in an adiabatic invariant equation for $\phi(v)$:
\begin{equation}
Q=\oint_{C}d\phi'\Pi[\phi',H(\phi(v),0,a_{\mathrm{slow}}(v)),a_{\mathrm{slow}}(v)],\label{eq:Q}
\end{equation}
Since this expression is a constant, we have a prediction for how
$\phi(v)$ varies as a function of $a_{\mathrm{slow}}(v)$.

Let us now obtain a more explicit expression for $\phi(v)$ using
the conservation of energy per integration cycle $\oint_{C}d\phi$.
First, canonically normalize the fields such that the Lagrangian is
$a_{\mathrm{slow}}^{3}(t)[\tfrac{1}{2}\dot{\phi}^{2}-V(\phi)].$ If
there are non-minimal gravitational couplings, such terms can be absorbed
into $V$. The momentum conjugate to $\phi$ is $\pi=a_{\mathrm{slow}}^{3}\dot{\phi}$,
and thus the Hamiltonian is
\begin{equation}
H(\phi,\pi,a_{\mathrm{slow}})=\frac{1}{2}\frac{\pi^{2}}{a_{\mathrm{slow}}^{3}}+a_{\mathrm{slow}}^{3}V(\phi).\label{eq:Hamiltonian}
\end{equation}
Solving for $\Pi$ using eqs.\,(\ref{eq:E(v) in terms of H}) and
(\ref{eq:Hamiltonian}), the adiabatic invariant can now be written
as
\begin{equation}
Q=2\sqrt{2}a_{\mathrm{slow}}^{3}(v)\int_{\phi_{C,-}(v)}^{\phi_{C,+}(v)}d\phi'\sqrt{V(\phi_{C,\pm}(v))-V(\phi')},\label{eq:Q in terms of phi_C}
\end{equation}
where $\phi_{C,-}(v)$ and $\phi_{C,+}(v)$ are respectively the smallest
and largest values $\phi$ takes during an oscillation cycle beginning
at time $v$. The turning points are determined by 
\begin{equation}
V(\phi_{C,+}(v))=V(\phi_{C,-}(v))\label{eq:phicminsol}
\end{equation}
where $\phi_{C,-}(v)$ can then be viewed as a function of $\phi_{C,+}(v)$.

The mass dimension of $Q$ is 3 which is the same as that of the phase
space density of canonically normalized scalar fields. This adiabatic
invariant is the phase space density multiplied by the time-dependent
volume scaling factor $a_{\mathrm{slow}}^{3}(v)$, which is intuitively
the reason why it is constant. The interesting nontrivial aspect is
that the phase space density here is not defined with respect to free
particles but with interactions turned on. Starting with an initial
condition $\phi_{C,+}(v_{i})$, one can replace $Q$ with this initial
data and thereby solve for the slowly varying $\phi_{C,+}(v)$ using
\begin{equation}
\int_{\phi_{C,-}[\phi_{C,+}(v)]}^{\phi_{C,+}(v)}d\phi'\sqrt{V(\phi_{C,+}(v))-V(\phi')}=\frac{a_{\mathrm{slow}}^{3}(v_{i})}{a_{\mathrm{slow}}^{3}(v)}\int_{\phi_{C,-}[\phi_{C,+}(v_{i})]}^{\phi_{C,+}(v_{i})}d\phi'\sqrt{V(\phi_{C,+}(v_{i}))-V(\phi')},\label{eq:solvable}
\end{equation}
where $v_{i}$ is the initial time (which we will often take to be
the time at the end of inflation in practice). This can also be rewritten
as 
\begin{equation}
\sqrt{1-\left(\partial_{\phi}I(\phi_{C,+}(v),\phi_{C,+}(v_{i}))\right)^{2}}I\left(\phi_{C,-}\left[\phi_{C,+}(v)\right],\phi_{C,+}(v)\right)=\frac{a_{\mathrm{slow}}^{3}(v_{i})}{a_{\mathrm{slow}}^{3}(v)}I\left(\phi_{C,-}\left[\phi_{C,+}(v_{i})\right],\phi_{C,+}(v_{i})\right),\label{eq:maineq}
\end{equation}
where
\begin{equation}
I(\phi,\phi_{C,+})\equiv\int_{\phi}^{\phi_{C,+}}d\phi'\sqrt{1-\frac{V(\phi)}{V(\phi_{C,+})}}.\label{eq:Iinteg}
\end{equation}
For some choice of potentials $V(\phi)$, $I(\phi,\phi_{C,+}(v))$
will be simple enough to compute such that $\phi_{+}(v)$ will be
easy to obtain from eq.\,(\ref{eq:maineq}). We will give some examples
below.

An alternative method is to compute $V_{m}\equiv V(\phi_{C,\pm})$
as a function of time using the formalism of appendix \ref{sec:A-perturbative-expansion},
and solve for $\phi_{C,\pm}(v)$ through
\begin{align}
\phi_{C,\pm} & =V^{-1}(V_{m})\\
 & =\phi_{\mathrm{min}}\pm M_{p}\sqrt{2\tilde{V}_{m}}\left(1\pm\phi_{3}\left(2\tilde{V}_{m}\right)^{\frac{1}{2}}+\phi_{4}\left(2\tilde{V}_{m}\right)^{1}\pm\phi_{5}\left(2\tilde{V}_{m}\right)^{\frac{3}{2}}+\phi_{6}\left(2\tilde{V}_{m}\right)^{2}+\dots\right),\label{eq:expandtech}
\end{align}
where the coefficients are defined in appendix \ref{sec:A-perturbative-expansion},
and $V_{m}$ is the potential value at the peak of the oscillations.
Let's turn to some examples.

\subsection{Quadratic potentials}

In this notation, we know for the quadratic potential $V(\phi)=m^{2}\phi^{2}/2$
such that eq.\,(\ref{eq:Iinteg}) becomes
\begin{equation}
I(\phi,\phi_{C,+}(v))=\frac{1}{2}\left(\phi_{C,+}(v)\left(\frac{\pi}{2}-\arcsin\left(\frac{\phi}{\phi_{C,+}(v)}\right)\right)-\phi\sqrt{1-\frac{\phi^{2}}{\phi_{C,+}^{2}(v)}}\right),
\end{equation}
and using eq.\,(\ref{eq:phicminsol}), we find that $\phi_{C,-}[\phi_{C,+}(v)]=-\phi_{C,+}(v)$.
Eq.\,(\ref{eq:maineq}) then becomes
\begin{equation}
\frac{\phi_{C,+}(v)}{\phi_{C,+}(v_{i})}\left[\frac{\pi}{2}\phi_{C,+}(v)\right]=\frac{a_{\mathrm{slow}}^{3}(v_{i})}{a_{\mathrm{slow}}^{3}(v)}\left[\frac{\pi}{2}\phi_{C,+}(v_{i})\right],
\end{equation}
which gives
\begin{equation}
\phi_{C,+}(v)=\left(\frac{a_{\mathrm{slow}}(v_{i})}{a_{\mathrm{slow}}(v)}\right)^{3/2}\phi_{C,+}(v_{i}).\label{eq:quadraticmain}
\end{equation}
This is one of the most important generic examples since most potentials
have the quadratic mass terms dominating as the amplitude of the oscillations
decreasing. That is why we separated this example from the next set
of examples. By taking $A=0$ in eq.\,(\ref{eq:hslowovm}) of appendix
\ref{sec:A-perturbative-expansion}, one can see the time scale of
$a_{\mathrm{slow}}/\dot{a}_{\mathrm{slow}}$ is determined entirely
by $H_{\mathrm{end}}^{-1}$, which is much larger than the oscillatory
period given by $2\pi/m_{\phi}$. This is because the solution of
eq.\,(\ref{eq:Q in terms of phi_C}) is $Q=6\pi M_{P}^{2}a_{\mathrm{end}}^{3}H_{\mathrm{end}}^{2}/m_{\phi}$
for a purely quadratic potential.

\subsection{$2n$ power potentials}

In the case of more general even powered potentials, for some integer
$n\geq1$ we have
\begin{equation}
V=\frac{\lambda}{\left(2n\right)!}\phi^{2n},\label{eq:evenpower}
\end{equation}
with eq.\,(\ref{eq:phicminsol}) giving $\phi_{C,-}[\phi_{C,+}]=-\phi_{C,+}.$
Now eq.\,(\ref{eq:Iinteg}) evaluates to
\begin{align}
I(\phi,\phi_{C,+}) & =\frac{1}{1+n}\left(-\phi\sqrt{1-\left[\frac{\phi}{\phi_{C,+}}\right]^{2n}}+n\phi_{C,+}\sqrt{\pi}\frac{\Gamma\left(1+\frac{1}{2n}\right)}{\Gamma\left(\frac{1+n}{2n}\right)}-n\phi\,_{2}F_{1}\left(\frac{1}{2},\frac{1}{2n};1+\frac{1}{2n},\frac{\phi^{2n}}{\phi_{C,+}^{2n}}\right)\right),
\end{align}
where $\,_{2}F_{1}$ is the hypergeometric function. Note that 
\begin{equation}
I(\phi_{C,-}[\phi_{C,+}],\phi_{C,+})=\frac{2n}{n+1}\phi_{C,+}\sqrt{\pi}\frac{\Gamma\left(1+\frac{1}{2n}\right)}{\Gamma\left(\frac{1+n}{2n}\right)},
\end{equation}
which allows eq.\,(\ref{eq:maineq}) to be evaluated as
\begin{equation}
\frac{\phi_{C,+}(v)}{\phi_{C,+}(v_{i})}=\left(\frac{a_{\mathrm{slow}}(v_{i})}{a_{\mathrm{slow}}(v)}\right)^{\frac{3}{n+1}}.\label{eq:evendimres}
\end{equation}

Consider $n=2$, which corresponds to a quartic potential. The field
$\phi_{C,+}(v)$ scales consistently with the conformal scaling dimension
of the scalar field and significantly different from the quadratic
power index $-3/2$ (see eq.\,(\ref{eq:quadraticmain})). That is
because even when the oscillation amplitude decreases, there is no
mass term for such potentials.

\subsection{Trigonometric integrability}

With the potential
\begin{equation}
V(\phi)=V_{0}\sin^{2}\left[\frac{\phi}{F}\pi\right]
\end{equation}
eq.\,(\ref{eq:phicminsol}) gives $\phi_{C,-}[\phi_{C,+}]=-\phi_{C,+}.$
We also find that as $|\phi|\rightarrow\infty$ the potential is bounded
from below. Therefore, eq.\,(\ref{eq:Iinteg}) evaluates to
\begin{align}
I(\phi,\phi_{C,+}) & =\frac{F}{\pi}\left[E\left(\frac{\pi\phi_{C,+}}{F},\csc^{2}\left[\frac{\phi_{C,+}}{F}\pi\right]\right)-E\left(\frac{\pi\phi}{F},\csc^{2}\left[\frac{\phi_{C,+}}{F}\pi\right]\right)\right],
\end{align}
where $E(\phi,m)$ is the elliptic integral of the second kind. Note
that 
\begin{equation}
I(\phi_{C,-}[\phi_{C,+}],\phi_{C,+})=\frac{2F}{\pi}E\left(\frac{\pi\phi_{C,+}}{F},\csc^{2}\left[\frac{\phi_{C,+}}{F}\pi\right]\right),
\end{equation}
which allows eq.\,(\ref{eq:maineq}) to be evaluated as
\begin{equation}
\frac{\sin\left[\frac{\phi_{C,+}(v)}{F}\pi\right]}{\sin\left[\frac{\phi_{C,+}(v_{i})}{F}\pi\right]}E\left(\frac{\pi\phi_{C,+}(v)}{F},\csc^{2}\left[\frac{\phi_{C,+}(v)}{F}\pi\right]\right)=\frac{a_{\mathrm{slow}}^{3}(v_{i})}{a_{\mathrm{slow}}^{3}(v)}E\left(\frac{\pi\phi_{C,+}(v_{i})}{F},\csc^{2}\left[\frac{\phi_{C,+}(v_{i})}{F}\pi\right]\right),\label{eq:ellipticexample}
\end{equation}
expressing a nontrivial scale factor dependence of $\phi_{C,+}(v)$.
One can easily plot the function to see that $\phi_{C+}(v)$ behaves
very close to $a_{\mathrm{slow}}^{-3/2}(v)$ except near $v=v_{i}$
for $\phi_{C+}(v_{i})\pi/F\sim O(1)$. That is simply because as $a_{\mathrm{slow}}(v)$
increases, the field samples the minimum of the potential which becomes
quadratic. This example also illustrates that in general, one cannot
find an expression for $\phi_{C,+}(v)$ as a function of $a_{\mathrm{slow}}(v)$
using elementary functions.

\subsection{Asymmetric potentials}

Consider the potential 
\begin{equation}
V(\phi)=\frac{m^{2}}{2}\phi^{2}-A\phi^{3},\label{eq:cubic}
\end{equation}
 which has asymmetric turning points
\begin{equation}
\phi_{C,-}[\phi_{C,+}]=\frac{1}{4}\left(\frac{m^{2}}{A}-2\phi_{C,+}-\sqrt{\frac{m^{4}}{A^{2}}+4\left(\frac{m^{2}}{A}-3\phi_{C,+}\right)\phi_{C,+}}\right)
\end{equation}
due to the cubic interaction term. Even though this potential is not
bounded from below, as long as one considers the initial oscillation
amplitude $\phi_{C,+}(v_{i})$ close to the minimum point of the potential
$\phi_{\mathrm{min}}=0$, there will be stable classical oscillations.
Because of $|\phi_{C,-}[\phi_{C,+}]|\neq|\phi_{C,+}|$ in the turning
points, this is an example that is not covered by techniques such
as that presented by ref.\,\citep{Turner:1983he}. Although eq.\,(\ref{eq:Iinteg})
can be evaluated giving a result similar to eq.\,(\ref{eq:ellipticexample}),
it is more instructive to use eq.\,(\ref{eq:expandtech}).

Using eqs.\,(\ref{eq:phicp1}), (\ref{eq:Vm_in_terms_of_Qtilde}),
and (\ref{eq:alpha23forcubic}), we find the long term field time
dependence of
\begin{align}
\phi_{C,\pm}(v) & =\pm m\sqrt{\frac{2V_{m}(v)}{m^{4}}}\left(1\pm\frac{A}{m}\sqrt{\frac{2V_{m}(v)}{m^{4}}}+\dots\right),\label{eq:phicpt}\\
V_{m}(v) & =m\mathscr{Q}\left(\frac{a_{\mathrm{slow}}(v_{i})}{a_{\mathrm{slow}}(v)}\right)^{3}\left(1-\frac{15}{4}\left[\frac{A}{m}\right]^{2}\frac{\mathscr{Q}}{m^{3}}\left(\frac{a_{\mathrm{slow}}(v_{i})}{a_{\mathrm{slow}}(v)}\right)^{3}+\dots\right),\label{eq:energytimebehavior}
\end{align}
where the physical, initial condition dependent charge is 
\begin{equation}
\mathscr{Q}\equiv\frac{Qa_{\mathrm{slow}}^{-3}(v_{i})}{2\pi}=\frac{\frac{m^{2}}{2}\phi_{C,+}^{2}(v_{i})-A\phi_{C,+}^{3}(v_{i})}{m}\left(1+\frac{15}{4}\left(\frac{A}{m}\right)^{2}\frac{\frac{m^{2}}{2}\phi_{C,+}^{2}(v_{i})-A\phi_{C,+}^{3}(v_{i})}{m^{4}}+\dots\right),
\end{equation}
as given by eq.\,(\ref{eq:begeq}).\footnote{Here we made a change of notation, using $v$ as the independent variable
by taking $u\rightarrow a_{\mathrm{slow}}(v)$ in the equations presented
in appendix \ref{sec:A-perturbative-expansion}.} Clearly, eq.\,(\ref{eq:phicpt}) shows that the asymmetric time
dependence of $\phi_{C,\pm}$ is suppressed parametrically by
\begin{equation}
\frac{A}{m}\sqrt{\frac{2V_{m}(v)}{m^{4}}}\sim\frac{A}{m}\frac{\phi}{m},
\end{equation}
which is intuitively expected since this is a measure of the cubic
potential strength to the quadratic potential strengths: i.e. the
deviation $\Delta\phi$ from a symmetric excursion $\phi_{\mathrm{sym}}\equiv\phi-\phi_{\mathrm{min}}$
is 
\begin{equation}
\frac{\Delta\phi}{\phi_{\mathrm{sym}}}\sim\frac{A\phi_{\mathrm{sym}}^{3}}{m^{2}\phi_{\mathrm{sym}}^{2}}
\end{equation}
at a linearization level. It is interesting to see explicitly in a
simple manner that field interactions lead to a non-scaling behavior
of the energy density, in contrast with other interaction results
such as eq.\,(\ref{eq:evendimres}) (or equivalently eq.\,(\ref{eq:Vmscaling})).

\section{\label{sec:Integration-approximation-of}Approximation of the Boltzmann
equation}

The Boltzmann collision equation allows number density computation
of a particle production process. The collision term includes both
a production rate and an annihilation rate. As noted in eq.\,(\ref{eq:original}),
an effective production rate was estimated in ref.\,\citep{Chung:2018ayg},
but as explained in section \ref{sec:Previous-work-on}, the effective
production rate can be improved by better time domain computation
of the inflaton field.

We present here an approximation of the quantum \textit{production}
rate induced by Bogoliubov transform, which models the $\chi$ particle
number density spectrum $f_{\chi}(k,t)=\left|\beta_{k}(t)\right|^{2}$
using a non-negative effective production rate spectrum. This is nontrivial
as the actual time evolution of $f_{\chi}(k,t)$ is not a monotonically
growing function unlike a production rate of a Boltzmann equation.\footnote{In $f_{\chi}(k,t)$, the variable $t$ represents the final time at
which the 1-particle state is defined with respect to an adiabatic
vacuum.} However, when coarse-grained over a short quantum fluctuation time
scale,\footnote{The quantum fluctuation time scale is fixed by the oscillation period
of the inflaton field.} the production rate is positive definite. Ultimately, this is because
the initial state is a vacuum, which by construction has no particles,
and the final state at time $t$ cannot have negative number of particles.
The main goal of this section is to set up the parameterization of
such a positive definite quantity approximation and compute the error
incurred by neglecting the fluctuations and coarse-graining.

Given the short time scale separation eq.\,(\ref{eq:betak massaged})
of the Bogoliubov coefficient, the Boltzmann collision equation will
be approximated as
\begin{equation}
\frac{d}{dt}\left(a^{3}(t)n_{\chi}(t)\right)=\int\frac{d^{3}k}{(2\pi)^{3}}\frac{\partial}{\partial t}\left|\beta_{k}(t)\right|^{2}\approx\int\frac{d^{3}k}{(2\pi)^{3}}\frac{1}{\Delta t(t)}\left|\tilde{b}_{k}(t)\right|^{2},\label{eq:Boltzmann equation approximation}
\end{equation}
where we defined
\begin{equation}
\tilde{b}_{k}(v)\equiv\int_{0}^{\Delta t(v)}dsB_{k}(v+s)e^{-2iE_{k}(v)s},\label{eq:b_k tilde definition}
\end{equation}
and $\Delta t$ is a characteristic coarse graining time that encompasses
many oscillations and will be chosen such that the error of this approximation
is minimized. After accounting for all errors, an expression for $\Delta t$
is given by eq.\,(\ref{eq:deltatresult}). The approximation eq.\,(\ref{eq:Boltzmann equation approximation})
of the Boltzmann rate equation is equivalent to the spectrum density
being modeled as
\begin{equation}
f_{\chi}(k,t)=\int_{t_{\mathrm{end}}}^{t}\frac{dv}{\Delta t(v)}\left|\tilde{b}_{k}(v)\right|^{2}+O(\mathcal{E}_{R_{1}})+O(\mathcal{E}_{\mathrm{Taylor}})+O(\mathcal{E}_{P})+O(\mathcal{E}_{I})+O(\mathcal{E}_{R_{2}}),\label{eq:fchi Boltzmann approximation}
\end{equation}
where the errors are estimated by
\begin{align}
O(\mathcal{E}_{R_{1}}) & =\left|\int_{t_{\mathrm{end}}}^{t}dsB_{k}(s)e^{-2i\Omega_{k}(s)}\right|^{2}-\left|\sum_{m=0}^{N_{t}-1}\int_{0}^{\Delta t(v_{m})}dsB_{k}(v_{m}+s)e^{-2i\Omega_{k}(v_{m}+s)}\right|^{2},\label{eq:remainder 1}\\
O(\mathcal{E}_{\mathrm{Taylor}}) & =\left|\sum_{m=0}^{N_{t}-1}\int_{0}^{\Delta t(v_{m})}dsB_{k}(v_{m}+s)e^{-2i\Omega_{k}(v_{m}+s)}\right|^{2}-\left|\sum_{m=0}^{N_{t}-1}\tilde{b}_{k}(v_{m})e^{-2i\Omega_{k}(v_{m})}\right|^{2},\label{eq:Taylor error}\\
O(\mathcal{E}_{P}) & \equiv\sum_{m\neq n}\tilde{b}_{k}^{*}(v_{m})\tilde{b}_{k}(v_{n})e^{2i\left(\Omega_{k}(v_{m})-\Omega_{k}(v_{n})\right)},\label{eq:phase interference error}\\
O(\mathcal{E}_{I}) & =\sum_{m=0}^{N_{t}-1}\left|\tilde{b}_{k}(v_{m})\right|^{2}-\int_{t_{\mathrm{end}}}^{v_{N_{t}}}\frac{dv}{\Delta t(v)}\left|\tilde{b}_{k}(v)\right|^{2},\label{eq:sum to integral error}\\
O(\mathcal{E}_{R_{2}}) & =-\int_{v_{N_{t}}}^{t}\frac{dv}{\Delta t(v)}\left|\tilde{b}_{k}(v)\right|^{2},\label{eq:remainder 2}
\end{align}
with $v_{m}$ in a discrete set of times whose spacing $v_{m+1}-v_{m}$
is $\Delta t(v_{m})$.

Let us now justify this result. First, to make uniform the kinematic
description of the fast time dynamics, we partition the real time
variable $t$ into a pair of variables $\{v_{m},s\}$ for $m\in\mathbb{X}\equiv\{0,1,2,...\}$:
\begin{equation}
b_{k}(v_{m},s)=B_{k}(v_{m}+s)\hspace{1em}s\in[0,\Delta t(v_{m})],\label{eq:bk(v,s) translation}
\end{equation}
where
\begin{equation}
v_{m+1}=v_{m}+\Delta t(v_{m})\hspace{1em}\hspace{1em}v_{0}=t_{\mathrm{end}}\hspace{1em}\hspace{1em}N_{t}=\mbox{minimum }m\in\mathbb{Z}\mbox{ such that }v_{m+1}>t,
\end{equation}
and where the finite time interval $\Delta t(v_{m})$ will be fixed
through an approximation scheme that minimizes the error terms of
eq.\,(\ref{eq:fchi Boltzmann approximation}). 

The estimation of the phase interference error eq.\,(\ref{eq:phase interference error}) will
be deferred to section \ref{sec:Estimation-of-Deltat} due to the
prerequisite results from sections \ref{sec:Time-model} and \ref{sec:Amplitude}.
It will be shown below that the other errors associated with eqs.\,(\ref{eq:remainder 1}),
(\ref{eq:Taylor error}), (\ref{eq:sum to integral error}), and (\ref{eq:remainder 2})
have upper bounds that approximately scale with positive powers of
$H(v)\Delta t(v)$, which is assumed to be a small expansion parameter.

\subsection{Taylor expansion error}

Here we will estimate the error of eq.\,(\ref{eq:Taylor error})
due to linearizing the Bogoliubov phase. Over the time interval $s\in[0,\Delta t(v)]$
from eq.\,(\ref{eq:bk(v,s) translation}), the phase $\Omega_{k}(v_{m}+s)$
is approximated by the first order Taylor expansion $\Omega_{k}(v_{m})+E_{k}(v_{m})s$.
The upper bound on the relative phase difference is
\begin{align}
\left|\frac{\Omega_{k}(v_{m}+s)-\Omega_{k}(v_{m})-E_{k}(v_{m})s}{E_{k}(v_{m})s}\right| & =\left|\frac{\frac{1}{2}\partial_{v}E_{k}(v_{m})s^{2}+\frac{1}{6}\partial_{v}^{2}E_{k}(v_{m})s^{3}+\dots}{E_{k}(v_{m})s}\right|\\
 & \leq\frac{1}{2}\left|\frac{\partial_{v}E_{k}(v_{m})}{E_{k}(v_{m})}\right|s+\frac{1}{6}\left|\frac{\partial_{v}^{2}E_{k}(v_{m})}{E_{k}(v_{m})}\right|s^{2}+\dots,
\end{align}
which is maximized at $s=\Delta t(v)$. As $E_{k}(v)$ is a slowly
varying function, its $n$-th derivative scales as $H^{n}(v)E_{k}(v)$.
Therefore the upper bound on the relative phase difference is
\begin{equation}
\left|\frac{\Omega_{k}(v_{m}+s)-\Omega_{k}(v_{m})-E_{k}(v_{m})s}{E_{k}(v_{m})s}\right|\leq\frac{E_{k}^{2}(v_{m})-m_{\chi}^{2}}{E_{k}^{2}(v_{m})}H(v_{m})\Delta t(v_{m})+O\left(H^{2}(v_{m})\Delta t^{2}(v_{m})\right),
\end{equation}
which is suppressed if $H(v_{m})\Delta t(v_{m})\ll1$.

\subsection{Sum to integral error}

This subsection will justifying neglecting the term eq.\,(\ref{eq:sum to integral error})
by showing that the relative sum to integral error, defined as
\begin{equation}
\mathcal{E}_{I}=\frac{\sum_{m=0}^{N_{t}-1}\left|\tilde{b}_{k}(v_{m})\right|^{2}-\int_{t_{\mathrm{end}}}^{v_{N_{t}}}\frac{dv}{\Delta t(v)}\left|\tilde{b}_{k}(v)\right|^{2}}{\sum_{m=0}^{N_{t}-1}\left|\tilde{b}_{k}(v_{m})\right|^{2}},\label{eq:sum to integral relative error}
\end{equation}
is suppressed by $H(v)\Delta t(v)$. With the Euler-Maclaurin formula,
the sum can be approximated with an integral as
\begin{align}
\sum_{m=0}^{N_{t}-1}\left|\tilde{b}_{k}(v_{m})\right|^{2} & -\int_{t_{\mathrm{end}}}^{v_{N_{t}}}\frac{dv}{\Delta t(v)}\left|\tilde{b}_{k}(v)\right|^{2}\approx\frac{\left|\tilde{b}_{k}(t_{\mathrm{end}})\right|^{2}-\left|\tilde{b}_{k}(v_{N_{t}})\right|^{2}}{2}+\int_{t_{\mathrm{end}}}^{v_{N_{t}}}dv\frac{d\left|\tilde{b}_{k}(v)\right|^{2}}{dv}P_{1}(m(v)),\label{eq:Euler Maclaurin}
\end{align}
where $P_{1}(x)$ is the first Bernoulli polynomial and we define
its argument as
\begin{equation}
m(v)\equiv\int_{t_{\mathrm{end}}}^{v}\frac{dv'}{\Delta t(v')}.
\end{equation}

Slowly varying functions such as $\tilde{b}_{k}(v)$ and $\Delta t(v)$
derive their time dependence from $a_{\mathrm{slow}}(t)$, and therefore
the derivative of such functions are typically on the order of the
function multiplied by the Hubble parameter. Thus the error terms
of the right hand side of eq.\,(\ref{eq:Euler Maclaurin}) are estimated
as
\begin{align}
\left|\tilde{b}_{k}(v_{N_{t}})\right|^{2}-\left|\tilde{b}_{k}(t_{\mathrm{end}})\right|^{2} & =\left|\tilde{b}_{k}(v_{N_{t}-1}+\Delta t(v_{N_{t}-1}))\right|^{2}-\left|\tilde{b}_{k}(t_{\mathrm{end}})\right|^{2}\\
 & \sim\left|\tilde{b}_{k}(v_{N_{t}-1})\right|^{2}+O(1)H(v_{N_{t}-1})\Delta t(v_{N_{t}-1}))\left|\tilde{b}_{k}(v_{N_{t}-1})\right|^{2}-\left|\tilde{b}_{k}(t_{\mathrm{end}})\right|^{2}\\
 & \sim\sum_{m=0}^{N_{t}-1}O(1)H(v_{m})\Delta t(v_{m})\left|\tilde{b}_{k}(v_{m})\right|^{2},
\end{align}
and
\begin{align}
\int_{t_{\mathrm{end}}}^{v_{N_{t}}}dv\frac{d\left|\tilde{b}_{k}(v)\right|^{2}}{dv}P_{1}(m(v)) & \sim\int_{t_{\mathrm{end}}}^{v_{N_{t}}}dvH(v)\left|\tilde{b}_{k}(v)\right|^{2}\\
 & \sim\sum_{m=0}^{N_{t}-1}O(1)\Delta t(v_{m})H(v_{m})\left|\tilde{b}_{k}(v_{m})\right|^{2},
\end{align}
respectively to leading order in $H(v_{m})\Delta t(v_{m})$. Therefore
the relative error is of the form
\begin{align}
\mathcal{E}_{I} & \sim\frac{\sum_{m=0}^{N_{t}-1}O(1)H(v_{m})\Delta t(v_{m})\left|\tilde{b}_{k}(v_{m})\right|^{2}}{\sum_{m=0}^{N_{t}-1}\left|\tilde{b}_{k}(v_{m})\right|^{2}},
\end{align}
with an upper bound given by
\begin{equation}
\left|\mathcal{E}_{I}\right|\lesssim\max_{m\in[0,N_{t}]}\left|c_{m}\right|H(v_{m})\Delta t(v_{m}),
\end{equation}
where $c_{m}$ are order one numbers. This is suppressed if $H(v_{m})\Delta t(v_{m})\ll1$.

\subsection{Remainder errors}

We will now estimate the relative remainder errors of eq.\,(\ref{eq:fchi Boltzmann approximation}).
These errors will vanishes as $t\rightarrow\infty$. For finite time,
they are suppressed by positive powers of $H(v_{N_{t}})\Delta t(v_{N_{t}})$.

The relative error associated with eq.\,(\ref{eq:remainder 1}) is
written as
\begin{align}
\mathcal{E}_{R_{1}} & =\left|\frac{\left|\int_{t_{\mathrm{end}}}^{t}dsB_{k}(s)e^{-2i\Omega_{k}(s)}\right|^{2}-\left|\int_{t_{\mathrm{end}}}^{v_{N_{t}}}dsB_{k}(s)e^{-2i\Omega_{k}(s)}\right|^{2}}{\left|\int_{t_{\mathrm{end}}}^{t}dsB_{k}(s)e^{-2i\Omega_{k}(s)}\right|^{2}}\right|\\
 & \approx2\left|\frac{\left|\int_{t_{\mathrm{end}}}^{t}dsB_{k}(s)e^{-2i\Omega_{k}(s)}\right|-\left|\int_{t_{\mathrm{end}}}^{v_{N_{t}}}dsB_{k}(s)e^{-2i\Omega_{k}(s)}\right|}{\left|\int_{t_{\mathrm{end}}}^{t}dsB_{k}(s)e^{-2i\Omega_{k}(s)}\right|}\right|,
\end{align}
Using the inequality
\begin{align}
\left|\int_{t_{\mathrm{end}}}^{t}dsB_{k}(s)e^{-2i\Omega_{k}(s)}\right|-\left|\int_{t_{\mathrm{end}}}^{v_{N_{t}}}dsB_{k}(s)e^{-2i\Omega_{k}(s)}\right| & \leq\left|\int_{v_{N_{t}}}^{t}dsB_{k}(s)e^{-2i\Omega_{k}(s)}\right|\leq\Delta t(v_{N_{t}})\left|B_{k}(v_{N_{t}})\right|,
\end{align}
we obtain an approximate upper bound of
\begin{equation}
\mathcal{E}_{R_{1}}\lesssim2\frac{\Delta t(v_{N_{t}})\left|B_{k}(v_{N_{t}})\right|}{\left|\int_{t_{\mathrm{end}}}^{t}dsB_{k}(s)e^{-2i\Omega_{k}(s)}\right|}.
\end{equation}
As will be shown in section \ref{sec:Amplitude}, the fast varying
component of the Bogoliubov integrand can be estimated as
\begin{equation}
\left|B_{k}(v)\right|\lesssim O(1)\frac{H^{2}(v)}{m_{\phi}},\label{eq:Bogoliubov integrand upper bound}
\end{equation}
 and therefore an approximate upper bound on this remainder error
is
\begin{equation}
\mathcal{E}_{R_{1}}\lesssim\frac{O(1)H(v_{N_{t}})/m_{\phi}}{\left|\int_{t_{\mathrm{end}}}^{t}dsB_{k}(s)e^{-2i\Omega_{k}(s)}\right|}H(v_{N_{t}})\Delta t(v_{N_{t}}).\label{eq:remainder 1 upper bound}
\end{equation}

The relative error arising from eq.\,(\ref{eq:remainder 2}) is given
by
\begin{equation}
\mathcal{E}_{R_{2}}=\frac{\int_{v_{N_{t}}}^{t}\frac{dv}{\Delta t(v)}\left|\tilde{b}_{k}(v)\right|^{2}}{\int_{t_{\mathrm{end}}}^{t}\frac{dv}{\Delta t(v)}\left|\tilde{b}_{k}(v)\right|^{2}}\leq\frac{\frac{t-v_{N_{t}}}{\Delta t(v_{N_{t}})}\left|\tilde{b}_{k}(v_{N_{t}})\right|^{2}}{\int_{t_{\mathrm{end}}}^{t}\frac{dv}{\Delta t(v)}\left|\tilde{b}_{k}(v)\right|^{2}}\leq\frac{\left|\tilde{b}_{k}(v_{N_{t}})\right|^{2}}{\int_{t_{\mathrm{end}}}^{t}\frac{dv}{\Delta t(v)}\left|\tilde{b}_{k}(v)\right|^{2}},
\end{equation}
The upper bound of eq.\,(\ref{eq:Bogoliubov integrand upper bound})
of the Bogoliubov integrand yields
\begin{equation}
\left|\tilde{b}_{k}(v)\right|\leq\Delta t(v)\left|B_{k}(v)\right|\lesssim O(1)\frac{H^{2}(v)}{m_{\phi}}\Delta t(v),
\end{equation}
and therefore an approximate upper bound on this remainder error is
\begin{equation}
\mathcal{E}_{R_{2}}\lesssim\frac{O(1)H^{2}(v_{N_{t}})/m_{\phi}^{2}}{\int_{t_{\mathrm{end}}}^{t}\frac{dv}{\Delta t(v)}\left|\tilde{b}_{k}(v)\right|^{2}}\left(H(v_{N_{t}})\Delta t(v_{N_{t}})\right)^{2}.\label{eq:remainder 2 upper bound}
\end{equation}
Eqs.\,(\ref{eq:remainder 1 upper bound}) and (\ref{eq:remainder 2 upper bound})
are both suppressed by positive powers of $H/m_{\phi}$ and $H\Delta t$.

\section{A time model of inflaton dynamics \label{sec:Time-model}}

Although Einstein equations give directly the Hubble rate $H$ in
eq.\,(\ref{eq:B_k(t)}), it is useful to compute its time derivative
since that will tend to suppress the slow frequency components such
that the fast frequency components become manifest. The time derivative
of $H$ gives
\begin{equation}
\dot{H}=-\frac{1}{2M_{P}^{2}}\dot{\phi}^{2}.\label{eq:hdoteq}
\end{equation}
Similarly, the Einstein equation for the Ricci scalar $R$ gives
\[
R=\frac{1}{M_{P}^{2}}\left(\dot{\phi}^{2}-4V(\phi)\right),
\]
which shows up in eq.\,(\ref{eq:B_k(t)}). To extract the high frequency
time variation of this term, it is also useful to take a derivative
\begin{align}
\dot{R}(t) & =\frac{3}{M_{P}^{2}}\left(\frac{d}{dt}\dot{\phi}^{2}+4H\dot{\phi}^{2}\right),\label{eq:rdot}
\end{align}
where one notes that the first term obtained a large contribution
from the potential term. In this form, the second term in eq.\,(\ref{eq:rdot})
is subdominant to the first term because $H$ is much smaller than
$m_{\phi}$, which is the inverse time scale of the time derivative.
Remarkably we have reduced the determination of the Bogoliubov integrand
$B_{k}(t)$ to modeling $\dot{\phi}^{2}$. In this section, we will
develop a model of $\dot{\phi}$ using an adiabatic invariant and
a fast-slow time decomposition.

Since $\phi_{C,\pm}$ are left-right bounds on the inflaton field
$\phi(t)$, we can parameterize the field time dependence in terms
of a phase $\Xi_{\phi}(v,s)$ such that
\begin{equation}
\phi(t)\approx\phi(v,s)=\bar{\phi}(v)+\Delta\phi(v)\cos\Xi_{\phi}(v,s),\label{eq:ansatz}
\end{equation}
where
\begin{align}
\bar{\phi}(v) & \equiv\frac{\phi_{C,+}(v)+\phi_{C,-}(v)}{2},\\
\Delta\phi(v) & \equiv\frac{\phi_{C,+}(v)-\phi_{C,-}(v)}{2},
\end{align}
and the trajectory is only over a time period $s\in\left[0,\Delta t\right].$
To finish defining the approximate model, we must give a map between
$t$ and $(v,s)$. As we will see, we will define this map through
$v$ only having $\Delta t$ resolution such that the set of points
$\{v\in(t_{1},t_{2}),s\in[0,\Delta t]\}$ approximately covers the
same domain as $t\in(t_{1},t_{2})$. This equivalence class definition
of $v$ is possible because all the quantities with $v$ dependence
will be slowly varying such that they are approximately constant in
the time interval $\Delta t$.

We define $\phi_{C,\pm}(v)$ to be the amplitude derived from the
adiabatic invariant. This automatically gives $v$ a resolution of
the period of the adiabatic invariant. In this case, we see $\Xi_{\phi}$
is real since $\phi_{C,\pm}$ are bounds of periodic motion in the
adiabatic invariant approximation. As a definition of the model, we
require
\begin{equation}
\Xi_{\phi}(v,0)=0\label{eq:phase1}
\end{equation}
 (modulo the usual solution periodicity which is unimportant here)
such that $\phi(v,s)$ satisfies $\phi(v,0)=\phi_{C,+}(v)$.

Now, we attribute the fast time behavior of $\phi(t)\approx\phi(v,s)$
in eq.\,(\ref{eq:ansatz}) through approximate energy conservation.
We can express $\partial_{s}\phi$ as
\begin{align}
\frac{1}{2}\left[\partial_{s}\phi(v,s)\right]^{2} & =V_{m}(v)-V\left(\phi(v,s)\right),\label{eq:approx energy con phase eq}\\
\partial_{s}\phi(v,s) & =-\Delta\phi(v)\frac{\partial\Xi_{\phi}(v,s)}{\partial s}\sin\Xi_{\phi}(v,s),\label{eq:derivative}
\end{align}
where
\begin{equation}
V_{m}(v)=V(\phi_{C,\pm}(v)).\label{Vmdef}
\end{equation}
Since $V(\phi_{C,-})=V(\phi_{C,+})$ by definition, the value of $V_{m}(v)$
can be obtained using either $\phi_{C,\pm}(v)$. Eqs.\,(\ref{eq:approx energy con phase eq})
and (\ref{eq:derivative}) lead to a quadrature integral that determines
the phase $\Xi_{\phi}$ by
\begin{equation}
s=\int_{\Xi_{\phi}(v,0)}^{\Xi_{\phi}(v,s)}\frac{\Delta\phi(v)\sin\Xi\,d\Xi}{\sqrt{2V_{m}(v)-2V(\bar{\phi}(v)+\Delta\phi(v)\cos\Xi)}}.\label{eq:quadrature}
\end{equation}

In this paper, we will consider only potentials where there is a single
dynamically relevant minimum $\phi_{\mathrm{min}}$ at the end of
inflation. If we keep only the leading asymmetric term of the potential,
we have the expansion
\begin{align}
V(\phi) & =m_{\phi}^{2}M_{P}^{2}\left(\frac{1}{2}\left(\frac{\phi-\phi_{\mathrm{min}}}{M_{P}}\right)^{2}+\alpha_{3}\left(\frac{\phi-\phi_{\mathrm{min}}}{M_{P}}\right)^{3}+\dots\right),\\
m_{\phi}^{2} & =V''(\phi_{\mathrm{min}}),\\
\alpha_{3} & \equiv M_{P}V'''(\phi_{\mathrm{min}})/(6m_{\phi}^{2}),
\end{align}
and if we define
\begin{align}
a_{3}(v) & =\alpha_{3}\sqrt{\frac{2V_{m}(v)}{m_{\phi}^{2}M_{P}^{2}}}=\sqrt{6}\alpha_{3}\frac{H_{\mathrm{slow}}(v)}{m_{\phi}},\label{eq:a3def}
\end{align}
this leads to the expansion
\begin{align}
\phi_{C,\pm}(v) & \approx\phi_{\text{min}}\pm\sqrt{\frac{2V_{m}(v)}{m_{\phi}^{2}}}-a_{3}(v)\sqrt{\frac{2V_{m}(v)}{m_{\phi}^{2}}}+O(\alpha_{3}^{2}),\\
V_{m}(v)-V(\phi(v,s)) & \approx V_{m}(v)\sin^{2}\Xi_{\phi}(v,s)\left(1+2a_{3}(v)\cos\Xi_{\phi}(v,s)\right)+O(\alpha_{3}^{2}),\label{eq:Vm expansion}
\end{align}
which makes manifest how $\alpha_{3}$ controls the asymmetric nature
of the maximum and minimum value of the field excursion represented
by $\phi_{C,\pm}$. Putting this into eq.\,(\ref{eq:quadrature})
gives
\begin{equation}
\sin\Xi_{\phi}(v,s)\approx\sin\tau(v,s)\left(1+a_{3}(v)\cos\tau(v,s)\right).\label{eq:sin}
\end{equation}
where
\begin{equation}
\tau(v,s)=\sqrt{\frac{2V_{m}(v)}{[\Delta\phi(v)]^{2}}}s.
\end{equation}
Using eqs.\,(\ref{eq:sin}) and (\ref{eq:Vm expansion}), the time
dependence of $[\partial_{s}\phi]^{2}$ is approximated (to leading
order in $\alpha_{3}$) as
\begin{align}
[\partial_{s}\phi(v,s)]^{2} & \approx V_{m}(v)\left(1-\cos2\tau(v,s)\right)\left(1+4a_{3}(v)\cos\tau(v,s)\right).
\end{align}
Note the squaring produced a slowly varying term even though $\Delta\phi(v)\cos\Xi_{\phi}$
naively looked to be fast varying only. The fast varying part is obtained
by throwing out the terms that depend on $v$ only. We will call this
UV:
\begin{equation}
\boxed{\left(\partial_{s}\phi(v,s)\right)_{\mathrm{UV}}^{2}=-V_{m}(v)\left[\cos2\tau(v,s)+2a_{3}(v)\left(\cos3\tau(v,s)-\cos\tau(v,s)\right)\right]}.\label{eq:UVasymm}
\end{equation}

\section{Amplitude of the production rate \label{sec:Amplitude}}

In the time interval of $s\in[0,\Delta t(v)]$, the relevant quantities
for evaluation of the Bogoliubov coefficient integrand are eqs.\,(\ref{eq:B_k(t)}),
(\ref{eq:hdoteq}), and (\ref{eq:rdot}). To leading order in $H/m_{\phi}$,
we therefore have
\begin{align}
b_{k}(v,s) & =\frac{1}{4M_{P}^{2}}\left(\frac{1-6\xi}{2m_{\chi}^{2}}\partial_{s}\left[\partial_{s}\phi(v,s)\right]^{2}-\left(\int_{0}^{s}ds'\left[\partial_{s'}\phi(v,s')\right]^{2}\right)_{\mathrm{UV}}\right)\frac{m_{\chi}^{2}}{E_{k}^{2}(v)},
\end{align}
which is essentially determined by the inflaton fast velocity squared
function $[\partial_{s}\phi(v,s)]_{\mathrm{UV}}^{2}$. For convenience
of interpretation later, this can be evaluated using eq.\,(\ref{eq:UVasymm})
rewritten in a complexified form as
\begin{equation}
\left(\partial_{s}\phi(v,s)\right)_{\mathrm{UV}}^{2}=-3M_{P}^{2}H_{\mathrm{slow}}^{2}(v)\Re\left\{ e^{2i\tau(v,s)}+2a_{3}(v)\left(e^{3i\tau(v,s)}-e^{i\tau(v,s)}\right)\right\} ,
\end{equation}
where we substituted $V_{m}=3M_{P}^{2}H_{\mathrm{slow}}^{2}$. The
key quantity eq.\,((\ref{eq:bk(v,s) translation})) can therefore
be written as
\begin{equation}
b_{k}(v,s)=\Re\left\{ A_{2}(k,v)e^{2i\omega_{*}(v)s}+A_{3}(k,v)e^{3i\omega_{*}(v)s}+A_{1}\left(k,v\right)e^{i\omega_{*}(v)s}\right\} ,\label{eq:bkfin}
\end{equation}
where we defined
\begin{align}
A_{2}(k,t) & =\frac{3H_{\mathrm{slow}}^{2}(t)}{4i\omega_{*}(t)}\frac{(1-6\xi)\omega_{*}^{2}(t)+\frac{1}{2}m_{\chi}^{2}}{E_{k}^{2}(t)},\label{eq:A2 definition}\\
A_{3}(k,t) & =+2a_{3}(t)\frac{3H_{\mathrm{slow}}^{2}(t)}{4i\omega_{*}(t)}\frac{\frac{3}{2}(1-6\xi)\omega_{*}^{2}(t)+\frac{1}{3}m_{\chi}^{2}}{E_{k}^{2}(t)},\label{eq:A3 definition}\\
A_{1}(k,t) & =-2a_{3}(t)\frac{3H_{\mathrm{slow}}^{2}(t)}{4i\omega_{*}(t)}\frac{\frac{1}{2}(1-6\xi)\omega_{*}^{2}(t)+m_{\chi}^{2}}{E_{k}^{2}(t)},\label{eq:A1 definition}\\
\omega_{*}(t) & =\sqrt{\frac{2V(\phi_{C,\pm}(t))}{[\Delta\phi(t)]^{2}}},\label{eq:omstardef}
\end{align}
and the order of terms in the brackets in eq.\,(\ref{eq:bkfin})
is in the order of typical importance in magnitude.

We now arrive at one of our main analytic results of this paper. Using
the Boltzmann equation approximation of section \ref{sec:Integration-approximation-of},
we write the spectrum as
\begin{align}
f_{\chi}(k,t) & =\int_{t_{\mathrm{end}}}^{t}\frac{dv}{\Delta t(v)}\left|\tilde{b}_{k}(v)\right|^{2}=\int_{t_{\mathrm{end}}}^{t}\frac{dv}{\Delta t(v)}\left|F(E_{k}(v))+F^{*}(-E_{k}(v))\right|^{2},\label{eq:fchi time model}\\
\tilde{b}_{k}(v) & =\int_{0}^{\Delta t}ds\,b_{k}(v,s)e^{-2iE_{k}(v)s}=F(E_{k}(v))+F^{*}(-E_{k}(v)),\label{eq:btilde_k result}\\
F(E_{k}) & \equiv\frac{A_{2}}{2}\frac{e^{i\left(2\omega_{*}-2E_{k}\right)\Delta t}-1}{i(2\omega_{*}-2E_{k})}+\frac{A_{3}}{2}\frac{e^{i\left(3\omega_{*}-2E_{k}\right)\Delta t}-1}{i(3\omega_{*}-2E_{k})}+\frac{A_{1}}{2}\frac{e^{i\left(\omega_{*}-2E_{k}\right)\Delta t}-1}{i(\omega_{*}-2E_{k})}.\label{eq:F(Ek) expression}
\end{align}
For brevity of notation $F(E_{k})$ has its $v$ dependences that
exist through $\{V_{m}(v),\omega_{*}(v),E_{k}(v),a_{3}(v)\}$ hidden
(see eqs.\,(\ref{Vmdef}), (\ref{eq:omstardef}), (\ref{eq:eksqdef}),
and (\ref{eq:a3def})). Note the denominator of eq.\,(\ref{eq:F(Ek) expression})
gives a clear interpretation of the various contributions. $\{A_{1},A_{2},A_{3}\}$
terms contribute to the processes $\{\delta\phi\rightarrow\chi\chi,\delta\phi\delta\phi\rightarrow\chi\chi,\delta\phi\delta\phi\delta\phi\rightarrow\chi\chi\}$.
The fact that $\mathrm{odd}\rightarrow\mathrm{even}$ processes are
mediated by the $m_{\phi}\alpha_{3}(\phi-\phi_{\mathrm{min}})^{3}\equiv m_{\phi}\alpha_{3}\delta\phi^{3}$
interaction vertex is clear since without the cubic vertex, the symmetry
$\delta\phi\rightarrow-\delta\phi$ would forbid all odd $\delta\phi$
number changing processes.

To obtain an intuition for $a_{3}$ consider the example potential
of
\begin{equation}
V(\phi)=M^{4}\left(1-\left(\frac{\phi}{v_{\phi}}\right)^{n}\right)^{2}.
\end{equation}
The inflaton mass will be $m_{\phi}=\sqrt{2}nM^{2}/v_{\phi}$. We
choose $v_{\phi}=0.5M_{P}$. The leading asymmetry term will be $\alpha_{3}=n-1$
and thus
\begin{align}
a_{3}(t_{\mathrm{end}}) & =\alpha_{3}\sqrt{\frac{2V(\phi_{\mathrm{end}})}{m_{\phi}^{2}M_{P}^{2}}}\simeq\frac{n-1}{2n},
\end{align}
which is an O(1) number. At later times, $a_{3}(t)$ will decay as
$H_{\text{slow}}(t)$ and therefore become a small expansion parameter
at large resonance times.

\section{Estimation of coarse graining time $\Delta t$ \label{sec:Estimation-of-Deltat}}

Until now, we still have not specified the time width $\Delta t(v)$.
Its specification dominantly affects the computation in several ways.
First, it fixes the errors of eq.\,(\ref{eq:fchi Boltzmann approximation}).
Second, it enters in the approximation eq.\,(\ref{eq:ansatz}) and
the trajectory interval $s\in[0,\Delta t(v)]$ specifying the adiabatic
invariant based time model. As far as most of the errors of eq.\,(\ref{eq:fchi Boltzmann approximation})
are concerned, the smallest error is obtained when $\Delta t$ is
the smallest. On the other hand, there is a preferred $\Delta t>0$
that makes the quantum interference error of eq.\,(\ref{eq:phase interference error})
negligible. The adiabatic invariant interpretation also sets a minimum
$\Delta t$. We will see that they can be made commensurate. We will
find below that the quantum interference error considerations lead
to a time width at an intermediate scale between those set by the
inflaton mass and Hubble expansion rate, i.e.\,$m_{\phi}^{-1}\ll\Delta t(v)\ll H^{-1}(v)$.
This scale is larger than the one set by the adiabatic invariant minimum
of a single oscillation.

We will first consider the quantum interference error of eq.\,(\ref{eq:phase interference error}).
The dominant interference term is
\begin{align}
\frac{|\mathcal{E}_{P}|}{\sum_{n=1}^{N(t)}\left|\tilde{b}_{k}(v_{n})\right|^{2}} & =\frac{\left|\sum_{m\neq n}e^{2i\left(\Omega_{k}(v_{m})-\Omega_{k}(v_{n})\right)}\tilde{b}_{k}^{*}(v_{m})\tilde{b}_{k}(v_{n})\right|}{\sum_{n}^{N(t)}\left|\tilde{b}_{k}(v_{n})\right|^{2}}\\
 & \simeq\frac{2\mathcal{R}\left\{ e^{2i\left(\Omega_{k}(v_{p})-\Omega_{k}(v_{p+1})\right)}\tilde{b}_{k}^{*}(v_{p})\tilde{b}_{k}(v_{p+1})+e^{2i\left(\Omega_{k}(v_{p})-\Omega_{k}(v_{p-1})\right)}\tilde{b}_{k}^{*}(v_{p})\tilde{b}_{k}(v_{p-1})\right\} }{\left|\tilde{b}_{k}(v_{p})\right|^{2}}\\
 & \lesssim2\frac{\left|\tilde{b}_{k}(v_{p+1})\right|+\left|\tilde{b}_{k}(v_{p-1})\right|}{\left|\tilde{b}_{k}(v_{p})\right|},
\end{align}
where $v_{p}$ is the nearest time to resonance $\bar{v}_{k}$, and
therefore dominates the denominator sum. Neglecting the cubic coupling
$a_{3}$, and for large frequencies $E_{k}\sim\omega_{*}\gg m_{\chi}$,
eqs.\,(\ref{eq:btilde_k result}) and (\ref{eq:F(Ek) expression})
lead to
\begin{align}
\left|\tilde{b}_{k}\left(v_{p}\right)\right| & \approx\frac{1}{2}\left|A_{2}(k,v_{p})\right|\Delta t(v_{p})\left|\text{sinc}\left[\left(E_{k}(v_{p})-\omega_{*}(v_{p})\right)\Delta t(v_{p})\right]\right|,
\end{align}
where $\mathrm{sinc}(x)=\frac{\sin x}{x}$, and $A_{2}$ is given
by eq.\,(\ref{eq:A2 definition}). As such, we have
\begin{equation}
\left|\left(E_{k}(v_{p})-\omega_{*}(v_{p})\right)\Delta t(v_{p})\right|<\pi,\label{eq:centralpeakcond}
\end{equation}
for times $v_{p}$ nearest the central peak of $\tilde{b}_{k}$. The
largest off-diagonal contributions will be
\begin{align}
\left|\tilde{b}_{k}\left(v_{p\pm1}\right)\right| & \approx\frac{1}{2}|A_{2}(k,v_{p\pm1})|\Delta t(v_{p\pm1})\left|\text{sinc}\left[\left(E_{k}(v_{p\pm1})-\omega_{*}(v_{p\pm1})\right)\Delta t(v_{p\pm1})\right]\right|\\
 & \approx\frac{1}{2}|A_{2}(k,v_{p})|\Delta t(v_{p})\left|\text{sinc}\left[\left(E_{k}(v_{p})-\omega_{*}(v_{p})\right)\Delta t(v_{p})\pm\partial_{v}E_{k}(v_{p})\left[\Delta t(v_{p})\right]^{2}\right]\right|,
\end{align}
where we made use of $\partial_{v}\omega_{*}(v)\sim H^{3}(v)/m_{\phi}$
and $\partial_{v}\Delta t(v)\sim H(v)\Delta t(v)\ll1$ to neglect
sub-leading terms. Therefore, the quantity
\begin{equation}
\frac{|\mathcal{E}_{P}|}{\sum_{n=1}^{N(t)}\left|\tilde{b}_{k}(v_{n})\right|^{2}}\sim\frac{\left|\text{sinc}\left[\left(E_{k}(v_{p})-\omega_{*}(v_{p})\right)\Delta t(v_{p})\pm\partial_{v}E_{k}(v_{p})\left[\Delta t(v_{p})\right]^{2}\right]\right|}{\left|\text{sinc}\left[\left(E_{k}(v_{p})-\omega_{*}(v_{p})\right)\Delta t(v_{p})\right]\right|}
\end{equation}
is an estimate of the relative error.

To minimize this error, we require the arguments of the sine-cardinal
functions in the numerator to lie outside the central peak:
\begin{equation}
\left|\partial_{v}E_{k}(v_{p})\right|\left[\Delta t(v_{p})\right]^{2}-\left|E_{k}(v_{p})-\omega_{*}(v_{p})\right|\Delta t(v_{p})>\pi,
\end{equation}
where the other sign equation is automatically satisfied if this condition
is satisfied. The above inequality will be satisfied if $\left|\partial_{v}E_{k}(v_{p})\right|\left[\Delta t(v_{p})\right]^{2}>2\pi$
for all $v_{p}$ that satisfy eq.\,(\ref{eq:centralpeakcond}). The
coarse graining time $\Delta t(v_{p})$ is therefore set to the lower
bound:
\begin{equation}
\Delta t(v_{p})=\sqrt{\frac{2\pi}{\left|\partial_{v}E_{k}(v_{p})\right|}}\simeq\sqrt{\frac{2\pi}{H(v_{p})m_{\phi}}},\label{eq:needpeak}
\end{equation}
where we have assumed $E_{k}\approx m_{\phi}$. For general times
$v$ away from the peak, we take $H(v_{p})\rightarrow H(v)$ to ensure
$H(v)\Delta t(v)$ remains a small expansion parameter. As such, we
set 
\begin{equation}
\Delta t(v)\simeq\sqrt{\frac{2\pi}{H(v)m_{\phi}}}\label{eq:deltatresult}
\end{equation}
as the coarse graining time width of the Boltzmann rate approximation.

This time interval is longer than the single period of the adiabatic
invariant construction oscillations which is of order $\omega_{*}^{-1}\sim m_{\phi}^{-1}$
since $H(v_{p})\ll m_{\phi}$ for these scenarios. However, we see
that $H(v)\Delta t(v)\rightarrow0$ as $v\rightarrow\infty$, which
allows for an accurate fast-slow decomposition. A slow varying function
can be treated as approximately constant over the time interval $\left(v,v+\Delta t(v)\right)$
because the Taylor expansion terms are suppressed by positive powers
of $H(v)\Delta t(v)$.

\section{\label{sec:Final-particle-number}Predicted spectrum and number density}

Using the time model described by eqs.\,(\ref{eq:fchi time model}),
(\ref{eq:btilde_k result}) and (\ref{eq:F(Ek) expression}), the
particle density spectrum\footnote{This can also be called the phase space density.}
$f_{\chi}(k,t)$ and the number density $n_{\chi}(t)$ can be computed
using slow varying quantities derived from the adiabatic invariant
formalism. We begin with the integration approximation
\begin{align}
f_{\chi}(k,t) & =\int_{t_{\mathrm{end}}}^{t}\frac{dv}{\Delta t(v)}\left|F(E_{k})+F^{*}(-E_{k})\right|^{2}\label{eq:FEgen}\\
 & \approx\int_{t_{\mathrm{end}}}^{t}dv\left|\frac{A_{2}(k,v)}{2}\right|^{2}\Delta t(v)\mathrm{sinc}^{2}\left[\left(\omega_{*}(v)-E_{k}(v)\right)\Delta t(v)\right],\label{eq:sincfunc}
\end{align}
with $F(E_{k})$ given in eq.\,(\ref{eq:F(Ek) expression}). Here
$F(-E_{k})$ and the cubic vertex amplitudes eqs.\,(\ref{eq:A1 definition})
and (\ref{eq:A3 definition}) were treated as subdominant to $F(E_{k})$
and $A_{2}$ of eq.\,(\ref{eq:A2 definition}) respectively.

It is evident that the particle production peaks at the resonance
time $\bar{v}_{k}$ when the energy matching condition $E_{k}(\bar{v}_{k})=\omega_{*}(\bar{v}_{k})$
is satisfied. As the inverse time width is much smaller than the energy
scale set by the inflaton mass, one can treat the limit $\Delta t^{-1}\rightarrow0$
as a good approximation. The spectrum of the particle production rate
then follows Fermi's golden rule:
\begin{equation}
\frac{\partial f_{\chi}(k,v)}{\partial v}\rightarrow2\pi\left|\frac{A_{2}(k,v)}{2}\right|^{2}\delta\left(2\omega_{*}(v)-2E_{k}(v)\right),\label{eq:FGR emerges}
\end{equation}
with $A_{2}/2$ acting as the transition matrix element, and the delta
function as the energy density of states. In analogy to refs.\,\citep{Tang:2017hvq,Mambrini:2021zpp},
this can be interpreted as the particle scattering process $\phi\phi\rightarrow\chi\chi$
of two inflatons with total energy $2\omega_{*}(v)$ under going gravitational
annihilation to produce two $\chi$ particles with total energy $2E_{k}(v)$
in the center of mass frame.

After integration of eq.\,(\ref{eq:FGR emerges}), the spectrum and
number density limit to
\begin{align}
f_{\chi}(k,t) & \rightarrow\frac{\pi}{4}\frac{\left|A_{2}(k,\bar{v}_{k})\right|^{2}}{\left|\partial_{v}\left(E_{k}(\bar{v}_{k})-\omega_{*}(\bar{v}_{k})\right)\right|}\Theta(t_{\mathrm{end}}\leq\bar{v}_{k}\leq t),\label{eq:spectrum limit}\\
a^{3}(t)n_{\chi}(t) & \rightarrow\frac{1}{8\pi}\int_{t_{\mathrm{end}}}^{t}dv\,a^{3}(v)\omega_{*}^{2}(v)\left|A_{2}(\bar{k}(v),v)\right|^{2}\sqrt{1-\frac{m_{\chi}^{2}}{\omega_{*}^{2}(v)}}\Theta\left(\omega_{*}(v)-m_{\chi}\right),\label{eq:density limit}
\end{align}
respectively, where $\bar{v}_{k}$ is the time such that 
\begin{equation}
E_{k}(\bar{v}_{k})=\omega_{*}(\bar{v}_{k}),\label{eq:vbar_k-resonance-condition}
\end{equation}
and we defined $\bar{k}(v)=a(v)\sqrt{\omega_{*}^{2}(v)-m_{\chi}^{2}}$.
The number density was obtained by resolving the delta function through
wave-vector $k$ integration. Integration of the spectrum eq.\,(\ref{eq:spectrum limit})
with bounds $k\in\left[\bar{k}(\bar{v}_{k}),\bar{k}(t)\right]$ yields
an alternative formula for the number density. Using the explicit
form of eq.\,(\ref{eq:A2 definition}) gives
\begin{align}
f_{\chi}(k,t) & \approx\frac{9\pi}{64}\frac{H_{\mathrm{slow}}^{3}(\bar{v}_{k})}{\omega_{*}(\bar{v}_{k})(\omega_{*}^{2}(u)-m_{\chi}^{2})}\left(1-6\xi+\frac{m_{\chi}^{2}}{2\omega_{*}^{2}(\bar{v}_{k})}\right)^{2}\Theta(t_{\mathrm{end}}\leq\bar{v}_{k}\leq t),\label{eq:spectrum A2 plugged in}\\
a^{3}(t)n_{\chi}(t) & \approx\frac{9}{128\pi}\int_{a_{\mathrm{end}}}^{a(t)}du\,u^{2}H_{\mathrm{slow}}^{3}(u)\left(1-6\xi+\frac{m_{\chi}^{2}}{2\omega_{*}^{2}(u)}\right)^{2}\sqrt{1-\frac{m_{\chi}^{2}}{\omega_{*}^{2}(u)}}\Theta\left(\omega_{*}(u)-m_{\chi}\right),\label{eq:number density A2 plugged in}
\end{align}
where we treated $\partial_{v}\omega_{*}(\bar{v}_{k})\sim H^{3}(\bar{v}_{k})/m_{\phi}$
as subdominant to $\partial_{v}E_{k}(\bar{v}_{k})\sim H(\bar{v}_{k})m_{\phi}$,
and made the change of variables $u=a(v)$ in the last line.

For a quadratic potential, the scale factor dependence of the Hubble
rate and oscillation frequency can be treated as 
\begin{equation}
H_{\mathrm{slow}}(a_{\mathrm{slow}}(t))=H_{\mathrm{end}}\left(\frac{a_{\mathrm{end}}}{a_{\mathrm{slow}}(t)}\right)^{3/2},\label{eq:Hslow quad potential}
\end{equation}
 and $\omega_{*}=m_{\phi}$ respectively, and this yields the estimates
\begin{align}
f_{\chi}(k,t) & =\frac{9\pi}{64}\frac{H_{\mathrm{end}}^{3}}{m_{\phi}(m_{\phi}^{2}-m_{\chi}^{2})}\left(1-6\xi+\frac{m_{\chi}^{2}}{2m_{\phi}^{2}}\right)^{2}\left(\frac{k/a_{\mathrm{end}}}{\sqrt{m_{\phi}^{2}-m_{\chi}^{2}}}\right)^{-9/2}\Theta(t_{\mathrm{end}}\leq\bar{v}_{k}\leq t),\label{eq:fchi quadratic}\\
\left(\frac{a(t)}{a_{\mathrm{end}}}\right)^{3}n_{\chi}(t) & =\frac{3H_{\mathrm{end}}^{3}}{64\pi}\left(1-6\xi+\frac{m_{\chi}^{2}}{2m_{\phi}^{2}}\right)^{2}\sqrt{1-\frac{m_{\chi}^{2}}{m_{\phi}^{2}}}\left(1-\left(\frac{a_{\mathrm{end}}}{a(t)}\right)^{3/2}\right)\Theta\left(m_{\phi}-m_{\chi}\right),\label{eq:nchi quadratic}
\end{align}
where $H_{\mathrm{end}}=H_{\mathrm{slow}}(t_{\mathrm{end}})$. The
number density scales as $n_{\chi}\propto a^{-3}(t)$ at large times,
as expected for non-relativistic massive particles. Note that the
spectrum scales as $f_{\chi}\propto k^{-9/2}$, matching the results
of ref.\,\citep{Ema:2018ucl}. These estimates of the number density
are larger by a factor of 2 from the results of the spectral model
of ref.\,\citep{Chung:2018ayg}. The spectral model suffered from
an ambiguity of setting the correct time scale of Fermi's golden rule.
The time model has no such ambiguity, with eq.\,(\ref{eq:FGR emerges})
emerging as a result without being put in by hand.

For a more general potential which is quadratic near its minimum,
we show in appendix \ref{sec:A-perturbative-expansion} that a perturbative
expansion in the adiabatic invariant charge $Q$ allows us to write
the Hubble expansion rate as
\begin{equation}
H_{\mathrm{slow}}(u)=\tilde{H}_{\mathrm{end}}\left(\frac{a_{\mathrm{end}}}{u}\right)^{3/2}\left(1+h_{1}\tilde{Q}_{\mathrm{end}}\left(\frac{a_{\mathrm{end}}}{u}\right)^{3}+h_{2}\tilde{Q}_{\mathrm{end}}^{2}\left(\frac{a_{\mathrm{end}}}{u}\right)^{6}+\dots\right),\label{eq:Hslow(u) non-quad potential}
\end{equation}
where we define
\begin{equation}
\tilde{Q}_{\mathrm{end}}\equiv\frac{Qa_{\mathrm{end}}^{-3}}{2\pi m_{\phi}M_{P}^{2}},\quad\text{and}\quad\tilde{H}_{\mathrm{end}}\equiv\sqrt{\frac{Qm_{\phi}a_{\mathrm{end}}^{-3}}{6\pi M_{P}^{2}}},
\end{equation}
and 
\begin{align}
h_{1} & =\frac{3}{8}\left(-5\alpha_{3}^{2}+2\alpha_{4}\right)\\
h_{2} & =\frac{1}{128}\left(-3045\alpha_{3}^{4}+3780\alpha_{3}^{2}\alpha_{4}-308\alpha_{4}^{2}-1120\alpha_{3}\alpha_{5}+160\alpha_{6}\right)\\
\alpha_{n} & \equiv\frac{1}{n!}\frac{M_{P}^{n-2}}{m_{\phi}^{2}}\left.\frac{\partial^{n}V}{\partial\phi^{n}}\right|_{\phi=\phi_{\min}},
\end{align}
are the relevant coefficients. The leading $u^{-3/2}$ term of eq.\,(\ref{eq:Hslow(u) non-quad potential})
quickly dominates at later times. The spectrum eq.\,(\ref{eq:spectrum A2 plugged in})
can therefore be estimated as 
\begin{align}
f_{\chi}(k,t) & =\frac{9\pi}{64}\frac{\tilde{H}_{\mathrm{end}}^{3}}{\omega_{*}(\bar{v}_{k})(\omega_{*}^{2}(\bar{v}_{k})-m_{\chi}^{2})}\left(1-6\xi+\frac{m_{\chi}^{2}}{2\omega_{*}^{2}(\bar{v}_{k})}\right)^{2}\left(\frac{k/a_{\mathrm{end}}}{\sqrt{\omega_{*}^{2}(\bar{v}_{k})-m_{\chi}^{2}}}\right)^{-9/2}\Theta(t_{\mathrm{end}}\leq\bar{v}_{k}\leq t).\label{eq:spectrum with Q}
\end{align}
for large $k$. Let us now neglect the time dependence of the frequency
$\omega_{*}$ as $H^{2}/m_{\phi}^{2}$ relative corrections to $m_{\phi}$.
The integral for the number density in eq.\,(\ref{eq:number density A2 plugged in})
is then proportional to
\begin{align}
\int_{a_{\mathrm{end}}}^{\infty}\frac{du\,u^{2}}{a_{\mathrm{end}}^{3}}H_{\mathrm{slow}}^{3}(u) & =\tilde{H}_{\mathrm{end}}^{3}\int_{1}^{\infty}dx\,x^{-\frac{5}{2}}\left(1+3h_{1}\tilde{Q}_{\mathrm{end}}x^{-3}+3\left(h_{1}^{2}+h_{2}\right)\tilde{Q}_{\mathrm{end}}^{2}x^{-6}+\dots\right)\\
 & =\frac{2}{3}\tilde{H}_{\mathrm{end}}^{3}\left(1+h_{1}\tilde{Q}_{\mathrm{end}}+\frac{3}{5}\left(h_{1}^{2}+h_{2}\right)\tilde{Q}_{\mathrm{end}}^{2}+\dots\right),\label{eq:Relevant u integral result}
\end{align}
and therefore the number density is estimated as 
\begin{align}
\left(\frac{a(t)}{a_{\mathrm{end}}}\right)^{3}n_{\chi}(t) & =\frac{3F}{64\pi}\tilde{H}_{\mathrm{end}}^{3}\left(1-6\xi+\frac{m_{\chi}^{2}}{2m_{\phi}^{2}}\right)^{2}\sqrt{1-\frac{m_{\chi}^{2}}{m_{\phi}^{2}}}\Theta\left(m_{\phi}-m_{\chi}\right)\label{eq:number density with Q}\\
F & =1+h_{1}\tilde{Q}_{\mathrm{end}}+\frac{3}{5}\left(h_{1}^{2}+h_{2}\right)\tilde{Q}_{\mathrm{end}}^{2}+\dots
\end{align}
at late times for $m_{\phi}>m_{\chi}$. Note the differences between
this number density and eq.\,(\ref{eq:nchi quadratic}) for a purely
quadratic potential. The effect of the non-quadratic terms of the
inflaton potential is summarized by the factors $F$ and $\tilde{H}_{\mathrm{end}}/H_{\mathrm{end}}$,
which are derived from the adiabatic invariant $Q$ and the $\alpha_{n}$
coefficients. In contrast with ref.\,\citep{Turner:1983he}, this
formalism is applicable for potentials that are asymmetric about their
minimum.
\begin{figure}
\centering{}\includegraphics[scale=0.25]{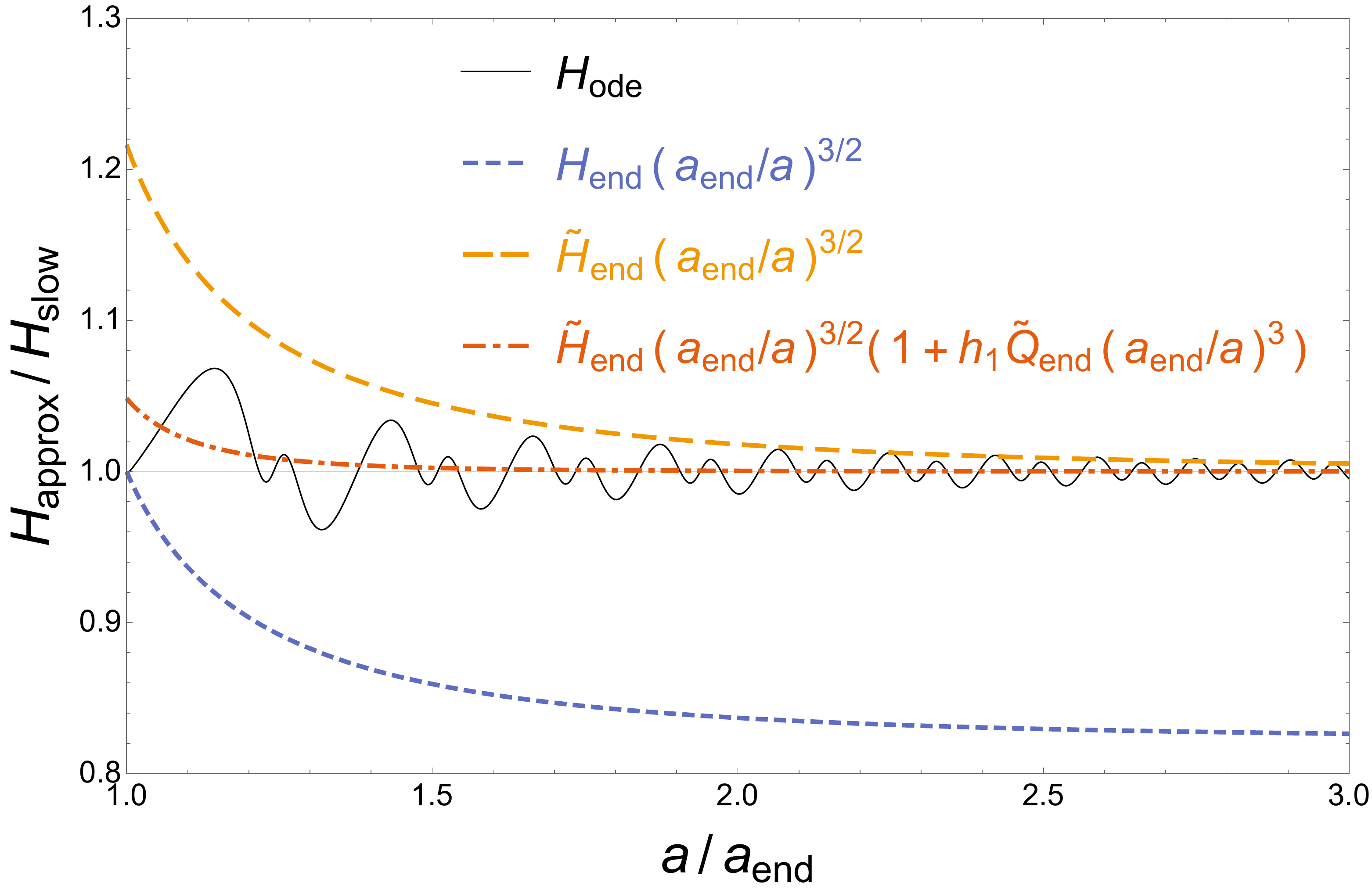}\includegraphics[scale=0.25]{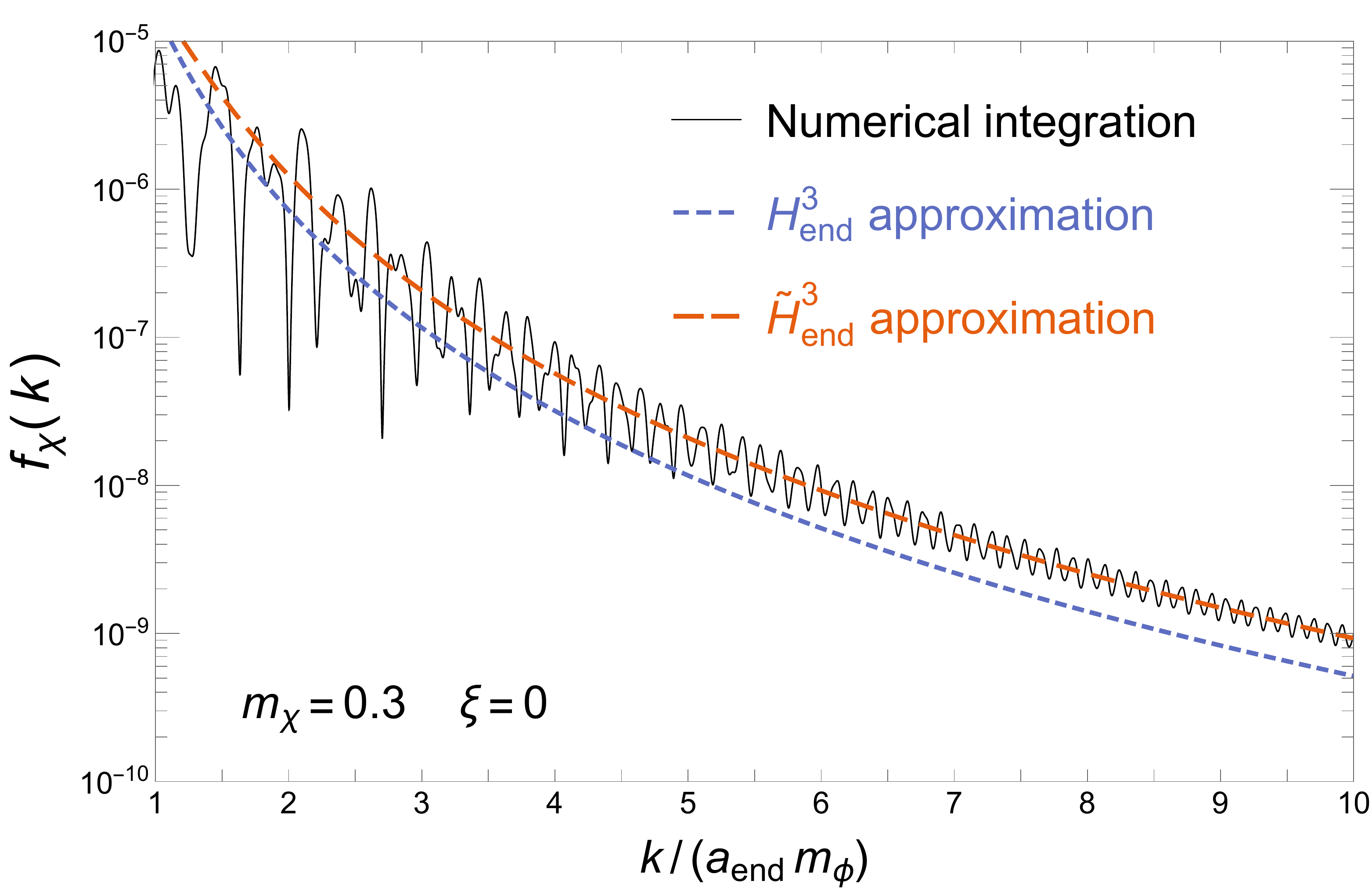}\caption{\label{fig:Comparing H and fchi approximations} Left: approximations
of the Hubble rate given by eqs.\,(\ref{eq:Hslow quad potential})
and (\ref{eq:Hslow(u) non-quad potential}) for the potential eq.\,(\ref{eq:example potential}).
Right: approximations of the spectrum given by eqs.\,(\ref{eq:fchi quadratic})
and (\ref{eq:number density with Q}).}
\end{figure}

As an example, consider the potential
\begin{equation}
V(\phi)=M^{4}\left(1-\left(\frac{\phi}{v_{\phi}}\right)^{6}\right)^{2},\label{eq:example potential}
\end{equation}
with $v_{\phi}=M_{P}/2$. The relevant parameters are
\begin{align}
m_{\phi} & =30.41H_{\mathrm{end}} & \tilde{Q}_{\mathrm{end}} & =4.875\times10^{-3}\\
\alpha_{3} & =5 & \alpha_{4} & =155/6\label{eq:a3a4}\\
\alpha_{5} & =260/3 & \alpha_{6} & =1844/9\label{a5a6}\\
h_{1} & =-55/2 & h_{2} & =-119973/128
\end{align}
 where the large numbers stem partly from the large $M_{p}/m_{\phi}$
ratio (see e.g.~eq.~(\ref{eq:alpha23forcubic}) and the comment
below (\ref{eq:hslowovm}) for further explanation). When the inflaton
potential is quadratic, the Hubble rate has the time behavior of eq.\,(\ref{eq:Hslow quad potential}).
For general potentials, the Hubble rate is more accurately modeled
as eq.\,(\ref{eq:Hslow(u) non-quad potential}), with the leading
order $u^{-3/2}$ term dominating at late times. The ratio of these
two estimates will deviate from unity for non-quadratic potentials,
and this will change the magnitude of predicted spectrum and number
density. For the potential eq.\,(\ref{eq:example potential}), this
ratio is
\begin{equation}
\frac{\tilde{H}_{\mathrm{end}}^{3}}{H_{\mathrm{end}}^{3}}=1.843,
\end{equation}
which deviates significantly from unity. The relative corrections
to the $u^{-3/2}$ dominant contribution are
\begin{align}
h_{1}\tilde{Q}_{\mathrm{end}} & =-0.134\qquad\text{and}\qquad\frac{3}{5}\left(h_{1}^{2}+h_{2}\right)\tilde{Q}_{\mathrm{end}}^{2}=-2.58\times10^{-3}.\label{eq:suppressionexample}
\end{align}
Overall, the predictions given by eqs.\,(\ref{eq:nchi quadratic})
and (\ref{eq:number density with Q}) differ by a factor of around
1.6 for this example. As shown in figure\,\ref{fig:Comparing H and fchi approximations},
numerical results demonstrate this for the example potential eq.\,(\ref{eq:example potential}),
with eq.\,(\ref{eq:spectrum with Q}) being a better numerical fit
of the spectrum than eq.\,(\ref{eq:fchi quadratic}) at large $k$.

\section{\label{sec:Comparison-of-different}Comparison of different computational
approaches}

In this section, we compare some different approaches to computing
the super-Hubble mass scale particle production. The second main result
of this paper presented in section \ref{subsec:Exact-numerical-integral}
is to note the $O(m_{\phi}/H)$ (which can easily be a factor of 100)
enhancement in numerical integration efficiency is achieved for a
formulation obtained by subtracting out the slowly varying component
of $H$. All of our numerical illustrations will be done with a prototype
inflationary model presented in subsection \ref{subsec:Inflationary-model-summary}.
In subsection \ref{subsec:Time-model-in}, we compare the particle
production results of the Gaussian spectral model of ref.\,\citep{Chung:2018ayg}
with a generalized Gaussian spectral model based on an adiabatic invariant
and an analogous time-model of \ref{sec:Time-model}. In subsection
\ref{subsec:Asymmetric-time-model}, we illustrate the time model
in the context of potentials where the maximum displacement from the
minimum of the potential is asymmetric on two sides of the minimum.

\subsection{\label{subsec:Inflationary-model-summary}Inflationary model used
for illustration}

As a prototype of an inflaton potential for low-scale inflation, we
test the time model developed in this paper by considering the hilltop
model :
\begin{equation}
V(\phi)=M^{4}\left(1-\left(\frac{\phi}{v_{\phi}}\right)^{n}\right)^{2},\label{eq:hiltop potential example}
\end{equation}
where $n>2$ is an integer and $v_{\phi}>0$ is the vacuum expectation
value of the inflaton at the minimum of its potential \citep{Ema:2017rkk,Barenboim:2013wra,Boubekeur:2005zm,Albrecht:1982wi,Linde:1981mu}.
The mass of the inflaton near the minimum is 
\begin{equation}
m_{\phi}=n\sqrt{2}M^{2}/v_{\phi}\,,
\end{equation}
and there is an effective cubic interaction of field displacements
from the minimum (as well as higher order interactions). The Hubble
expansion rate at the end of inflation is estimated as $H_{\mathrm{end}}\simeq\left(\sqrt{3}M_{P}\right)^{-1}M^{2}$.
By calculating the large-scale curvature perturbations, the scalar
spectral index and the tensor-to-scalar ratio is found to be
\begin{equation}
n_{s}=1-\frac{2}{N}\frac{n-1}{n-2},\qquad r=\frac{16n}{N(n-2)}\left[\frac{1}{2Nn(n-2)}\frac{v_{\phi}^{2}}{M_{P}^{2}}\right]^{\frac{n}{n-2}},
\end{equation}
where the e-folding number $N$ of the CMB lies between 50 and 60
\citep{Liddle:2000cg}. As observed by the Planck satellite, the overall
normalization of the curvature perturbation implies
\begin{equation}
P_{\zeta}\simeq\frac{M^{4}}{12\pi^{2}}\left[\frac{2n(N(n-2))^{n-1}}{v_{\phi}^{n}M_{P}^{n-4}}\right]^{\frac{2}{n-2}}\simeq2.2\times10^{-9},
\end{equation}
which relates $M$ and $v_{\phi}$. Hence there is one free parameter
left, which can be taken as $v_{\phi}$ \citep{Ema:2017rkk}.

The measured range of the spectral index ($n_{s}=0.968(6)$ at 1$\sigma$
level \citep{Planck:2015sxf}) can be made consistent with $n\geq6$
\citep{Ema:2017rkk}. We take $n=6$ in the potential eq.\,(\ref{eq:hiltop potential example})
and $v_{\phi}=0.5M_{P}$ to compare to the numerical results of ref.\,\citep{Ema:2018ucl}.
This is a somewhat tuned model that could be destabilized by loop
generated Planck suppressed operators, but it serves as an algebraically
simple demonstration of the adiabatic invariant formalism.\footnote{Since most models of inflation are tuned to some extent and since
the current UV physics picture in the context of landscape most likely
suggests some tuning is possible, we will still consider this example
to be not completely unrealistic.} It is interesting that the cosmological data driven phenomenology
favoring $n\geq6$ also enforces $m_{\phi}/H\gg1$ since 
\begin{equation}
\frac{m_{\phi}}{H_{\mathrm{end}}}\approx\frac{n\sqrt{2}M^{2}/v_{\phi}}{\sqrt{M^{4}/(3M_{P}^{2})}}=2n\sqrt{6},
\end{equation}
which is equivalent to the condition $m_{\phi}\approx30H_{\mathrm{end}}$
with $n=6$. As we saw in eqs.\,(\ref{eq:a3a4}) and (\ref{a5a6}),
the expansion in non-quadratic parameters $\alpha_{n}\propto\partial_{\phi}^{n}V$
defined in eq.\,(\ref{eq:alphadef}) allows us to capture many different
models \citep{Ling:2021zlj,Bose:2013kya,Brax:2010ai} even though
we focused for numerical illustrations on the model of eq.\,(\ref{eq:hiltop potential example}).

\subsection{\label{subsec:Exact-numerical-integral}Exact versus fast component
numerical integration}

In this section, we will summarize our numerical procedure to compute
$\beta_{k}$ using the brute force exact integration eq.\,(\ref{eq:beta exact definition})
and the fast only component integration eq.\,(\ref{eq:betak massaged}).
We will then discuss figures that illustrate the differences between
these methods, and explain observed features.

Both methods first required solving for the solution $\phi_{\mathrm{ode}}(t)$
to the inflaton equation of motion, given by the non-linear ordinary
differential equation
\begin{equation}
\ddot{\phi}_{\mathrm{ode}}+3\sqrt{\frac{\frac{1}{2}\dot{\phi}_{\mathrm{ode}}^{2}+V(\phi_{\mathrm{ode}})}{3M_{P}^{2}}}\dot{\phi}_{\mathrm{ode}}+V'(\phi_{\mathrm{ode}})=0,
\end{equation}
where we assumed a background homogeneous inflaton field that dominates
the energy density. Here the initial conditions and parameters were
chosen such that at least $N\approx55$ e-folds in the scale factor
occurred before the end of inflation. We used the initial conditions
$\phi(t_{i})=0.17\phi_{\mathrm{end}}$ and $\dot{\phi}(t_{i})=0$,
and found that $t_{i}-t_{\mathrm{end}}\approx-63H_{\mathrm{end}}^{-1}$.\footnote{The fractional difference in the $\phi_{\mathrm{ode}}$ trajectory
that would arise from the standard slow-roll boundary condition of
$3H(t_{i})\dot{\phi}_{\mathrm{ode}}(t_{i})=-V'(\phi_{\mathrm{ode}}(t_{i}))$
is $O(10^{-8})$.} It was found that the relevant portions of the ODE solution were
nearly independent of the slow roll initial conditions.

The time $t_{\mathrm{end}}$ when the quasi-dS era ends is given by
the solution to

\begin{equation}
\phi_{\mathrm{ode}}(t_{\mathrm{end}})=\phi_{\mathrm{end}}\equiv v_{\phi}\left(\frac{v_{\phi}}{\sqrt{2}nM_{P}}\right)^{1/(n-1)},
\end{equation}
following the criterion used in ref.\,\citep{Ema:2018ucl}. For the
sake of convenience, we shifted time such that $t_{\mathrm{end}}=0$.
We also set our time scale such that $M=5297H_{\mathrm{end}}$ and
$M_{P}=1.565\times10^{7}H_{\mathrm{end}}$, where we have defined
\begin{equation}
H_{\mathrm{end}}\equiv\sqrt{\frac{V(\phi_{\mathrm{end}})}{3M_{P}^{2}}},
\end{equation}
which is approximately the usual Hubble expansion rate at the end
of inflation.

The Hubble rate and Ricci scalar were determined by the Einstein equations
$3M_{P}^{2}H_{\mathrm{ode}}^{2}=$ $\frac{1}{2}\dot{\phi}_{\mathrm{ode}}^{2}+V(\phi_{\mathrm{ode}})$,
and $M_{P}^{2}R_{\mathrm{ode}}=\dot{\phi}_{\mathrm{ode}}^{2}-4V(\phi_{\mathrm{ode}})$,
respectively. The scalar factor was then computed by solving $\dot{a}_{\mathrm{ode}}=H_{\mathrm{ode}}a_{\mathrm{ode}}$.
With these quantities in hand, we evaluated the exact $\beta_{k}$
integration as
\begin{align}
\beta_{k}^{(\mathrm{exact})}(t_{f}) & =\frac{1}{2}\int_{t_{i}}^{t_{f}}dt\frac{H_{\mathrm{ode}}(t)m_{\chi}^{2}+\frac{1}{12}(1-6\xi)\left(\dot{R}_{\mathrm{ode}}(t)+2H_{\mathrm{ode}}(t)R_{\mathrm{ode}}(t)\right)}{\frac{k^{2}}{a_{\mathrm{ode}}^{2}(t)}+m_{\chi}^{2}+\frac{1}{6}(1-6\xi)R_{\mathrm{ode}}(t)}e^{-2i\Phi_{k}(t)}\label{eq:exact}\\
\Phi_{k}(t) & \equiv\int_{t_{i}}^{t}dt'\sqrt{\frac{k^{2}}{a_{\mathrm{ode}}^{2}(t')}+m_{\chi}^{2}+\frac{1}{6}(1-6\xi)R_{\mathrm{ode}}(t')},
\end{align}
where $t_{i}$ was chosen sufficiently in the past compared to $t_{\mathrm{end}}$
to approximate the adiabatic initial conditions of the Bunch-Davies
vacuum. In our computations, we chose $H_{\mathrm{end}}(t_{i}-t_{\mathrm{end}})=-5$.

To obtain the necessary components for the fast only integration,
we solved the adiabatic invariant equation to obtain the slow time
behavior of the Hubble expansion parameter. We first defined
\begin{equation}
J(V_{m})=2\int d\phi\Theta\left(V_{m}-V(\phi)\right)\sqrt{2V_{m}-2V(\phi)},
\end{equation}
and computed it for various values of $V_{m}$ starting with $V(\phi_{\mathrm{end}})$
and proceeding to smaller values as needed. The step function ensured
that finding the turning points was numerically unnecessary, and the
smooth monotonic nature of the integration ensured that only a few
sampling points of $V_{m}$ were needed to obtain a good interpolation.

The monotonic nature of $J(V_{m})$ also guarantees the existence
of its inverse. The adiabatic invariant equation $Q=a_{\mathrm{slow}}^{3}J(V_{m})$
can then be written as
\begin{equation}
V_{m}(a_{\mathrm{slow}})=\mathrm{inverse\,of\,}J\left(Qa_{\mathrm{slow}}^{-3}\right),
\end{equation}
which determines $V_{m}$ in terms of $a_{\mathrm{slow}}$. We computed
the adiabatic invariant using the initial conditions as $Q=a_{\mathrm{end}}^{3}J\left(V(\phi_{\mathrm{end}})\right)$.
The time dependence of $a_{\mathrm{slow}}$ was then obtained by integrating
\begin{equation}
\dot{a}_{\mathrm{slow}}=a_{\mathrm{slow}}\sqrt{\frac{V_{m}(a_{\mathrm{slow}})}{3M_{P}^{2}}},
\end{equation}
with the initial condition being $a_{\mathrm{slow}}(t_{\mathrm{end}})=a_{\mathrm{end}}$.
For convenience, we chose for our numerical work the normalization
$a_{\mathrm{end}}=1$.

With $a_{\mathrm{slow}}(t)$ in hand, we arrived at the desired quantities
\begin{equation}
H_{\mathrm{slow}}(t)\equiv\sqrt{\frac{V_{m}(a_{\mathrm{slow}}(t))}{3M_{P}^{2}}},\quad\text{and}\quad H_{\mathrm{fast}}(t)\equiv H_{\mathrm{ode}}(t)-H_{\mathrm{slow}}(t),
\end{equation}
where $H_{\mathrm{fast}}$ is defined to be zero before $t_{\mathrm{end}}$.
One can compute other slowly varying quantities by taking $\langle\frac{1}{2}\dot{\phi}^{2}\rangle_{\mathrm{slow}}=\langle V(\phi)\rangle_{\mathrm{slow}}\approx\frac{1}{2}V_{m}(a_{\mathrm{slow}}(t))$.
For example, $\rho_{\mathrm{slow}}=V_{m}$ and $R_{\mathrm{slow}}=-V_{m}/M_{P}^{2}$.
However, only $H_{\mathrm{slow}}$ was necessary for our purposes
here as $\dot{R}_{\mathrm{ode}}$ is dominated by its oscillatory
components.

The fast only integration refers to subtracting out the slow components
and sub-leading contributions, such as assuming that $\dot{R}\gg HR$
and $m_{\chi}^{2}\gg R$, as indicated by eq.\,(\ref{eq:betak massaged}).
We also assumed $\dot{R}_{\mathrm{slow}}\ll\dot{R}_{\mathrm{fast}}$
and kept $R_{\mathrm{ode}}$ for the sake of computational simplicity
and accuracy. For similar reasons, we avoided replacing $a_{\mathrm{ode}}$
with $a_{\mathrm{slow}}$. In summary, we evaluated
\begin{align}
\beta_{k}^{(\mathrm{fast})}(t_{f}) & =\frac{1}{2}\int_{t_{\mathrm{end}}}^{t_{f}}dt\left(H_{\text{fast}}(t)+\frac{1}{12}\left(1-6\xi\right)\frac{\dot{R}_{\mathrm{ode}}(t)}{m_{\chi}^{2}}\right)\frac{m_{\chi}^{2}}{E_{k}^{2}(t)}e^{-2i\int_{t_{\text{end}}}^{t}E_{k}(t')dt'},\label{eq:beta fast def}
\end{align}
where $E_{k}(t)=\sqrt{k^{2}/a_{\mathrm{ode}}^{2}(t)+m_{\chi}^{2}}$.
The integration time $t_{f}$ should be chosen long enough such that
the dominant resonance occurs for a given $k$ mode i.e.\,$t_{f}>\bar{v}_{k}$,
where $\bar{v}_{k}$ is defined in eq.\,(\ref{eq:vbar_k-resonance-condition}).
This is particularly important for particle mass $m_{\chi}$ close
to the $m_{\phi}$ threshold. The scale factor at resonance is $a(\bar{v}_{k})\approx k/\sqrt{m_{\phi}^{2}-m_{\chi}^{2}}$,
and therefore $\bar{v}_{k}\rightarrow\infty$ as $m_{\chi}\rightarrow m_{\phi}$.
In practice, one uses a cutoff $k_{c}$ for the wave-vector integration
of the number density. The relative error incurred in the number density
can be estimated to be
\begin{equation}
\left(\frac{k_{c}/a_{\mathrm{end}}}{\sqrt{m_{\phi}^{2}-m_{\chi}^{2}}}\right)^{-3/2},
\end{equation}
as implied by eq.\,(\ref{eq:nchi quadratic}).

\begin{figure}
\begin{centering}
\includegraphics[scale=0.25]{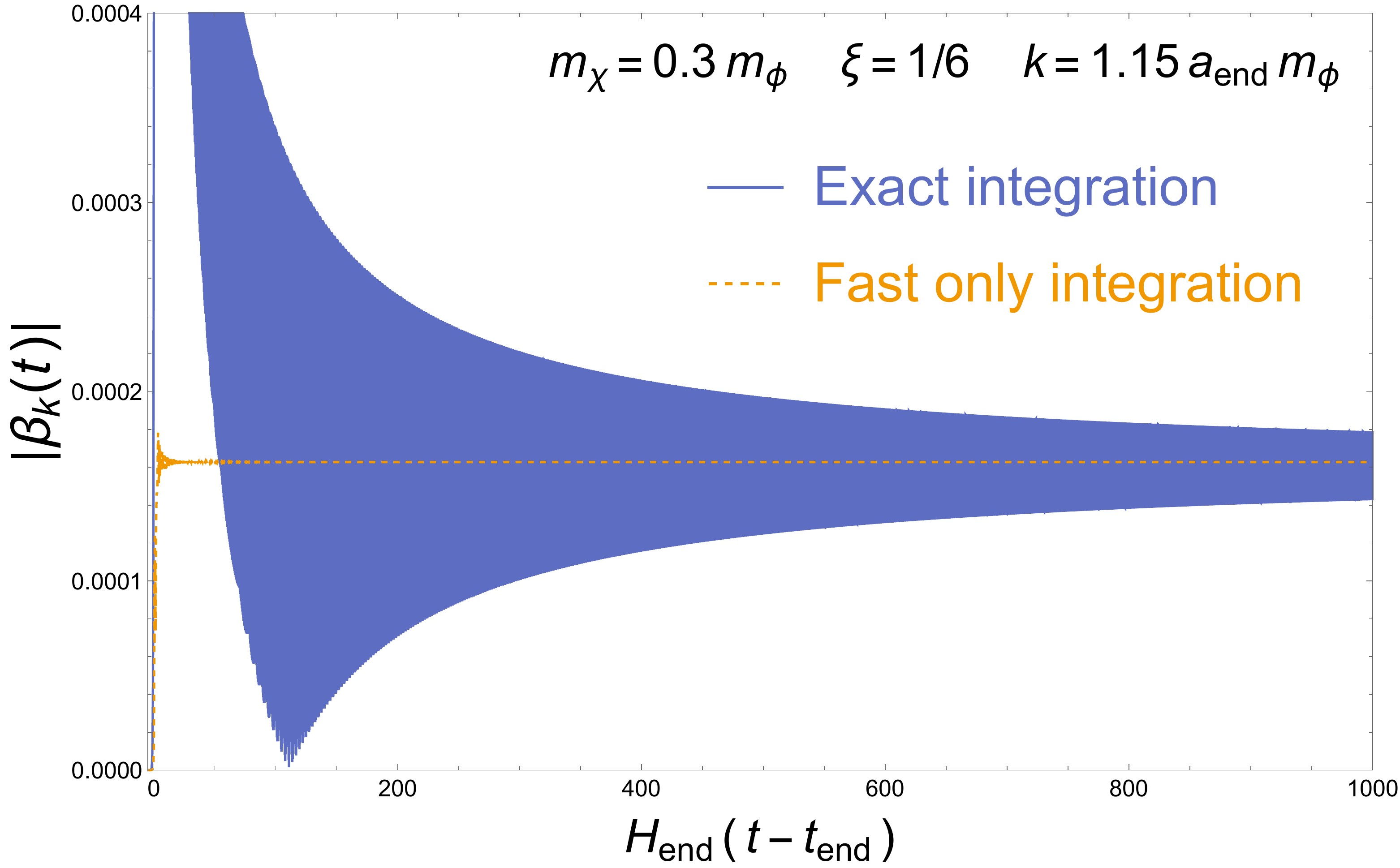}\includegraphics[scale=0.25]{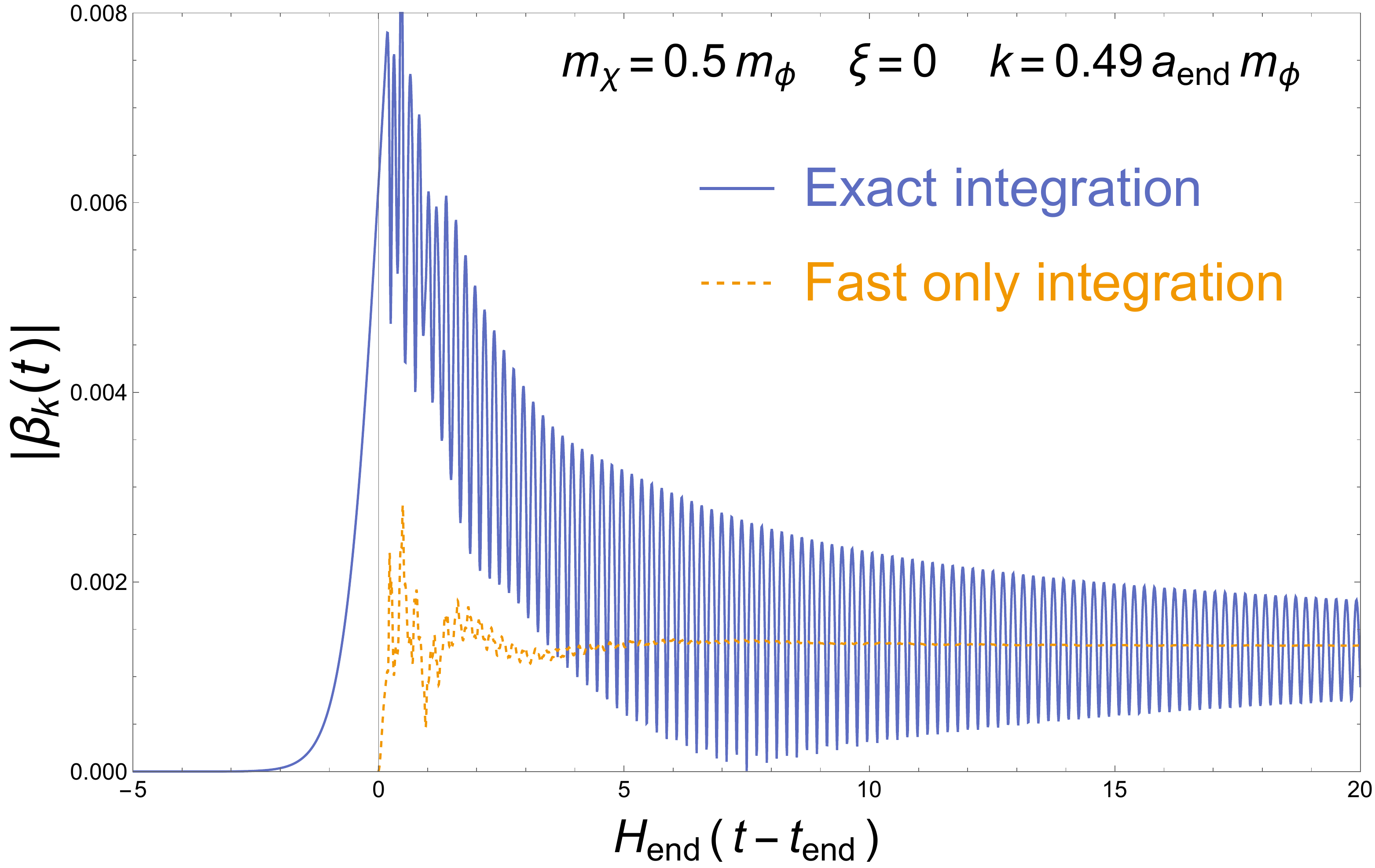}
\par\end{centering}
\caption{\label{fig:Comparison-of-convergence}The exact (slow+fast) vs fast
only integrations for computing $\beta_{k}(t)$, given by eqs.\,(\ref{eq:exact})
and (\ref{eq:beta fast def}), respectively, are compared for the
inflationary model of eq.\,(\ref{eq:hiltop potential example}).
Although both methods tend towards nearly identical asymptotic values,
the fast integration converges 3 orders of magnitude faster in integration
time than the exact integration. The long convergence time of the
latter is due to the slowly varying component in eq.\,(\ref{eq:beta exact definition}),
which damps down only as $t^{-1}$ for large times. On the right,
we see that the exact integration is sensitive to the initial conditions
before the end of inflation, while the fast integration is not. Starting
the exact integration at $t_{\text{end}}$ would lead to a large deviation
in the asymptotic limit, which is due to the adiabatic condition not
being satisfied at that time.}
\end{figure}

Figure \ref{fig:Comparison-of-convergence} illustrates the efficiency
advantage of fast integration formulation of eq.\,(\ref{eq:beta fast def})
over the exact integration formulation of eq.\,(\ref{eq:beta exact definition}).
For this illustration, the parameters were chosen such that the $2\rightarrow2$
resonance at time (see eq.\,(\ref{eq:FGR emerges})) in the $m_{\phi}/H_{\mathrm{end}}\approx30$
model of section \ref{subsec:Inflationary-model-summary} is at $t$
satisfying
\begin{equation}
\frac{H_{\mathrm{end}}t_{\mathrm{end}}}{H_{\mathrm{end}}t}\approx\left(\frac{a_{\text{end}}}{a_{\mathrm{slow}}(t)}\right)^{3/2}\sim O(1),
\end{equation}
which explains the spike in the dashed curve at near unity time. The
fast integration formulation of eq.\,(\ref{eq:beta fast def}) occurs
because the integration over the sinc function in eq.\,(\ref{eq:sincfunc})
converges on a time scale 
\begin{equation}
\Delta t\approx\frac{14}{m_{\phi}}\approx\frac{14}{30}H_{\mathrm{end}}^{-1},
\end{equation}
as given by \,eq.\,(\ref{eq:deltatresult}).

The comparatively slow convergence of the ``exact'' numerical integration
stems from 
\begin{equation}
\beta_{k}(t)=\frac{1}{2}\int^{t}dt'\frac{m_{\chi}^{2}H(t')}{k^{2}/a^{2}+m_{\chi}^{2}}e^{-2i\int_{-\infty}^{t'}dt''\frac{\omega_{k}(t'')}{a(t'')}},
\end{equation}
which has a fluctuation amplitude $\Delta\beta_{k}$ scaling as $t^{-1}$
if $k/a\ll1$ after the resonance is reached. This requires an integration
time $t$ of
\begin{equation}
H_{\mathrm{end}}t\sim10^{-1}|\Delta\beta|^{-1}\label{eq:integrationerror}
\end{equation}
for $|\Delta\beta|$ is the accuracy of the amplitude that is desired.
For figure \ref{fig:Comparison-of-convergence}, the final $|\beta_{k}|$
is $O(10^{-4})$, which means we require an accuracy of $|\Delta\beta|\sim|\beta_{k}|\sim10^{-4}$.
This gives a convergence time of
\begin{equation}
H_{\mathrm{end}}t\gtrsim10^{3}.
\end{equation}
for the brute force numerical integration (``exact integration'').
As shown by eq.\,(\ref{eq:integrationerror}), the advantage of the
fast only integration method increases for small $|\beta_{k}|$ which
is typical for the scenarios of physical interest.

Figure \ref{fig:Comparison-of-convergence} illustrates another advantage
of the fast only integration technique over the exact integration
for a mode slightly off resonance at the initial time $t_{\mathrm{end}}$.
The exact integration is more sensitive to satisfying the adiabatic
vacuum condition during the quasi-dS era because imposing the standard
adiabatic vacuum condition at a non-adiabatic time of $t_{\mathrm{end}}$
is apparently an excited state which contains a sizable amount late
time high frequency particle modes. On the other hand the fast only
integration started at $t_{\mathrm{end}}$ has by construction subtracted
out the leading nonadiabaticity governed by $O(H_{\mathrm{end}})$
dynamical scales, giving rise to an apparently acceptable level of
consistency with the implicitly assumed adiabaticity (with respect
to high frequency modes) at $t_{\mathrm{end}}$.

To put this another way, assuming an adiabatic boundary condition
at a nonadiabatic time with an adiabaticity violating scale of $O(H_{\mathrm{end}})$
is equivalent to assuming the extra presence of late-time high frequency
(i.e.\,$\omega_{k}/a_{\mathrm{end}}\gg H_{\mathrm{end}}$) modes,
even though naively $O(H_{\mathrm{end}})$ nonadiabaticity should
not contain any high frequency components. This somewhat surprising
result may be due to the fact that imposing the adiabatic vacuum boundary
condition at a nonadiabatic time $t_{\mathrm{end}}$ implicitly contains
a step function in time with a non-negligible amplitude, which necessarily
includes non-negligible high frequency components.

\begin{figure}
\begin{centering}
\includegraphics[scale=0.25]{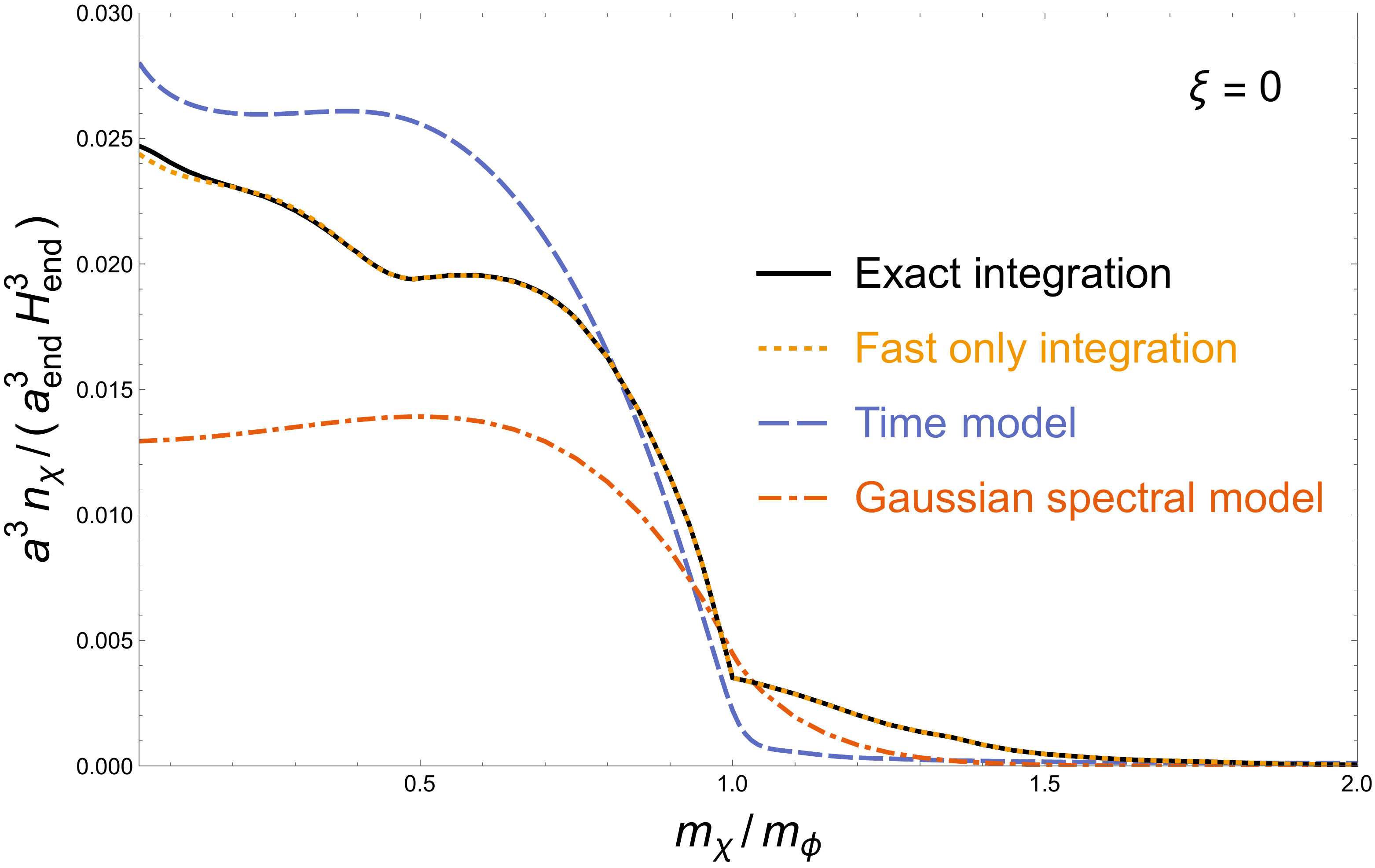}\includegraphics[scale=0.25]{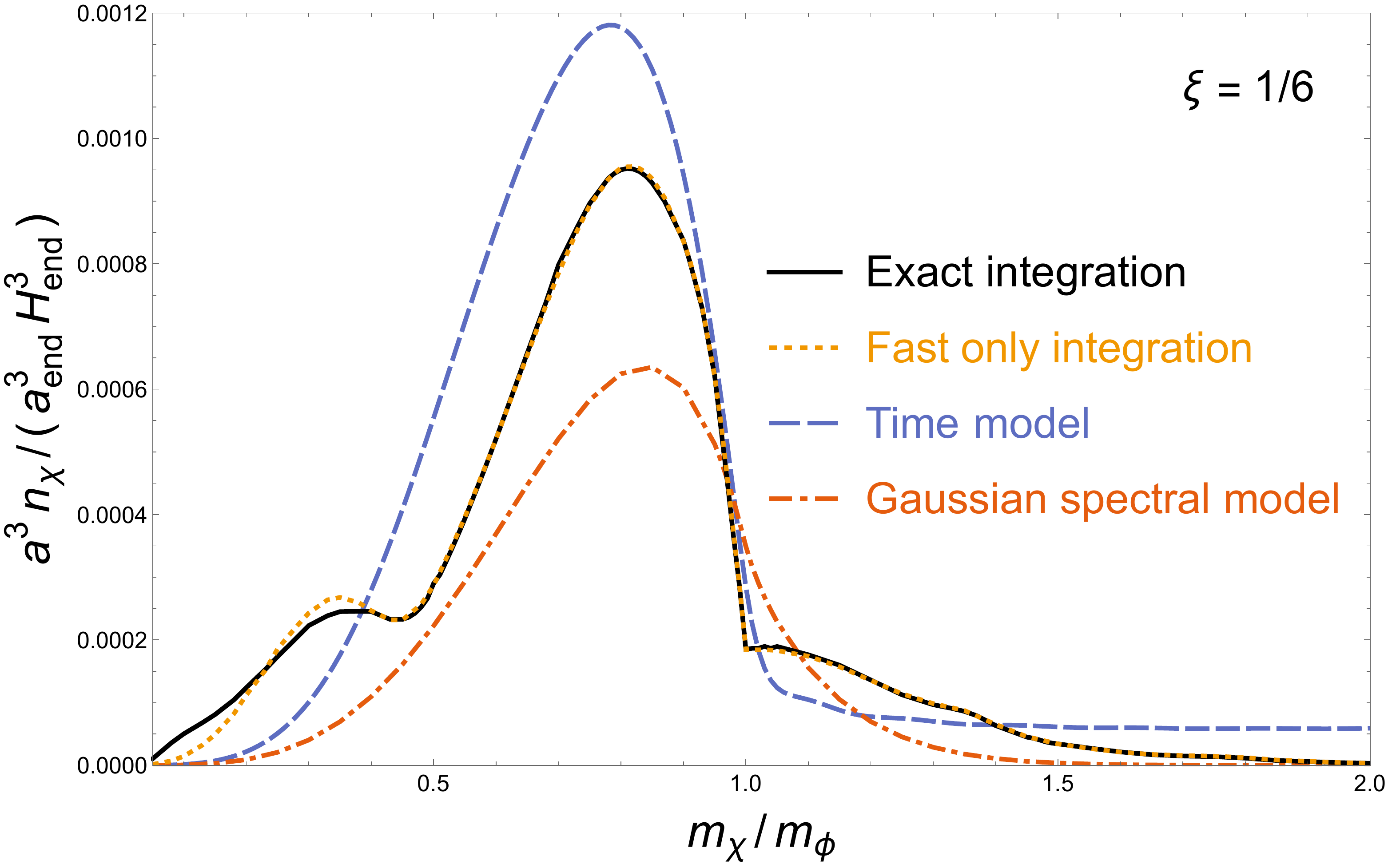}
\par\end{centering}
\caption{\label{fig:Number-density}Different approximations of the number
density $a^{3}n_{\chi}$ as a function of the mass $m_{\chi}$ are
compared. In both the exact and fast only integrations, there is a
clear threshold behavior at $m_{\chi}=m_{\phi}$. For $m_{\chi}<m_{\phi}$,
there is an approximate factor of $2$ difference between the time
model and the Gaussian spectral model, as noted below eq.\,(\ref{eq:nchi quadratic}).
The \textquotedblleft Time model\textquotedblright{} here refers to
the symmetric time model of eq.~(\ref{eq:time_model}). The \textquotedblleft Gaussian
spectral model\textquotedblright{} is defined according to eq.~(\ref{eq:gaussian}). }
\end{figure}

The produced $\chi$ number density as a function of $m_{\chi}$ is
shown using various approximations in figure \ref{fig:Number-density}.
The ``Exact integration'' and ``Fast only integration'' use eqs.~(\ref{eq:exact})
and (\ref{eq:beta fast def}), respectively, along with eq.~(\ref{eq:numdensitydef}).
The figure clearly shows that the number density reaches a significant
threshold at $m_{\chi}=m_{\phi}$. This is due to the $\delta\phi\delta\phi\rightarrow\chi\chi$
resonance becoming kinematically suppressed. A similar suppression
of the $\delta\phi\rightarrow\chi\chi$ resonance causes the slope
change around the $m_{\chi}=\frac{1}{2}m_{\phi}$ threshold. The largest
deviation between the fast only and exact integrations occurs at low
$m_{\chi}$, where the assumption of $m_{\chi}^{2}\gg R$ made in
eq,\,(\ref{eq:eksqdef}) becomes less accurate.

\subsection{\label{subsec:Time-model-in}Time model versus Gaussian model}

In this subsection, we generalize the Gaussian spectral model of ref.\,\citep{Chung:2018ayg}
and compare its computational accuracy with that of the symmetric
time model of eq.\,(\ref{eq:sincfunc}). In both models, we compute
the spectrum using the Boltzmann equation approximation form of eq.\,(\ref{eq:fchi Boltzmann approximation}):
\begin{equation}
f_{\chi}(k,t)=\int_{t_{\mathrm{end}}}^{t}\frac{dv}{\Delta t(v)}\left|\tilde{b}_{k}(v)\right|^{2},\label{eq:fchiherekt}
\end{equation}
where $\Delta t(v)$ and $\tilde{b}_{k}(v)$ will be defined differently
based on the model: i.e.\,$\{\Delta t\rightarrow\Delta t^{(\mathrm{GS})},\tilde{b}_{k}\rightarrow\tilde{b}_{k}^{(\mathrm{GS})}\}$
or $\{\Delta t\rightarrow\Delta t^{(\mathrm{TM})},\tilde{b}_{k}\rightarrow\tilde{b}_{k}^{(\mathrm{TM})}(v)\}$.

\begin{figure}
\begin{centering}
\includegraphics[scale=0.15]{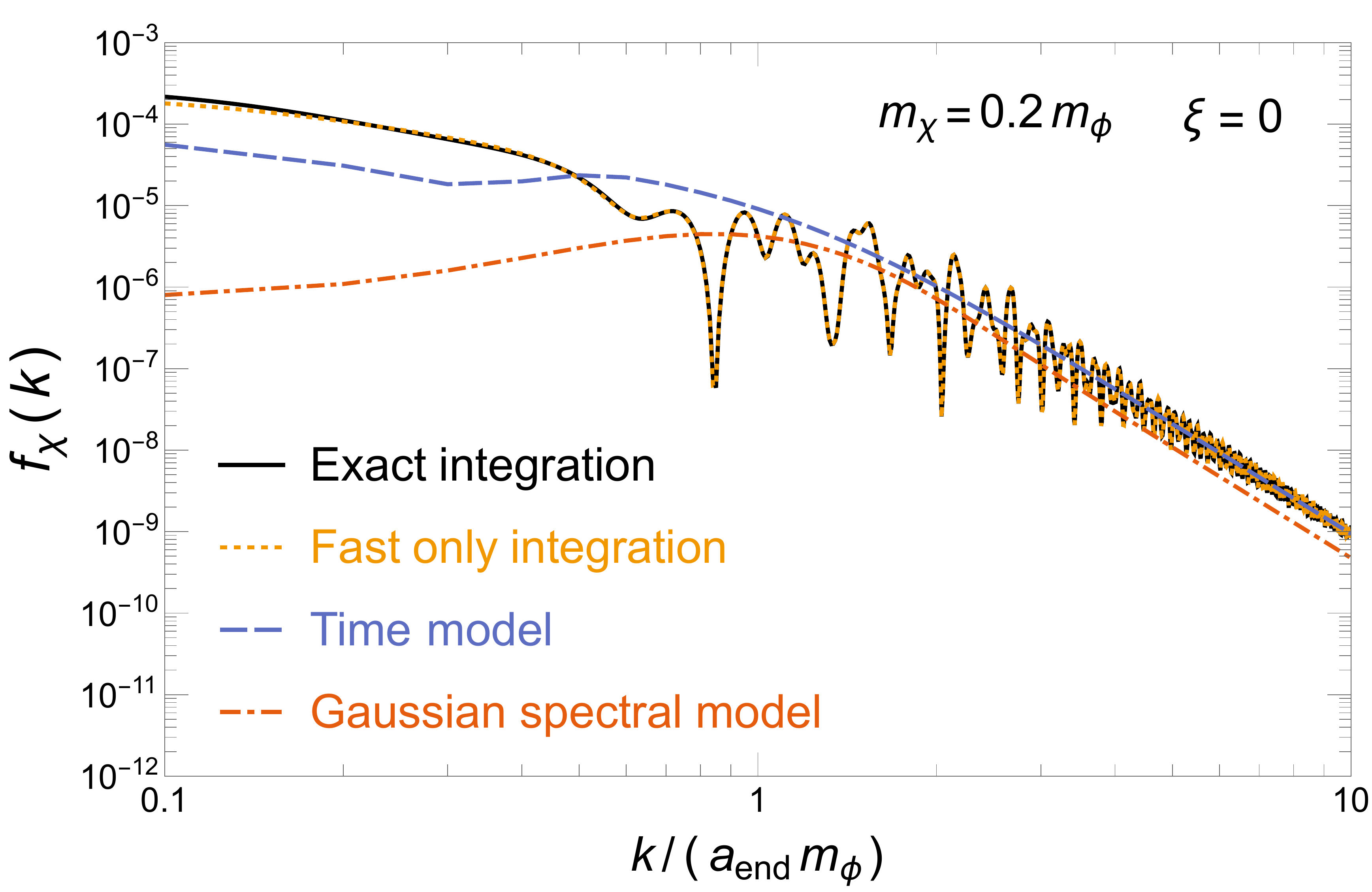}\includegraphics[scale=0.15]{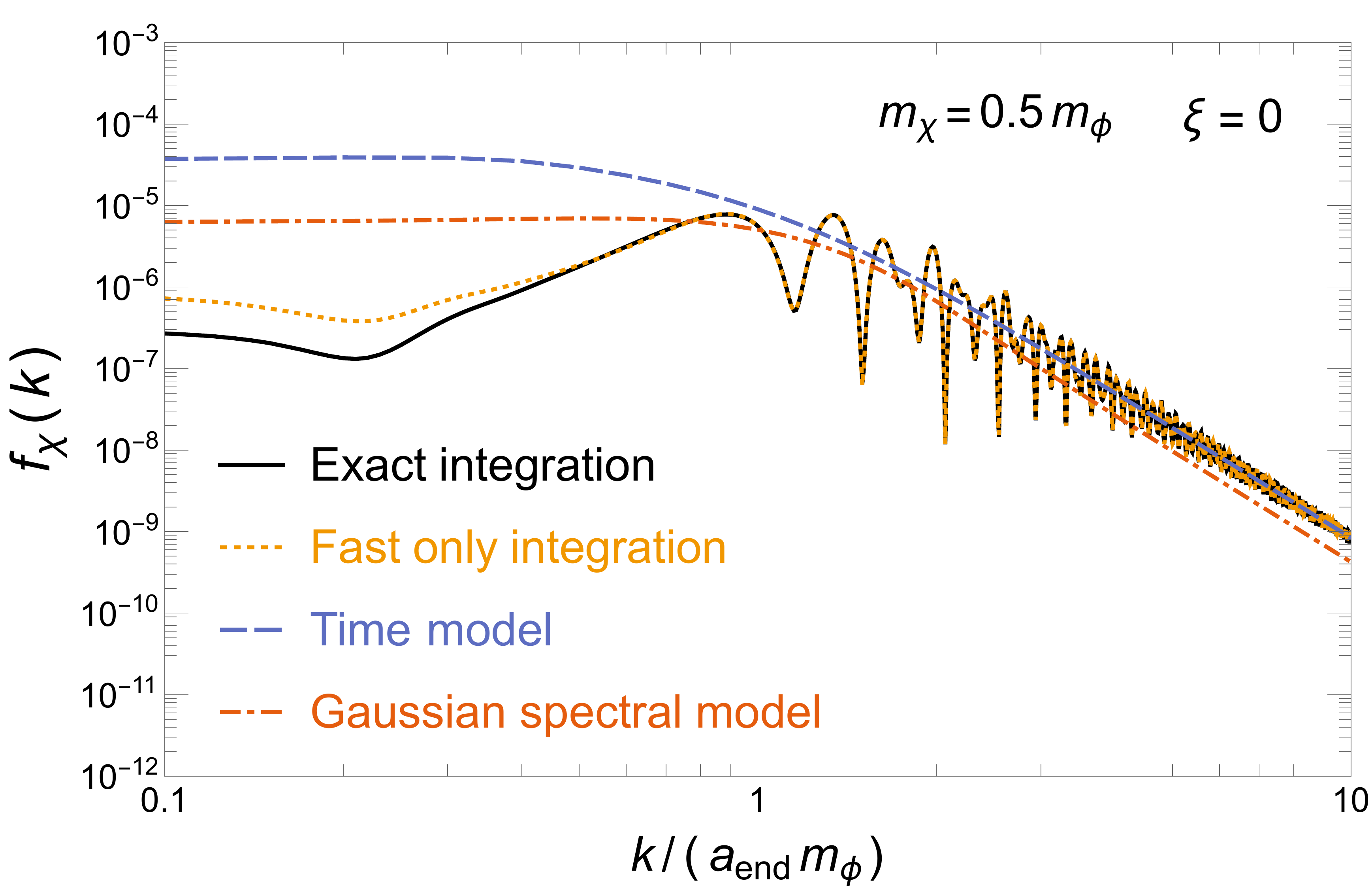}\includegraphics[scale=0.15]{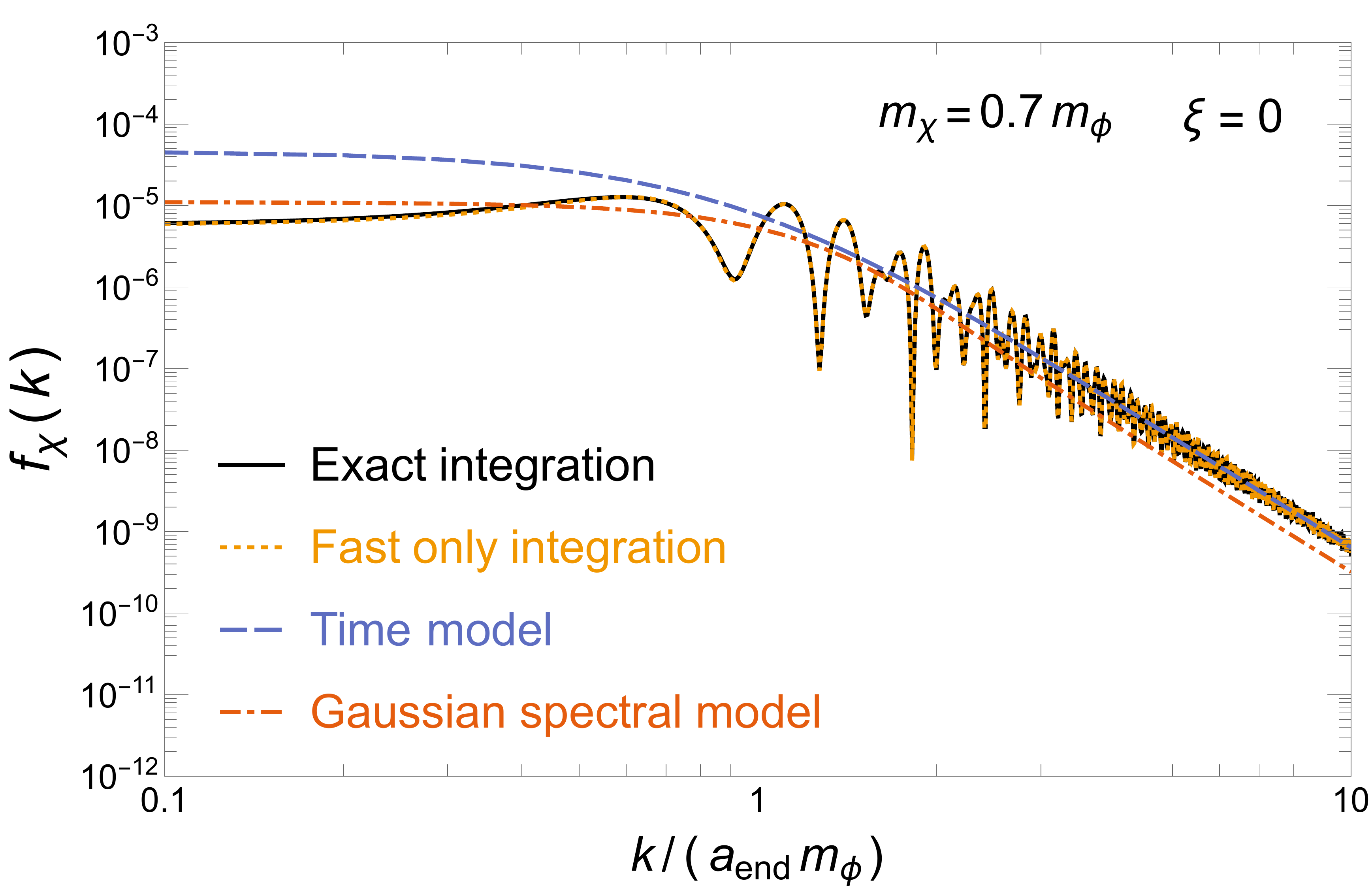}
\par\end{centering}
\begin{centering}
\includegraphics[scale=0.15]{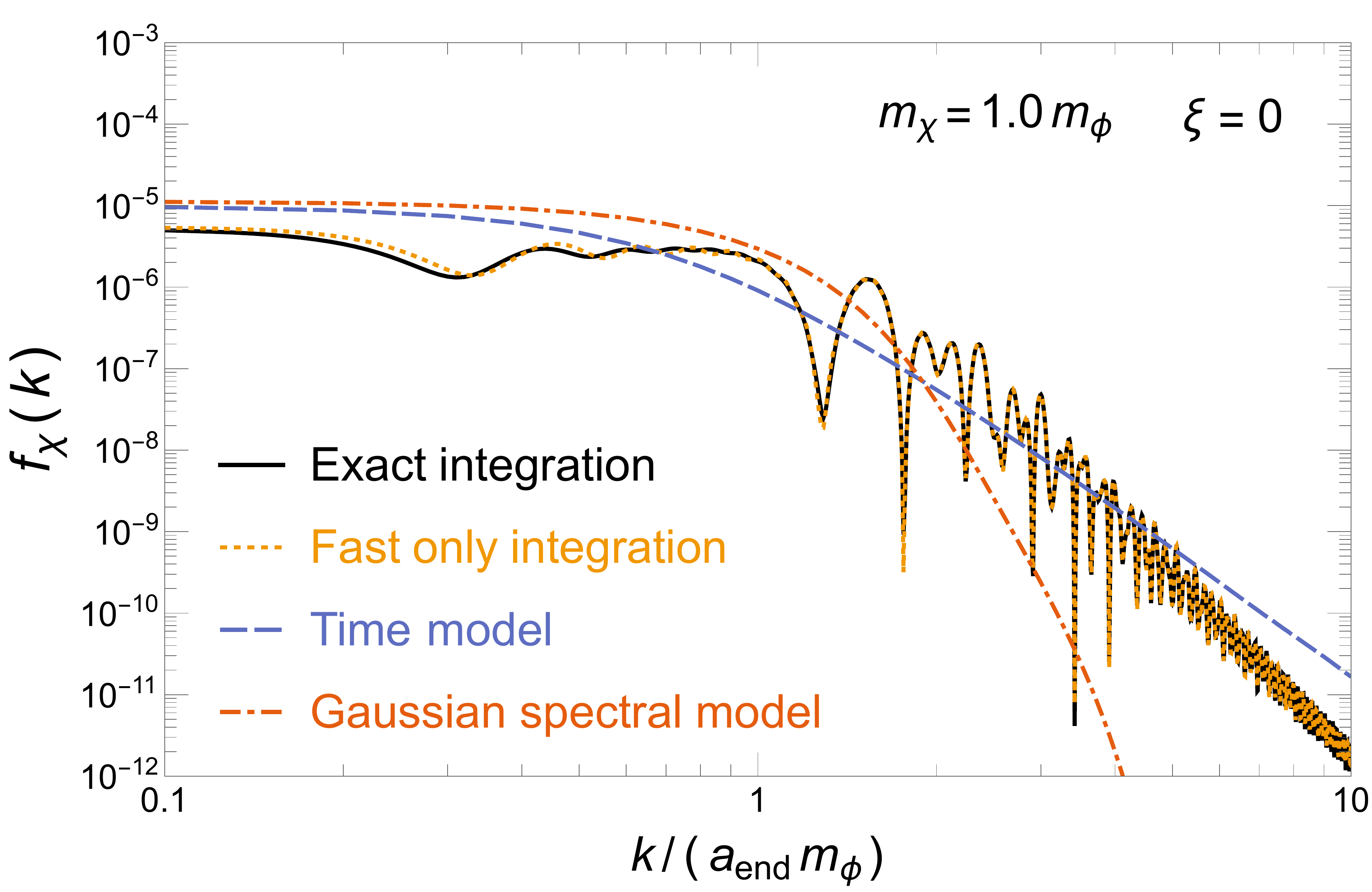}\includegraphics[scale=0.15]{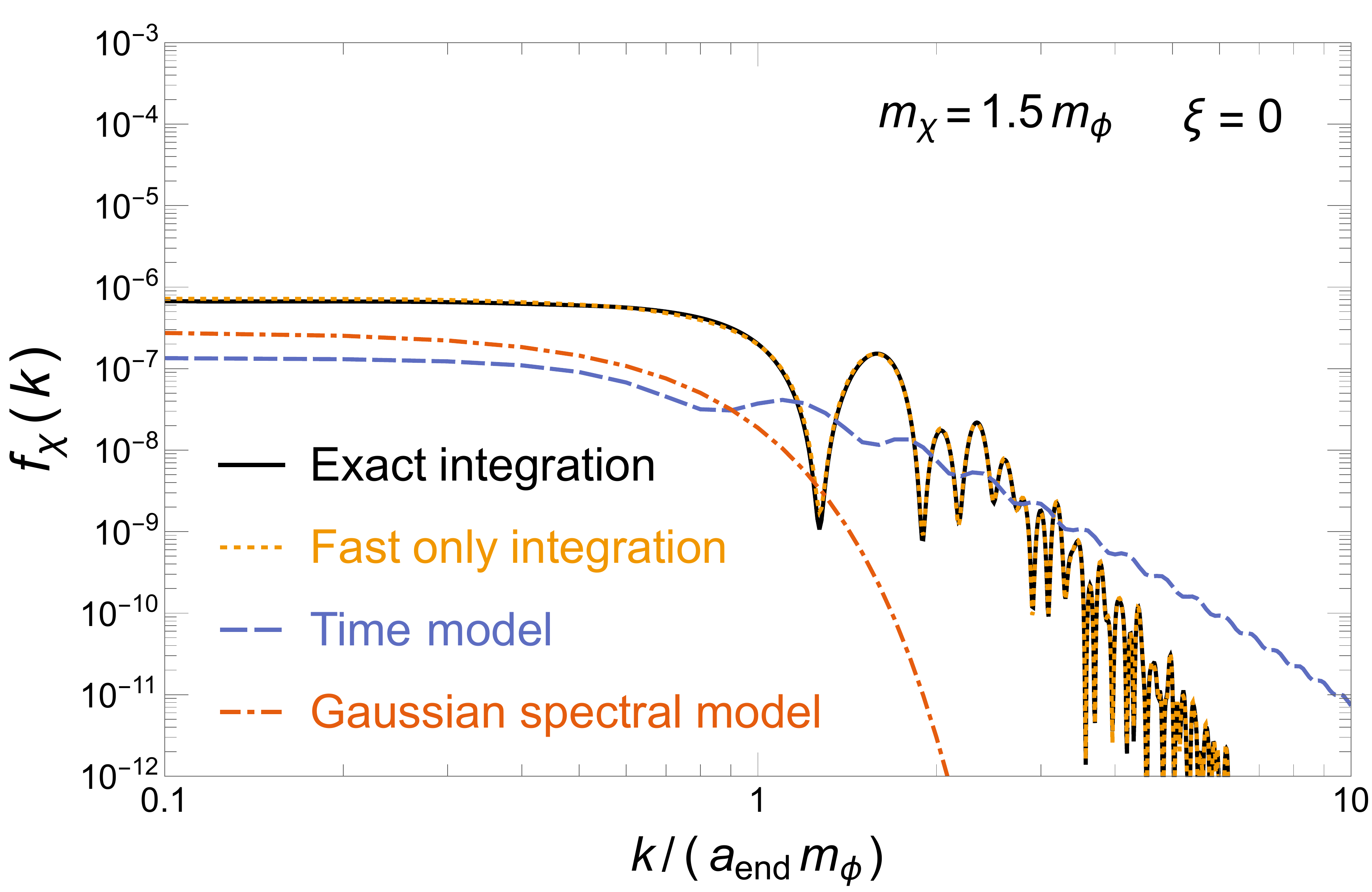}\includegraphics[scale=0.15]{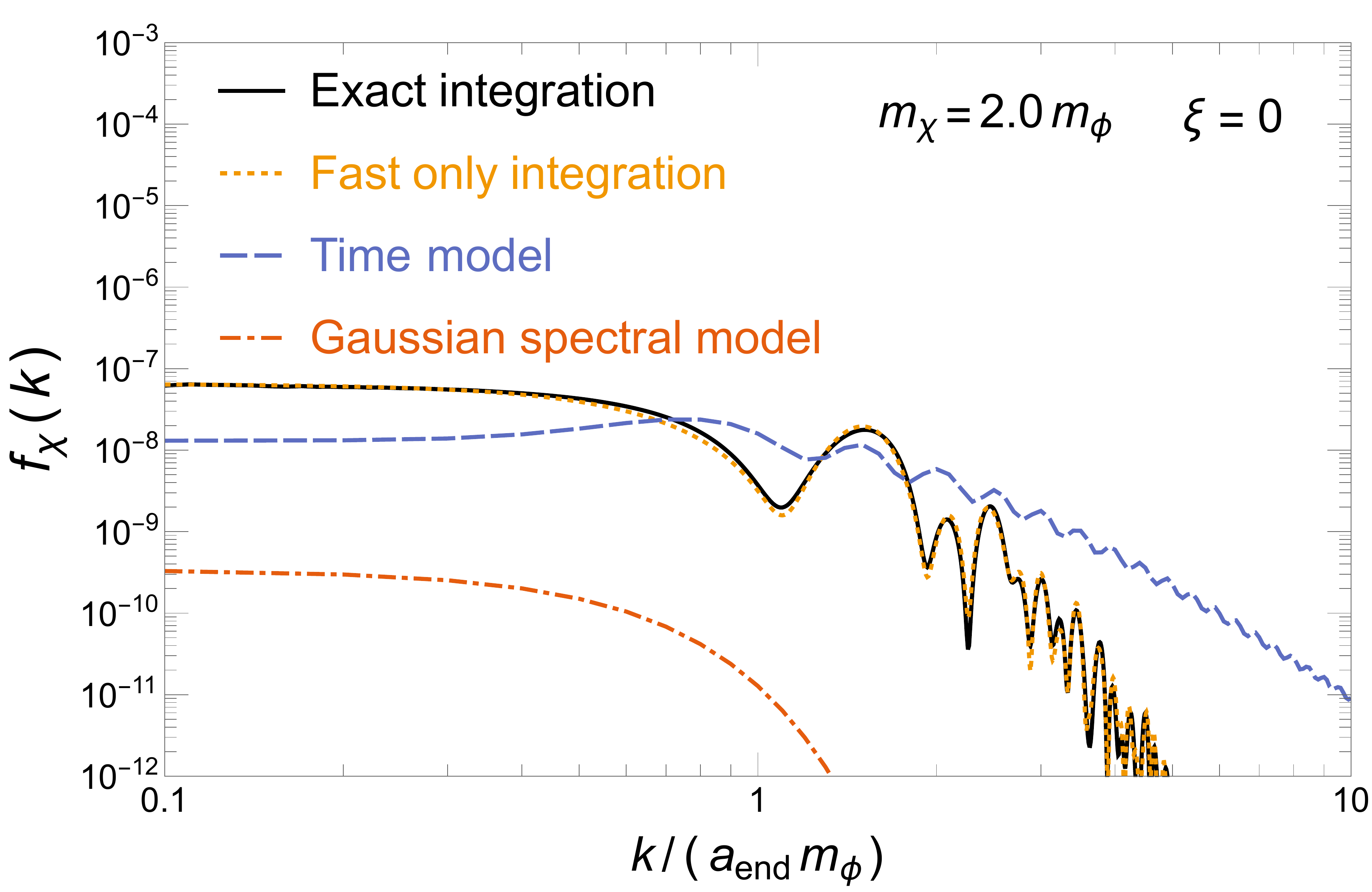}
\par\end{centering}
\caption{\label{fig:Time-model-vs} This figure compares the produced particle
spectrum computed using the Bogoliubov integration eq.\,(\ref{eq:exact}),
fast only integration eq.\,(\ref{eq:beta fast def}), Gaussian spectral
model eq.\,(\ref{eq:gaussian}), and the time model eq.\,(\ref{eq:time_model}).
Note that for $m_{\chi}<m_{\phi}$, and $k>m_{\phi}a_{\mathrm{end}}$,
the time model and exact integration match while the spectral model
differs from both by a factor of 2. The spectrum in this regime scales
like $k^{-9/2}$, which appears as linear behavior on this log-log
plot.}
\end{figure}

As discussed in section \ref{sec:Previous-work-on}, the Gaussian
model had several shortcomings partly because of the ambiguities associated
with mapping to Boltzmann equations as well as not having a long time
understanding of oscillating field amplitudes. The Gaussian spectral
model with the adiabatic invariant evolved fields resolving these
deficiencies is defined by generalizing the Gaussian model of ref.\,\citep{Chung:2018ayg}
as
\begin{align}
\Delta t^{(\mathrm{GS})}(v) & =\frac{\sqrt{2\pi}}{\sigma(v)},\\
\tilde{b}_{k}^{(\mathrm{GS})}(v) & \approx\frac{\pi}{8i}\left(\frac{\Delta\phi(v)}{M_{P}}\right)^{2}\left(\left(1-6\xi\right)E_{k}(v)+\frac{m_{\chi}^{2}}{2E_{k}(v)}\right)\frac{1}{\sqrt{4\pi\sigma^{2}(v)}}e^{-\left(E_{k}(v)-\omega_{*}(v)\right)^{2}/\sigma^{2}(v)},\label{eq:gaussian}\\
\omega_{*}(v) & =\frac{2\pi}{\oint_{C}d\phi/\sqrt{2V_{m}(v)-2V(\phi)}}
\end{align}
where the width $\sigma(v)$ is defined as
\begin{equation}
\sigma^{2}(v)=\left|\frac{V'\left(\phi_{C,+}(v)\right)-V'\left(\phi_{C,-}(v)\right)}{\phi_{C,+}(v)-\phi_{C,-}(v)}-\omega_{*}^{2}(v)\right|+\epsilon,
\end{equation}
with $\epsilon$ as a small positive parameter to ensure computational
convergence. This model still has the limitation that the spectrum
has a limited shape, but it will be able to reproduce the high $k$
part of the spectrum as can be seen in figure \ref{fig:Time-model-vs}.
This analytic fit to the high $k$ part of the spectrum was used as
the primary guidance in the generalization of the Gaussian model.

In the time model, we have instead from eq.\,(\ref{eq:sincfunc})
the expressions
\begin{align}
\Delta t^{(\mathrm{TM})}(v) & =\sqrt{\frac{2\pi}{m_{\phi}H_{\text{slow}}(v)}},\\
\tilde{b}_{k}^{(\mathrm{TM})}(v) & =\frac{3\pi}{4i}\frac{H_{\text{slow}}^{2}(v)}{\omega_{*}(v)}\left(\frac{\left(1-6\xi\right)\omega_{*}^{2}(v)+\frac{1}{2}m_{\chi}^{2}}{E_{k}^{2}(v)}\right)\frac{\Delta t(v)}{2\pi}\text{sinc}\left[\Delta t(v)\left(E_{k}(v)-\omega_{*}(v)\right)\right],\label{eq:time_model}\\
\omega_{*}(v) & =\sqrt{\frac{2V_{m}(v)}{[\Delta\phi(v)]^{2}}},
\end{align}
which are similar to eq.\,(\ref{eq:gaussian}) in having a peaked
function, but different in how fast that peaked function falls off
with $E_{k}-\omega_{*}$. The $\mathrm{sinc}$ function falls off
much slower and has multiple peaks. Furthermore, the prefactors dependent
on the non-minimal coupling parameter $\xi$ are different and can
become significant for large $\omega_{*}/E_{k}\sim m_{\phi}/m_{\chi}$
applicable for small $k/(a_{\mathrm{end}}m_{\phi})$.\footnote{For another perspective on the non-minimal coupling's role in the
$m_{\chi}/H\gg1$ dark matter gravitational production, see ref.\,\citep{Babichev:2020yeo}.} These features can be seen in figure \ref{fig:Time-model-vs}.

Shown in figure \ref{fig:Number-density} is also the integrated number
density of the ``Time model'' (eq.\,(\ref{eq:time_model})) and
the ``Gaussian spectral model'' (eq.\,(\ref{eq:gaussian})). The
``Time model'' plot does not fall off with large $m_{\chi}/m_{\phi}$
 because eq.\,(\ref{eq:time_model}) grows as $m_{\chi}^{2}$ and
therefore the rate is boosted by an $m_{\chi}^{4}$ factor. Even though
the minimally coupled case plot seems to indicate that the large $m_{\chi}/m_{\phi}$
region has a vanishing asymptote, this is an illusion associated with
the vertical resolution of the figure. This mismatch between the ``Time
model'' approximation and the better approximation of the ``Fast
only integration'' is due to the loss of validity in the kinematic
region in which eq.~(\ref{eq:onshellstart}) does not have a solution
(e.g.~see eq.(\ref{eq:needpeak})). 

In fact, for the symmetric time model, one can interpret the nonzero
$a^{3}n_{\chi}$ for the $m_{\chi}>m_{\phi}$ parametric region as
an error on the particle production computation for any finite $v$.
Using eqs.~(\ref{eq:numdensitydef}), ~(\ref{eq:fchi Boltzmann approximation})
and (\ref{eq:time_model}), we can estimate this error as
\begin{equation}
\Delta\left(a^{3}(t)n_{\chi}(t)\right)\sim\left.\frac{9}{512\pi^{2}}m_{\chi}\int_{t_{\mathrm{end}}}^{t}\frac{dv}{\Delta t(v)}\left(\frac{H_{\text{slow}}^{2}(v)}{m_{\phi}}\right)^{2}a_{\mathrm{slow}}^{3}\int_{0}^{\infty}dw\frac{w^{2}}{\left(w^{2}+1\right)^{3}}\right|_{m_{\chi}=m_{\phi}}
\end{equation}
which with
\begin{equation}
H_{\mathrm{slow}}(v)\approx\frac{H_{i}}{1+\frac{3}{2}H_{i}(v-t_{i})}
\end{equation}
evaluates to 
\begin{equation}
\Delta\left(\frac{a^{3}}{a_{i}^{3}}n_{\chi}\right)\sim\frac{1}{2048\sqrt{2}\pi^{3/2}}\frac{H_{i}^{7/2}}{m_{\phi}^{1/2}}
\end{equation}
which is suppressed by $\sqrt{H_{i}/m_{\phi}}$ and the small prefactor.
The Gaussian model in contrast has an exponentially suppressed error
as $m_{\chi}/m_{\phi}$ becomes larger than unity.
\begin{figure}
\begin{centering}
\includegraphics[scale=0.4]{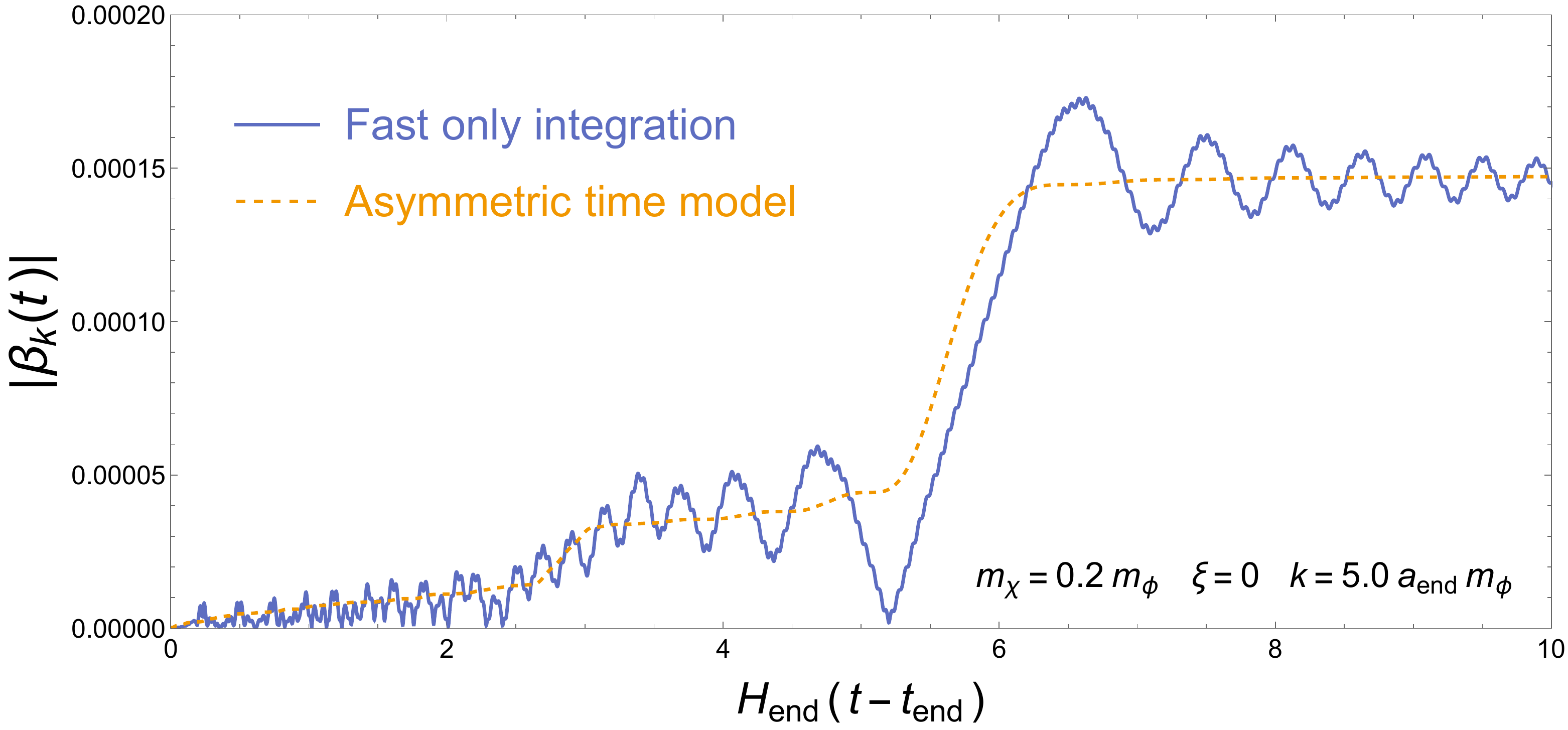}
\par\end{centering}
\caption{\label{fig:Asymmetric-time-model}The Bogoliubov coefficient approximation
results are compared for the asymmetric time model with the fast integration
method. There are two spikes in the rate corresponding to the $\delta\phi\delta\phi\delta\phi\rightarrow\chi\chi$
and $\delta\phi\delta\phi\rightarrow\chi\chi$ resonances. In the
case of the time model, $\left|\beta_{k}(t)\right|$ was obtained
by taking the square root of $f_{\chi}(k,t)$, which was computed
via eqs.\,(\ref{eq:fchiherekt}), (\ref{eq:Deltat_ASTM}), and (\ref{eq:bkastm}).}
\end{figure}

\subsection{\label{subsec:Asymmetric-time-model}Asymmetric time model}

The time model in subsection \ref{subsec:Time-model-in} did not account
for the $\delta\phi\delta\phi\delta\phi\rightarrow\chi\chi$ and $\delta\phi\rightarrow\chi\chi$
scattering. Using eq.\,(\ref{eq:fchi time model}), we can make the
substitution $\{\Delta t\rightarrow\Delta t^{\mathrm{ASTM}},\tilde{b}_{k}\rightarrow\tilde{b}_{k}^{\mathrm{ASTM}}\}$
in eq.\,(\ref{eq:fchiherekt}) where
\begin{align}
\Delta t^{\mathrm{ASTM}} & =\sqrt{\frac{2\pi}{m_{\phi}H_{\text{slow}}}},\label{eq:Deltat_ASTM}\\
\tilde{b}_{k}^{\mathrm{ASTM}} & =\frac{A_{1}}{2}\frac{e^{i\left(\omega_{*}-2E_{k}\right)\Delta t}-1}{i(\omega_{*}-2E_{k})}+\frac{A_{2}}{2}\frac{e^{i\left(2\omega_{*}-2E_{k}\right)\Delta t}-1}{i(2\omega_{*}-2E_{k})}+\frac{A_{3}}{2}\frac{e^{i\left(3\omega_{*}-2E_{k}\right)\Delta t}-1}{i(3\omega_{*}-2E_{k})},\label{eq:bkastm}
\end{align}
and $A_{n}$ are amplitudes given by eqs.\,(\ref{eq:A1 definition}),
(\ref{eq:A2 definition}), and (\ref{eq:A3 definition}). Notice that
because $\omega_{*}\approx m_{\phi}$ while $E_{k}$ redshifts, one
can see from the denominators of eq.\,(\ref{eq:bkastm}) that $\delta\phi\delta\phi\delta\phi\rightarrow\chi\chi$
amplitude $A_{3}$ will be reached first before the $\delta\phi\delta\phi\rightarrow\chi\chi$
amplitude as a function of time for $2m_{\chi}<3m_{\phi}$

The time evolution of the particle spectrum for an example parametric
point is shown in figure \ref{fig:Asymmetric-time-model}. In the
following argument, we will set $t_{\mathrm{end}}=0$ for convenience.
Using the resonance condition $3\omega_{*}=2E_{k}$, we would predict
a step in particle spectrum time evolution at
\begin{equation}
a(t_{*})=\frac{k}{\sqrt{\left(\frac{3}{2}\omega_{*}\right)^{2}-m_{\chi}^{2}}}\qquad\Rightarrow\qquad H_{\mathrm{end}}t_{*}\approx\frac{2}{3}\left[\left(\frac{k/(m_{\phi}a_{\mathrm{end}})}{\sqrt{9/4-m_{\chi}^{2}/m_{\phi}^{2}}}\right)^{3/2}-1\right],
\end{equation}
which is consistent with the step feature in the figure near $H_{\mathrm{end}}t\approx3$.
The step at $H_{\mathrm{end}}t\approx5.5$ is due to the dominant
$2\omega_{*}=2E_{k}$ resonance.

\section{\label{sec:Summary}Summary}

In this paper, we developed analytic and numerical techniques to compute
the Bogoliubov coefficient $\beta_{k}$ more accurately and efficiently.
Using the adiabatic invariant equation, we computed the slowly varying
components of the $\beta_{k}$ integrand necessary to implement the
fast-slow decomposition formalism of ref.\,\citep{Chung:2018ayg}.
In the process, a numerical integration technique was created that
is O(1000) faster than a brute force computation.

To compute the particle production rate and number density, an approximation
of the Boltzmann equation was presented, and a coarse graining time
$\Delta t$ was introduced. We demonstrated that $\Delta t$ minimizes
the error of this approximation at an intermediate scale between the
inflaton mass and Hubble rate, and that the overall error is suppressed
by powers of $H/m_{\phi}$. This simplified the integration by linearizing
the phase of the Bogoliubov integrand, yielding a form reminiscent
of a Fourier transform.

We derived a time model of inflaton dynamics and used it simplify
the $\beta_{k}$ integrand to a form amenable to exact analytical
integration. This was done using the adiabatic turning points to create
an envelope of the oscillatory motion. A differential equation for
the fast oscillatory phase was then obtained using approximate energy
conservation. By treating the slowly varying components as constant,
and using a perturbative expansion of the inflaton potential, the
phase was obtained as a function of adiabatic invariants.

A closed form production rate proportional to a sine-cardinal function
was derived, with a resonant peak when the inflaton oscillatory frequency
$\omega_{*}$ equals the time derivative of the Bogoliubov phase $E_{k}$.
We then integrated the production rate to obtain the spectrum and
number density. For $m_{\chi}<m_{\phi}$ and large $k$, these results
matched the exact Bogoliubov integration.

For $k>m_{\phi}a_{\mathrm{end}}$, the rate limits to the form of
Fermi\textquoteright s golden rule for the tree-level graviton-mediated
process $\phi\phi\rightarrow\chi\chi$ in a frame where both inflatons
are at rest. In this case, $\omega_{*}$ and $E_{k}$ can be interpreted
as the energies of the $\phi$ and $\chi$ particles, respectively.
The delta function in the production rate allowed an analytical evaluation
of the time integral, which yielded simple closed form approximations
of the spectrum and number density. These made O(1) corrections to
the predictions of the Gaussian spectral model of ref.\,\citep{Chung:2018ayg}.

This formalism demonstrates a correspondence between the Bogoliubov
and scattering methods of computing gravitational particle production.
The predicted spectrum and number density are the same, up to factors
of 2, as the results of papers that considered a gas of inflatons
at rest undergoing gravitational annihilation \citep{Tang:2017hvq,Mambrini:2021zpp}.
These factors of 2 are not currently understood, but we suspect that
the inflaton gas must be treated as an appropriate coherent state
in the scattering picture to obtain an exact correspondence.

The noisy behavior of the numerical spectrum as a function of $k$
(see for example numerical results of ref.\,\citep{Ema:2018ucl}
and figure \ref{fig:Time-model-vs} of the present work) remains unexplained
and is a subject of ongoing research by our group. Preliminary results
suggest this is due to quantum interference between different resonances
of the Bogoliubov integral, and have interesting implications for
the particle scattering method of computing gravitational particle
production. It would also be of great interest to find cosmological
observables that are sensitive to the details in the momentum spectrum
of the gravitationally produced particles, which is one of the main
results of this paper.
\begin{acknowledgments}
This work was supported in part by the Ray MacDonald Fund at UW-Madison.
\end{acknowledgments}

\appendix

\section{\label{sec:Adiabatic-invariant}Adiabatic invariant}

Here we give a clarification of the adiabatic invariant construction
presented in ref.\,\citep{Landau:1975pou}. In addition to the derivation
of the conserved adiabatic invariant expression of eq.\,(\ref{eq:charge}),
the error incurred over a long time period $T/\epsilon$ (where $T$
is the period of fast periodic motion and $\epsilon$ is a small adiabaticity
parameter) is explicitly evaluated in eq.\,(\ref{eq:finalerror})
using eq.\,(\ref{eq:longtermerror}). We also compare it to the canonical
transformation approach of ref.\,\citep{goldstein2002classical}.

Consider an external influence on the system represented by the function
$\lambda(t)$. The Hamiltonian of the system is denoted as $H(\vec{q},\vec{\pi},\lambda(t))$,
where $\vec{\pi}$ is the conjugate momentum to $\vec{q}$. Suppose
this function is slowly (adiabatically) changing such that 
\begin{equation}
T\frac{d\lambda}{dt}+T^{2}\frac{d^{2}\lambda}{dt^{2}}+...\ll\lambda
\end{equation}
on a time scale $T$. The energy $E(t)\equiv H(\vec{q}(t),\vec{\pi}(t),\lambda(t))$
is not conserved because of the time dependence of $\lambda(t)$.
We write the time average of $dE/dt$ at time $v$ over a time period
$T$ as
\begin{align}
\left\langle \frac{dE(t)}{dt}\right\rangle _{v} & =\left\langle \frac{d\lambda}{dt}\partial_{\lambda}H\right\rangle _{v}\approx\frac{1}{T}\left.\frac{d\lambda}{dt}\right|_{v}\int_{v}^{v+T}dt\,\partial_{\lambda}H(\vec{q}(t),\vec{\pi}(t),\lambda(v))+\Delta_{1}(v),\label{eq:start}
\end{align}
where the leading order error $\Delta_{1}$ is estimated as
\begin{align}
\Delta_{1}(v) & \equiv\frac{1}{T}\int_{v}^{v+T}dt(t-v)\left(\left.\frac{d^{2}\lambda}{dt^{2}}\right|_{v}\partial_{\lambda}H(\vec{q}(t),\vec{\pi}(t),\lambda(v))+\left(\left.\frac{d\lambda}{dt}\right|_{v}\right)^{2}\partial_{\lambda}^{2}H(\vec{q}(t),\vec{\pi}(t),\lambda(v))\right)
\end{align}
through a Taylor expansion.

Next, restrict to a 1-dimensional motion which becomes periodic in
$t$ when $\lambda(t)$ becomes independent of $t$ as\textbf{ $\lambda(t)=\lambda(v)$}.
Define $\Pi(q,E,\lambda)$ to be the solution to
\begin{equation}
E=H\left(q,\Pi(q,E,\lambda),\lambda\right)\label{eq:Eidentity}
\end{equation}
as an algebraic identity. Using this, we can define the time period
to be
\begin{eqnarray}
T(v) & \equiv & \int_{v}^{v+T}dt\label{eq:period}\\
 & \approx & \oint_{C_{T}(v)}\frac{dq'}{\partial_{\pi}H\left(q',\Pi\left[q',E(v),\lambda(v)\right],\lambda(v)\right)}\left(1+\Delta_{2}\right)\label{eq:Tv}\\
\Delta_{2} & \equiv & \frac{O\left([q(v+T)-q(v)]\left.\frac{dt}{dq}\right|_{v}\right)}{\oint_{C_{T}(v)}\frac{dq'}{\partial_{\pi}H(q',\Pi\left[q',E(v),\lambda(v)\right],\lambda(v))}},\label{eq:delta2def}
\end{eqnarray}
where $C_{T}(v)$ represents a chosen 1-dimensional path where $q_{C}(t)$
returns to itself periodically with $\lambda(t)=\lambda(v)$ fixed,
i.e.\,$q_{C}(t)$ is a solution to the equation of motion with $\lambda(t)=\lambda(v)$.

From eq.\,(\ref{eq:Eidentity}), we conclude
\begin{align}
\frac{dE}{d\lambda} & =\partial_{\lambda}H\left(q,\Pi\left(q,E,\lambda\right),\lambda\right)+\partial_{\lambda}\Pi\left(q,E,\lambda\right)\partial_{\pi}H\left(q,\Pi\left(q,E,\lambda\right),\lambda\right)=0,
\end{align}
as $E$ and $\lambda$ are independent variables. One can use this
to write
\begin{equation}
dt=-dq\frac{\partial_{\lambda}\Pi[q,E(t),\lambda(t)]}{\partial_{\lambda}H\left(q,\Pi\left(q,E(t),\lambda(t)\right),\lambda(t)\right)},\label{eq:dt_in_terms_of_dq}
\end{equation}
where we made use of $\dot{q}=\partial_{\pi}H\left(q,\Pi\left(q,E,\lambda\right),\lambda\right)$
just as in eq.~(\ref{eq:Tv}). Since there are 2 values of $t$ for
a given value of $q$ in the time interval $[v,v+T]$, we will call
one branch of eq.\,(\ref{eq:start}) integral ``branch 1'' and
the other ``branch 2''. Putting eq.~(\ref{eq:dt_in_terms_of_dq})
into eq.\,(\ref{eq:start}) gives
\begin{align}
\left\langle \frac{dE(t)}{dt}\right\rangle _{v} & \approx\frac{-1}{T}\left.\frac{d\lambda}{dt}\right|_{v}\left[\int_{\mathrm{branch\,1}}dq'+\int_{\mathrm{branch\,2}}dq'\right]\partial_{\lambda}\Pi\left[q',E(\tilde{t}(q')),\lambda(\tilde{t}(q'))\right]+\Delta_{1}(v),\label{eq:dedtdlamdtstillin}
\end{align}
where $\tilde{t}(q')$ is the implicit solution of eq.\,(\ref{eq:dt_in_terms_of_dq})
such that $\tilde{t}(q(t))=t$.

As the external parameter is fixed at $\lambda=\lambda(v)$ without
time variation for which $q_{C}(t)$ is defined below eq.~(\ref{eq:delta2def}),
a small error is incurred by putting in approximate solutions with
$\lambda$ fixed. Therefore, we define the error $\Delta_{4}$ as
\begin{equation}
\partial_{\lambda}\Pi(q(t),E(t),\lambda(t))=\partial_{\lambda}\Pi(q_{C}(t),E(v),\lambda(v))\left(1+\Delta_{4}(t)\right),
\end{equation}
where
\begin{equation}
\Delta_{4}(t)\equiv\frac{\partial_{\lambda}\Pi(q(t),E(t),\lambda(t))}{\partial_{\lambda}\Pi(q_{C}(t),E(v),\lambda(v))}-1
\end{equation}
which is still a function of time. Putting this in to eq.\,(\ref{eq:dedtdlamdtstillin})
gives
\begin{align}
\left\langle \frac{dE(t)}{dt}\right\rangle _{v} & =-\frac{1}{T(v)}\left.\frac{d\lambda}{dt}\right|_{v}\oint_{C_{T}(v)}dq'\partial_{\lambda}\Pi[q',E(v),\lambda(v)]\left(1+\Delta_{4}(\tilde{t}(q'))\right)+\Delta_{1}(v),
\end{align}
which implies
\begin{align}
0 & =T(v)\left(\left\langle \frac{dE(t)}{dt}\right\rangle _{v}-\Delta_{1}(v)\right)+\left.\frac{d\lambda}{dt}\right|_{v}\oint_{C_{T}(v)}dq'\partial_{\lambda}\Pi[q',E(v),\lambda(v)]\left(1+\Delta_{4}(\tilde{t}(q'))\right)\\
 & =\oint_{C_{T}(v)}dq'\left[\frac{\left(1+\Delta_{2}(q',v)\right)\left(\left\langle \frac{dE(t)}{dt}\right\rangle _{v}-\Delta_{1}(v)\right)}{\partial_{\pi}H\left(q',\Pi\left[q',E(v),\lambda(v)\right],\lambda(v)\right)}+\left.\frac{d\lambda}{dt}\right|_{v}\partial_{\lambda}\Pi[q',E(v),\lambda(v)]\left(1+\Delta_{4}(\tilde{t}(q'))\right)\right],\label{eq:intermedzeroeq}
\end{align}
where we replaced $T(v)$ in the second line using eq.\,(\ref{eq:period}).
Taking a derivative of eq.\,(\ref{eq:Eidentity}) with respect to
$E$ yields
\begin{equation}
1=\partial_{E}\Pi\left(q,E,\lambda\right)\partial_{\pi}H\left(q,\Pi[q,E,\lambda],\lambda\right).
\end{equation}
 Combining this with eq.~(\ref{eq:intermedzeroeq}) results in
\begin{equation}
0=\oint_{C_{T}(v)}dq'\left[\frac{\left(1+\Delta_{2}(q',v)\right)\left(\left\langle \frac{dE(t)}{dt}\right\rangle _{v}-\Delta_{1}(v)\right)}{1/\partial_{E}\Pi\left(q',E(v),\lambda(v)\right)}+\left.\frac{d\lambda}{dt}\right|_{v}\partial_{\lambda}\Pi\left(q',E(v),\lambda(v)\right)\left(1+\Delta_{4}(\tilde{t}(q'))\right)\right],\label{eq:almost_there}
\end{equation}
which almost brings us to the desired result after neglecting the
$\Delta_{n}$ errors.

Given that $v$-dependent quantities do not depend on $t$, we expand
the time-averaging in eq.\,(\ref{eq:almost_there}) to the entire
integral as
\begin{equation}
0=\left\langle \oint_{C_{T}(v)}dq'\left[\frac{\left(1+\Delta_{2}(q',v)\right)\left(\frac{dE(t)}{dt}-\Delta_{1}(v)\right)}{1/\partial_{E}\Pi\left(q',E(v),\lambda(v)\right)}+\left.\frac{d\lambda}{dt}\right|_{v}\partial_{\lambda}\Pi\left(q',E(v),\lambda(v)\right)\left(1+\Delta_{4}(\tilde{t}(q'))\right)\right]\right\rangle _{v}
\end{equation}
such that there is now a simple motivation to approximate judicious
objects inside the more inclusive time average as a function of $t$.
Taylor expanding $v$ dependent quantities $E(v)$ and $\lambda(v)$
about the point $t$, the integrand becomes
\begin{align}
0 & =\left\langle \oint_{C_{T}(v)}dq'\left[\frac{\left(1+\Delta_{2}(q',v)\right)\left(\left.\frac{dE}{dt}\right|_{t}-\Delta_{1}(v)\right)}{1/\partial_{E}\Pi\left(q',E(t)-(t-v)\left.\frac{dE}{dt}\right|_{v},\lambda(t)-(t-v)\left.\frac{d\lambda}{dt}\right|_{v}\right)}+\left(\left.\frac{d\lambda}{dt}\right|_{t}-(t-v)\left.\frac{d^{2}\lambda}{dt}\right|_{v}\right)\right.\right.\nonumber \\
 & \left.\left.\times\partial_{\lambda}\Pi\left(q',E(t)-(t-v)\left.\frac{dE}{dt}\right|_{v},\lambda(t)-(t-v)\left.\frac{d\lambda}{dt}\right|_{v}\right)\left(1+\Delta_{4}(\tilde{t}(q'))\right)\right]\right\rangle _{v},
\end{align}
and the function $\Pi$ can be expanded further as
\begin{align}
 & \partial_{E}\Pi\left(q',E(t)-(t-v)\left.\frac{dE}{dt}\right|_{v},\lambda(t)-(t-v)\left.\frac{d\lambda}{dt}\right|_{v}\right)\nonumber \\
 & =\partial_{E}\Pi\left(q,E(t),\lambda(t)\right)-(t-v)\left(\left.\frac{dE}{dt}\right|_{t}\partial_{E}^{2}\Pi\left(q,\lambda(t),E(t)\right)+\left.\frac{d\lambda}{dt}\right|_{t}\partial_{\lambda}\partial_{E}\Pi\left(q,\lambda(t),E(t)\right)\right)+\dots,\\
 & \partial_{\lambda}\Pi\left(q',E(t)-(t-v)\frac{dE}{dt}|_{t},\lambda(t)-(t-v)\frac{d\lambda}{dt}|_{t}\right)\nonumber \\
 & =\partial_{\lambda}\Pi\left(q,E(t),\lambda(t)\right)-(t-v)\left(\left.\frac{dE}{dt}\right|_{t}\partial_{E}\partial_{\lambda}\Pi\left(q,\lambda(t),E(t)\right)+\left.\frac{d\lambda}{dt}\right|_{t}\partial_{\lambda}^{2}\Pi\left(q,\lambda(t),E(t)\right)\right)+\dots,
\end{align}
to obtain more small error quantities. We therefore define
\begin{align}
\Delta_{3}(q,t,v) & \equiv-(t-v)\left[\left(\left.\frac{dE}{dt}\right|_{t}\right)^{2}\partial_{E}^{2}\Pi\left(q,\lambda(t),E(t)\right)+2\left.\frac{dE}{dt}\right|_{t}\left.\frac{d\lambda}{dt}\right|_{t}\partial_{E}\partial_{\lambda}\Pi\left(q,\lambda(t),E(t)\right)\right.\nonumber \\
 & \left.+\left(\left.\frac{d\lambda}{dt}\right|_{t}\right)^{2}\partial_{\lambda}^{2}\Pi\left(q,\lambda(t),E(t)\right)+\left.\frac{d^{2}\lambda}{dt^{2}}\right|_{v}\partial_{\lambda}\Pi\left(q,E(t),\lambda(t)\right)\right]\\
 & =O\left(\left(\frac{dE}{dt}\right)^{2},\frac{dE}{dt}\frac{d\lambda}{dt},\left(\frac{d\lambda}{dt}\right)^{2},\frac{d^{2}\lambda}{dt^{2}}\right),
\end{align}
as the error due to these last set of expansions.

Neglecting all $\Delta_{n}$, we find
\begin{align}
0 & =\left\langle \oint_{C_{T}(v)}dq'\left[\left.\frac{dE}{dt}\right|_{t}\frac{\partial\Pi\left(q',E(t),\lambda(t)\right)}{\partial E}+\left.\frac{d\lambda}{dt}\right|_{t}\frac{\partial\Pi[q',E(t),\lambda(t)]}{\partial\lambda}\right]\right\rangle _{v},
\end{align}
or equivalently, the statement is that
\begin{equation}
\boxed{Q(t)\equiv\oint_{C_{T}(t)}dq'\Pi\left(q',E(t),\lambda(t)\right)}\label{eq:charge}
\end{equation}
is conserved when $\Delta_{n}$ are all neglected. To be more explicit,
the identity is 
\[
\left\langle \dot{Q}\right\rangle _{v}=0
\]
 when errors $\Delta_{n}$ are neglected.

The correction to $Q(t)$ is 
\begin{align}
\Delta\left\langle \dot{Q}\right\rangle _{v} & =\left\langle \oint_{C_{T}(v)}dq'\left[\frac{\Delta_{2}(q',v)\left.\frac{dE}{dt}\right|_{t}-\Delta_{1}(v)}{1/\partial_{E}\Pi\left(q',E(t),\lambda(t)\right)}+\left.\frac{d\lambda}{dt}\right|_{t}\partial_{\lambda}\Pi\left(q',E(t),\lambda(t)\right)\Delta_{4}(\tilde{t}(q'))+\Delta_{3}(q',t,v)\right]\right\rangle _{v}\label{eq:longtermerror}
\end{align}
to leading order in $\Delta_{n}$. Let us compute the change in $Q$
over a long time period over which $\lambda$ changes significantly,
denoted as
\begin{equation}
\Delta t(v)\sim\frac{\lambda(v)}{\dot{\lambda}(v)}\equiv\frac{T(v)}{\epsilon(v)},\label{eq:deltTdef}
\end{equation}
to leading order in $\epsilon\ll1$.\footnote{Note that the small parameter $\epsilon$ that is on the order of
$H/m_{\phi}$ in the context of the gravitational particle production
studied in this paper.} For a generic non-adiabatic-invariant quantity $\mathcal{O}_{\mathrm{NAI}}(t)$
that derives time dependence from $\lambda(t)$, the change over the
long time period $\Delta t(v)$ is 
\begin{equation}
\Delta\mathcal{O_{\mathrm{NAI}}}\sim\Delta t\left\langle \dot{\mathcal{O}}_{\mathrm{NAI}}\right\rangle _{v}\sim\mathcal{O}_{\mathrm{NAI}}\label{eq:naichange}
\end{equation}
where the time average has been estimated as $\left\langle \dot{\mathcal{O}}_{\mathrm{NAI}}\right\rangle _{v}\sim\epsilon\mathcal{O}_{\mathrm{NAI}}/T(v)$
which exactly vanishes in the limit that $\lambda(t)$ becomes time
independent. Note regardless of how small the time variation $\epsilon$
is, the change in $\mathcal{O}_{\mathrm{NAI}}$ over the long time
period $\Delta t$ is large.

This is in contrast with the case of the adiabatic invariant for which
the change is 
\begin{align}
\Delta Q & =\Delta\left\langle \dot{Q}\right\rangle _{v}\Delta t\sim\left(\frac{O\left([q(v+T)-q(v)]\frac{dt}{dq}\right)}{T(v)}Q+QO(\epsilon)+\frac{T(v)}{\epsilon}\oint_{C_{T}(v)}dq'\Delta_{3}(q',T(v),v)\right)\label{eq:DeltaQcontrast}
\end{align}
where we have assumed, for example, $\dot{E}(t)\sim\dot{\lambda}(t)E/\lambda,$
$\ddot{\lambda}(t)/\lambda\sim O(\epsilon^{2})/T^{2}$, $Q\sim ET$,
and $\partial_{\lambda}^{2}H(\vec{q},\vec{\pi},\lambda(v))\sim H/\lambda^{2}$,
based on smoothness of these functions. Simplifying the large piece
$\Delta_{3}$, we find
\begin{align}
\frac{T(v)}{\epsilon}\oint_{C_{T}(v)}dq'\Delta_{3}(q',T(v),v) & =QO(\epsilon)
\end{align}
and thus
\begin{equation}
\Delta Q\sim\left(\frac{O\left([q(v+T)-q(v)]\frac{dt}{dq}\right)}{T}+O(\epsilon)\right)Q\label{eq:deltaqeq}
\end{equation}
to be the variation in the adiabatic invariant charge $Q$ over a
long time period $\Delta t\sim T/\epsilon$. This makes manifest the
importance of the periodicity of $q$ motion. If the motion were not
periodic, then we would conclude
\begin{equation}
\frac{O\left([q_{\mathrm{nonperiodic}}(v+T)-q_{\mathrm{nonperiodic}}(v)]\frac{dt}{dq_{\mathrm{nonperiodic}}}\right)}{T}\sim O(1)\label{eq:nonperiodic}
\end{equation}
leading to a result analogous to eq.~(\ref{eq:naichange}). However,
for periodic motion relevant to the adiabatic invariant construction,
the first term in eq.~(\ref{eq:deltaqeq}) turns into
\begin{equation}
\frac{q(v+T)-q(v)}{\dot{q}}\sim\epsilon T,
\end{equation}
 resulting in
\begin{equation}
\frac{\Delta Q}{Q}\sim O(\epsilon),\label{eq:finalerror}
\end{equation}
which shows that $Q$ is conserved to leading order even on long time
scales over which $\lambda$ changes.

This matches the more subtle analysis of ref.\,\citep{goldstein2002classical}
where the key idea is to first derive the leading order approximation
\begin{equation}
\langle\dot{Q}\rangle_{T}=-\frac{\epsilon\dot{\lambda}(0)}{T}\int_{0}^{T}dt\frac{\partial}{\partial\Omega}\left(\left[\partial_{\lambda}\underline{F}_{1}(q(t),\Omega(t),\lambda(0))\right]_{q=q(\Omega,J,\lambda(0))}\right)\label{eq:Qdotactionangle}
\end{equation}
where $\underline{F}_{1}(q,\Omega,\lambda(0))$ is the generating
function of the canonical transformation to a special set of variables:
action angle variables $(\Omega,Q)$ for the time independent problem
except with the constant parameter $\lambda(0)$ lifted to a function
of time $\lambda(t)$.\footnote{Here $Q$ is the conjugate momentum to the angle variable $\Omega$
in contrast with the usual notation of calling $Q$ as $J$. This
notation change was natural since $Q$ is often used to denote a conserved
charge.} More explicitly, one first solves the phase space symplectic form
preserving map $(q,p;\lambda(0))\leftrightarrow(\Omega,Q;\lambda(0))$
for the time independent problem,\footnote{Symplectic form preservation is a sufficient condition for the construction
of canonical transformations for a time-dependent Hamiltonian, even
if the form-preserving transformation does not lead to any special
simplification in the general time-dependent case: i.e.\,when one
uses the time-independent problem derived action-angle canonical transformations
on the time-dependent system, the action variable $Q$ is not constant,
but the transformation is still canonical.} use this to find $p(q,\Omega;a(0))$ and $Q\left(q,\Omega;a(0)\right)$,
make the replacement $a(0)\rightarrow a(t)$, and solve the differential
equations
\begin{align}
p(q,\Omega;a(t)) & =\partial_{q}\underline{F_{1}}(q,\Omega,a(t)),\\
Q\left(q,\Omega;a(t)\right) & =-\partial_{\Omega}\underline{F_{1}}(q,\Omega,a(t)),
\end{align}
to construct $\underline{F_{1}}(q,\Omega,a(t))$. Because of the action-angle
variable choice, one can show that 
\begin{equation}
\left[\partial_{\lambda}\underline{F}_{1}(q(t),\Omega(t),\lambda(0))\right]_{q=q(\Omega,J,\lambda(0))}\mbox{ and }\frac{\partial}{\partial\Omega}\left(\left[\partial_{\lambda}\underline{F}_{1}(q(t),\Omega(t),\lambda(0))\right]_{q=q(\Omega,J,\lambda(0))}\right),
\end{equation}
are periodic with period $T$ such that eq.\,(\ref{eq:Qdotactionangle})
vanishes.

\section{A perturbative expansion of the Hubble rate time dependence \label{sec:A-perturbative-expansion}}

In this section we will develop a formalism to estimate the time dependence
of the adiabatic energy $V_{m}=V(\phi_{C,\pm})$ by using an expansion
of the inflation potential from leading order behavior at its minimum.
As the time dependence is given by the first few terms of this expansion,
this formalism is broadly applicable to many models of inflation.

\subsection{Integral equation}

If we want to solve for $V_{m}$ as a function of scale factor $a_{\mathrm{slow}}$,
we want to solve 
\begin{equation}
\frac{Q}{2\sqrt{2}a_{\mathrm{slow}}^{3}V_{m}^{3/2}}=\int_{0}^{1}ds\frac{\sqrt{1-s}}{\left[-V'(V_{<}^{-1}(sV_{m}))\right]}+\int_{0}^{1}ds\frac{\sqrt{1-s}}{V'(V_{>}^{-1}(sV_{m}))}\label{eq:simpleintg}
\end{equation}
obtained from eq.\,(\ref{eq:Q in terms of phi_C}) for the potential
at the oscillation maximum $V_{m}$ where
\begin{align}
V_{>}^{-1}(V) & \equiv V^{-1}(V)>\phi_{\mathrm{min}}\\
V_{<}^{-1}(V) & \equiv V^{-1}(V)<\phi_{\mathrm{min}},
\end{align}
and we have shifted the potential such that minimum of the potential
is at $V(\phi_{\mathrm{min}})=0$. For potentials $V(\phi)$ where
the inverse function $V^{-1}(V)$ is simple enough for the integration
to be executed in a closed form (such as monomial potentials), eq.\,(\ref{eq:simpleintg})
is useful. Since $Q$ is a constant, we can take ratios to write
\begin{equation}
\frac{a_{\mathrm{slow}}^{3}V_{m}^{3/2}}{a_{\mathrm{end}}^{3}V_{m}^{3/2}|_{\mathrm{end}}}=\frac{\int_{0}^{1}ds\frac{\sqrt{1-s}}{\left[-V'(V_{<}^{-1}(sV_{m}|_{\mathrm{end}}))\right]}+\int_{0}^{1}ds\frac{\sqrt{1-s}}{V'(V_{>}^{-1}(sV_{m}|_{\mathrm{end}}))}}{\int_{0}^{1}ds\frac{\sqrt{1-s}}{\left[-V'(V_{<}^{-1}(sV_{m}))\right]}+\int_{0}^{1}ds\frac{\sqrt{1-s}}{V'(V_{>}^{-1}(sV_{m}))}}.\label{eq:ratio}
\end{equation}
An interesting information offered by this expression is that even
though $V_{m}$ will generically not scale as a power law with $a_{\mathrm{slow}}$,
there is always some function of $V_{m}$ only that will scale as
$a_{\mathrm{slow}}^{3}/a_{\mathrm{end}}^{3}$.

As an example of using eq.\,(\ref{eq:simpleintg}), consider the
potential of eq.\,(\ref{eq:evenpower}). We can easily work out
\begin{equation}
V_{\gtrless}^{-1}(V)=\pm\left(\frac{(2n)!}{\lambda}V\right)^{\frac{1}{2n}},
\end{equation}
making eq.\,(\ref{eq:simpleintg}) evaluate to
\begin{equation}
\frac{Q}{2\sqrt{2}\left(a_{\mathrm{slow}}\right)^{3}V_{m}^{3/2}}=\frac{\sqrt{\pi}}{2n}\left(\frac{\lambda}{(2n)!}\right)^{-1/(2n)}V_{m}^{-1+\frac{1}{2n}}\frac{\Gamma\left(\frac{1}{2n}\right)}{\Gamma\left(\frac{3}{2}+\frac{1}{2n}\right)},
\end{equation}
which allows us to write by taking the ratio as in eq.\,(\ref{eq:ratio})
a simple expression
\begin{equation}
V_{m}=V_{m}|_{\mathrm{end}}\left(\frac{a_{\mathrm{end}}}{a}\right)^{\frac{6n}{1+n}},\label{eq:Vmscaling}
\end{equation}
matching the result of eq.\,(\ref{eq:evendimres}). In the limit
$n\rightarrow\infty$, the potential is infinitely flat near the minimum,
corresponding to an effectively massless scalar energy density behavior.
Indeed, it is well known that for kination dominated fields, the equation
of state is such that the energy density scales as $a^{-6}$ consistently
with eq.\,(\ref{eq:Vmscaling}).

As a second example, consider eq.\,(\ref{eq:example potential})
for which
\begin{equation}
V_{\gtrless}^{-1}(V)=v_{\phi}\left(1\pm\sqrt{\frac{V}{M^{4}}}\right)^{1/6},\label{eq:Vinv1}
\end{equation}
and
\begin{equation}
V'(\phi)=-12M^{4}\left(1-\left(\frac{\phi}{v_{\phi}}\right)^{6}\right)\left(\frac{\phi^{5}}{v_{\phi}^{6}}\right),\label{eq:vpr}
\end{equation}
which can be inserted into eq.\,(\ref{eq:simpleintg}) to evaluate
$V_{m}$ implicitly in terms of hypergeometric functions. However,
writing the formal expressions are not as illuminating as the original
integral expression of eq.\,(\ref{eq:ratio}) with the substitutions
of Eqs.\,(\ref{eq:Vinv1}) and (\ref{eq:vpr}). This in turn is not
as illuminating as the Taylor expansion derived expression of eq.\,(\ref{eq:Hslow(u) non-quad potential})
whose derivation we turn to next.

\subsection{\label{subsec:Small--expansion}Small $\sqrt{V_{m}}$ expansion}

The equation (\ref{eq:Q in terms of phi_C}) for the adiabatic invariant
can be written as
\begin{equation}
\tilde{Q}(u)=\frac{1}{\sqrt{2}}\sqrt{\tilde{V}_{m}(u)}\oint_{C}\frac{d\phi}{M_{P}\pi}\sqrt{1-\frac{V(\phi)}{m_{\phi}^{2}M_{P}^{2}\tilde{V}_{m}(u)}},\label{eq:adiabatic equation}
\end{equation}
where we defined the dimensionless parameters
\begin{equation}
\tilde{Q}(u)=\frac{Qu^{-3}}{2\pi m_{\phi}M_{P}^{2}},\qquad\text{and}\qquad\tilde{V}_{m}(u)=\frac{V_{m}(u)}{m_{\phi}^{2}M_{P}^{2}}=\frac{3H_{\mathrm{slow}}^{2}(u)}{m_{\phi}^{2}}.
\end{equation}
and $u=a_{\mathrm{slow}}(t)$ is the independent variable of the slow
time dependence. Let us now solve eq.\,(\ref{eq:adiabatic equation})
for $\tilde{V}_{m}(u)$ by expanding the potential about its local
minimum $\phi_{\min}$ in the path $C$ as
\begin{align}
\frac{V(\phi_{\min}+\delta\phi)}{m_{\phi}^{2}M_{P}^{2}} & =\sum_{n=2}\alpha_{n}\left(\frac{\delta\phi}{M_{P}}\right)^{n}\label{eq:taylor}\\
\alpha_{n} & \equiv\frac{M_{P}^{n-2}}{m_{\phi}^{2}n!}\left.\partial_{\phi}^{n}V(\phi)\right|_{\phi=\phi_{\min}}\label{eq:alphadef}
\end{align}
which allows one to compute the integral of eq.\,(\ref{eq:adiabatic equation})
with the bounds of integration $\phi_{C,\pm}$ computed by inverting
$\tilde{V}_{m}=V(\phi_{C,\pm})/(m_{\phi}^{2}M_{P}^{2})$:
\begin{align}
\frac{\phi_{C,\pm}}{M_{P}} & =\frac{\phi_{\min}}{M_{P}}\pm\sqrt{2\tilde{V}_{m}}\left(1\pm\phi_{3}\left(2\tilde{V}_{m}\right)^{\frac{1}{2}}+\phi_{4}\left(2\tilde{V}_{m}\right)^{1}\pm\phi_{5}\left(2\tilde{V}_{m}\right)^{\frac{3}{2}}+\phi_{6}\left(2\tilde{V}_{m}\right)^{2}+\dots\right)\label{eq:phicp1}\\
\phi_{3} & =-\alpha_{3}\\
\phi_{4} & =\frac{5}{2}\alpha_{3}^{2}-\alpha_{4}\\
\phi_{5} & =-8\alpha_{3}^{2}+6\alpha_{3}\alpha_{4}-\alpha_{5}\\
\phi_{6} & =\frac{231}{8}\alpha_{3}^{4}-\frac{63}{2}\alpha_{3}^{2}\alpha_{4}+\frac{7}{2}\alpha_{4}^{2}+7\alpha_{3}\alpha_{5}-\alpha_{6}\\
\phi_{7} & =\dots
\end{align}
which is effectively an expansion in smallness of $\sqrt{\hat{V}}\sim H/m_{\phi}$.

We then parameterize the integral over $\phi$ as
\begin{equation}
\phi=\bar{\phi}(\tilde{V}_{m})+\Delta\phi(\tilde{V}_{m})\cos\Xi,
\end{equation}
where
\begin{align}
\bar{\phi} & =\frac{\phi_{C,+}+\phi_{C,-}}{2}=\phi_{\min}+M_{P}\sqrt{2\tilde{V}_{m}}\left(\phi_{3}\left(2\tilde{V}_{m}\right)^{\frac{1}{2}}+\phi_{5}\left(2\tilde{V}_{m}\right)^{\frac{3}{2}}+\dots\right),\\
\Delta\phi & =\frac{\phi_{C,+}-\phi_{C,-}}{2}=M_{P}\sqrt{2\tilde{V}_{m}}\left(1+\phi_{4}\left(2\tilde{V}_{m}\right)^{1}+\phi_{6}\left(2\tilde{V}_{m}\right)^{2}+\dots\right),
\end{align}
and the integration is over $\Xi\in[0,2\pi]$. Eq.\,(\ref{eq:adiabatic equation})
can then be expressed as
\begin{equation}
\tilde{Q}=\sqrt{\frac{\tilde{V}_{m}}{2}}\frac{\Delta\phi(\tilde{V}_{m})}{M_{P}}\oint\frac{d\left(\cos\Xi\right)}{\pi}\sqrt{1-\frac{V\left(\bar{\phi}(\tilde{V}_{m})+\Delta\phi(\tilde{V}_{m})\cos\Xi\right)}{m_{\phi}^{2}M_{P}^{2}\tilde{V}_{m}}}.
\end{equation}
The first terms of the expansion of the integrand are
\begin{align}
 & \sqrt{1-\frac{V\left(\bar{\phi}(\tilde{V}_{m})+\Delta\phi(\tilde{V}_{m})\cos\Xi\right)}{m_{\phi}^{2}M_{P}^{2}\tilde{V}_{m}}}\\
 & =\left|\sin\Xi\right|\left(1+\left(\sqrt{2}\alpha_{3}\cos\Xi\right)\tilde{V}_{m}^{1/2}+\left(-\alpha_{3}^{2}\left(1+\cos^{2}\Xi\right)+2\alpha_{4}\cos^{2}\Xi\right)\tilde{V}_{m}+\dots\right),
\end{align}
where we have assumed that $\tilde{V}_{m}$ is small enough to avoid
any branch points. This automatically holds given that $V_{m}\geq V(\phi)$
along the entire integration path. The integrand will always be proportional
to $\sqrt{1-\cos^{2}\Xi}$ at each order in the perturbative series
as one can always factor out the roots of $1-V(\phi)/V_{m}$ at $\cos\Xi=\pm1$.
This expansion can be integrated to find
\begin{align}
 & \oint\frac{dx}{\pi}\sqrt{1-\frac{V\left(\bar{\phi}(\tilde{V}_{m})+\Delta\phi(\tilde{V}_{m})x\right)}{m_{\phi}^{2}M_{P}^{2}\tilde{V}_{m}}}\nonumber \\
 & =\frac{2}{\pi}\int_{-1}^{+1}dx\sqrt{1-x^{2}}\left(1+\sqrt{2}\alpha_{3}\tilde{V}_{m}^{1/2}x+\left(-\alpha_{3}^{2}\left(1+x^{2}\right)+2\alpha_{4}x^{2}\right)\tilde{V}_{m}+\dots\right)\\
 & =1+\frac{-5\alpha_{3}^{3}+2\alpha_{4}}{4}\tilde{V}_{m}+\frac{-593\alpha_{3}^{4}+676\alpha_{3}^{2}\alpha_{4}-68\alpha_{4}^{2}-168\alpha_{3}\alpha_{5}+24\alpha_{6}}{16}\tilde{V}_{m}^{2}+\dots
\end{align}
By multiplying this with $\Delta\phi(\tilde{V}_{m})\sqrt{\tilde{V}_{m}/2}$,
we find $\tilde{Q}$ in terms of an expansion in $\tilde{V}_{m}$:
\begin{align}
\tilde{Q} & =\tilde{V}_{m}\left(1+q_{1}\tilde{V}_{m}+q_{2}\tilde{V}_{m}^{2}+q_{3}\tilde{V}_{m}^{3}+q_{4}\tilde{V}_{m}^{4}+\dots\right)\label{eq:begeq}\\
q_{1} & =\frac{3}{4}\left(5\alpha_{3}^{2}-2\alpha_{4}\right)\label{eq:q1eq}\\
q_{2} & =\frac{5}{16}\left(231\alpha_{3}^{4}-252\alpha_{3}^{2}\alpha_{4}+28\alpha_{4}^{2}+56\alpha_{3}\alpha_{5}-8\alpha_{6}\right)\\
q_{3} & =\dots
\end{align}

The inversion of this result gives
\begin{align}
\tilde{V}_{m} & =\tilde{Q}\left(1+v_{1}\tilde{Q}+v_{2}\tilde{Q}^{2}+v_{3}\tilde{Q}^{3}+v_{4}\tilde{Q}^{4}+\dots\right)\label{eq:Vm_in_terms_of_Qtilde}\\
v_{1} & =-q_{1}\\
v_{2} & =2q_{1}^{2}-q_{2}\\
v_{3} & =-5q_{1}^{3}+5q_{1}q_{2}-q_{3}\\
v_{4} & =14q_{1}^{4}-21q_{1}^{2}q_{2}+3q_{2}^{2}+6q_{1}q_{3}-q_{4}\\
v_{5} & =\dots
\end{align}
Using the relation $H_{\mathrm{slow}}=\sqrt{V_{m}/3M_{P}^{2}}$ yields
a perturbative expansion of the Hubble rate as
\begin{align}
\frac{H_{\mathrm{slow}}(u)}{m_{\phi}} & =\sqrt{\frac{\tilde{Q}(u)}{3}}\left(1+h_{1}\tilde{Q}(u)+h_{2}\tilde{Q}^{2}(u)+h_{3}\tilde{Q}^{3}(u)+h_{4}\tilde{Q}^{4}(u)+\dots\right)\\
h_{1} & =\tfrac{1}{2}v_{1}\\
h_{2} & =\tfrac{1}{8}\left(-v_{1}^{2}+4v_{2}\right)\\
h_{3} & =\tfrac{1}{16}\left(v_{1}^{3}-4v_{1}v_{2}+8v_{3}\right)\\
h_{4} & =\tfrac{1}{128}\left(-5v_{1}^{4}+24v_{1}^{2}v_{2}-16v_{2}^{2}-32v_{1}v_{3}+64v_{4}\right)
\end{align}
We can summarize in terms of the potential coefficients as
\begin{align}
\frac{H_{\mathrm{slow}}(u)}{m_{\phi}} & =\sqrt{\frac{\tilde{Q}_{\mathrm{end}}}{3}}\frac{a_{\mathrm{end}}^{3/2}}{u^{3/2}}\left(1+h_{1}\tilde{Q}_{\mathrm{end}}\frac{a_{\mathrm{end}}^{3}}{u^{3}}+h_{2}\tilde{Q}_{\mathrm{end}}^{2}\frac{a_{\mathrm{end}}^{6}}{u^{6}}+\dots\right)\label{eq:Hslow expansion summary}\\
\tilde{Q}_{\mathrm{end}} & =\frac{Qa_{\mathrm{end}}^{-3}}{2\pi m_{\phi}M_{P}^{2}}\sim\frac{3H_{\mathrm{end}}^{3}}{m_{\phi}^{2}}\\
h_{1} & =\frac{3}{8}\left(-5\alpha_{3}^{2}+2\alpha_{4}\right)\\
h_{2} & =\frac{1}{128}\left(-3045\alpha_{3}^{4}+3780\alpha_{3}^{2}\alpha_{4}-308\alpha_{4}^{2}-1120\alpha_{3}\alpha_{5}+160\alpha_{6}\right)\\
h_{3} & =\dots
\end{align}

To gain intuition for this formalism, consider a toy potential of
eq.\,(\ref{eq:cubic}) with $m_{\phi}=m=M$ giving 
\begin{equation}
V=\frac{m^{2}}{2}\phi^{2}-A\phi^{3},\label{eq:cubicpot}
\end{equation}
and $\phi_{\mathrm{min}}=0$. The potential Taylor expansion coefficients
are
\begin{equation}
\alpha_{2}=\frac{1}{2}\hspace{1em}\alpha_{3}=-\frac{AM_{P}}{m^{2}},\label{eq:alpha23forcubic}
\end{equation}
which gives
\begin{equation}
\phi_{C,\pm}=\pm\frac{\sqrt{2V_{m}}}{m}+\frac{2AV_{m}}{m^{4}}+...\label{eq:phicpasymexp}
\end{equation}
consistently with an easily obtainable exact solution.\footnote{The quantity $V_{m}$ is defined in eq.\,(\ref{Vmdef}).}
Note that in situations where $A\phi^{3}$ vanish, $\alpha_{3}$ and
$2V_{m}/m^{3}$ would be absent. This also illustrates that $\alpha_{n}$
is not necessarily small, partly due to the hierarchy between $M_{P}$
and the non-gravitational dynamics scales $m$ and $A$. As long as
the potential is analytic, the Taylor expansion of eq.\,(\ref{eq:taylor})
can be exact, which means that $\alpha_{n}$ has no requirement of
being small. Eq.\,(\ref{eq:Hslow expansion summary}) becomes
\begin{equation}
\frac{H_{\mathrm{slow}}(u)}{m}=\sqrt{\frac{Qa_{\mathrm{end}}^{-3}}{6\pi mM_{P}^{2}}}\frac{a_{\mathrm{end}}^{3/2}}{u^{3/2}}\left(1-\frac{15}{8}\frac{A^{2}Q}{2\pi m^{5}}\frac{a_{\mathrm{end}}^{3}}{u^{3}}-\frac{3045}{128}\left[\frac{A^{2}Q}{2\pi m^{5}}\right]^{2}\frac{a_{\mathrm{end}}^{6}}{u^{6}}+\dots\right),\label{eq:hslowovm}
\end{equation}
where one sees that the corrections to the $u^{-3/2}$ coming from
the cubic interactions do not depend on the $M_{P}/m$ hierarchy of
the intermediate steps coming from $\alpha_{n}$. Although the growing
pure numbers such as $3045/128$ seem alarming, the extra power of
$A^{2}Q/(2\pi m^{5}u^{3})$ (which can be orders of magnitude smaller
than $10^{-1}$ in practice) leads to a suppression, particularly
at large times when $u^{3}$ grows large. For a cosmologically relevant
example, see eq.\,(\ref{eq:suppressionexample}).

\bibliographystyle{JHEP2}
\bibliography{analytic-formalism-particle-prod}

\end{document}